\documentclass[12pt]{article}

\usepackage{textcomp}
\usepackage{chetnew}
\usepackage{amssymb,amsmath,amsfonts,mathrsfs}
\usepackage[utf8]{inputenc}
\usepackage{tikz}
\usetikzlibrary{arrows,decorations.markings}

\usepackage[colorinlistoftodos,textsize=tiny]{todonotes}

\def\one{{\,\hbox{1\kern-.8mm l}}}

\newcommand{\SU}{\mathrm {SU}}
\newcommand{\U}{\mathrm U}
\newcommand{\Tr}{\mathrm{Tr}}

\newcommand{\QQ}{\mathcal Q}
\newcommand{\QS}{\mathcal S}
\allowdisplaybreaks

\def\makeatletter{\catcode`\@=11}
\makeatletter
\def\mathbox#1{\hbox{$\m@th#1$}}%
\def\math@ccstyles#1#2#3#4#5#6#7{{\leavevmode
      \setbox0\mathbox{#6#7}%
      \setbox2\mathbox{#4#5}%
      \dimen@ #3%
      \baselineskip\z@\lineskiplimit#1\lineskip\z@
      \vbox{\ialign{##\crcr
             \hfil \kern #2\box2 \hfil\crcr
             \noalign{\kern\dimen@}%
             \hfil\box0\hfil\crcr}}}}
\def\mathaccstyles{\math@ccstyles\maxdimen}
\def\maththroughstyles{\math@ccstyles{-\maxdimen}}
\def\unity%
 {\maththroughstyles{.45\ht0}\z@\displaystyle {\mathchar"006C}\displaystyle 1}

\preprint{QMUL-PH-17-06\\DESY-17-030}

\title{Exact Deconstruction of the 6D (2,0) Theory}

\author{J.~Hayling$^{1\heartsuit}$, C.~Papageorgakis$^{1\spadesuit}$, E.~Pomoni$^{2\diamondsuit}$ and D.~Rodríguez-Gómez$^{3\clubsuit}$}

\affiliation{\\$^1$ CRST and School of Physics and Astronomy\\ Queen
  Mary University of London, E1 4NS, UK\\ $ $ \\$^2$ DESY Theory Group\\
  Notkestraße 85, 22607 Hamburg, Germany\\ $ $ \\$^3$ Department of Physics,
  University of Oviedo\\Avda. Calvo Sotelo 18, 33007, Oviedo, Spain

\emails{$^{\heartsuit}$j.a.hayling@qmul.ac.uk,
  $^{\spadesuit}$c.papageorgakis@qmul.ac.uk,
  $^{\diamondsuit}$elli.pomoni@desy.de, $^{\clubsuit}$d.rodriguez.gomez@uniovi.es}
}

\abstract{The dimensional-deconstruction prescription of Arkani-Hamed,
  Cohen, Kaplan, Karch and Motl provides a mechanism for recovering
  the $A$-type (2,0) theories on $T^2$, starting from a
  four-dimensional $\mathcal N=2$ circular-quiver theory. We put this
  conjecture to the test using two exact-counting arguments: In the
  decompactification limit, we compare the Higgs-branch Hilbert series
  of the 4D $\mathcal N=2$ quiver to the ``half-BPS'' limit of the
  (2,0) superconformal index. We also compare the full partition function
  for the 4D quiver on $S^4$ to the (2,0) partition function on
  $S^4 \times T^2$. In both cases we find exact agreement. The partition function calculation sets up a dictionary between exact results in 4D and 6D.}

\date{\today}

\begin{document}

\maketitle

\setcounter{tocdepth}{2}

\toc

\section{Introduction and Summary}

Over the last few years, there has been significant progress in the study of six-dimensional (2,0)-superconformal field theories (SCFTs). This progress relied on advances in the exact calculation of protected quantities, such as the superconformal index \cite{Romelsberger:2005eg,Kinney:2005ej,Bhattacharya:2008zy}, using e.g. the method of supersymmetric localisation \cite{Pestun:2007rz}.\footnote{For a recent, comprehensive review of
  supersymmetric localisation applied to diverse setups see
  \cite{Pestun:2016zxk}.} Since the $ADE$ (2,0) theories have no known Lagrangian descriptions, various supersymmetric partition functions have been calculated by appealing to the relationship between 5D maximally-supersymmetric Yang--Mills theory (MSYM) and the 6D (2,0) theory on a circle $S^1_{R_6}$ of radius $R_6 = g_{5}^2/2 \pi$. The prototypical example of such a protected quantity is the supersymmetric partition function of 5D MSYM on $S^5$ \cite{Kallen:2012cs,Kallen:2012va,Kim:2012ava,Kim:2012qf,Kallen:2012zn,Jafferis:2012iv} (or $\mathbb{CP}^2\times S^1$ \cite{Kim:2013nva}), which computes the superconformal index of the (2,0) theory;\footnote{This is despite the
  fact that the 5D MSYM theory is perturbatively non-renormalisable
  \cite{Bern:2012di}, although it is also not sufficient evidence to
  rule in favour of the conjecture of
  \cite{Douglas:2010iu,Lambert:2010iw, Papageorgakis:2014dma}.} for a more complete list of references see \cite{Kim:2016usy}. The (2,0) superconformal index was also recovered in \cite{Lockhart:2012vp,Haghighat:2013gba} using the ``refined topological vertex'' formalism for constructing topological string amplitudes \cite{Aganagic:2003db,Iqbal:2007ii}. Some of these exact tools were subsequently used to help establish a striking connection between BPS subsectors of the (2,0)$_{ADE}$ theory and 2D $\mathcal W_{ADE}$-algebras. By doing so, the authors of \cite{Beem:2014kka} solved for said subsector of the (2,0) theory, since e.g. 3-point functions of associated operators can immediately be obtained from the $\mathcal W$-algebra literature. Finally, a significant complementary approach was initiated in \cite{Beem:2015aoa}, aiming to constrain the (2,0) theory as an abstract SCFT through the conformal-bootstrap programme, which is solely based on the system's symmetries and a minimal set of initial assumptions.

These new developments join some older proposals for attacking the (2,0) theory, most notably employing Discrete Light-Cone Quantisation (DLCQ) \cite{Aharony:1997an} and dimensional deconstruction \cite{ArkaniHamed:2001ie}. In this paper we would like to revisit the latter armed with some modern non-perturbative techniques.\footnote{We would like to thank G.~Moore, T.~Nishinaka and D.~Shih for early collaboration in this direction.} Towards that end, we remind the reader of the work of \cite{ArkaniHamed:2001ie}, where it was postulated that starting from a superconformal $\mathcal N=2$ four-dimensional circular-quiver gauge theory with $\SU(k)$ gauge-group nodes, and upon taking a specific limit of parameters that takes the theory to the Higgs phase, one recovers the corresponding $A_{k-1}$ (2,0) theory on a torus of fixed (but arbitrary) size. Evidence for this claim included estimating the Kaluza--Klein (KK) spectrum of the 6D theory on $T^2$, as well as a string-duality argument where the field theories involved were geometrically engineered using branes. In follow-up work \cite{Lambert:2012qy}, it was confirmed that the circular-quiver theory explicitly deconstructs 5D MSYM on a finite circle $S^1_{R_5}$, for values of the parameters corresponding to weak coupling (for fixed energies), $g_{5}^2 = 2\pi R_6 $, where the theory is well defined.\footnote{It naturally follows that for large values of $R_6$ the UV-finite 4D $\mathcal N=2$ theory
  provides a quantum definition of 5D MSYM through deconstruction \cite{Lambert:2012qy}.}

Although the existing qualitative evidence is highly suggestive, to our knowledge there is no quantitative test for the proposal of \cite{ArkaniHamed:2001ie}. Our main aim in this work is to address precisely this point. We perform a detailed comparison between the part of the circular-quiver spectrum that survives deconstruction, and that of the $(2,0)_k$ theory that captures the low-energy dynamics of $k$ M5-branes on a torus.\footnote{This is the $A_{k-1}$
  (2,0) theory plus an additional free tensor multiplet accounting for
  the centre-of-mass degrees of freedom.} This is carried out via two independent calculations, both of which are only indirectly sensitive to the choice of Higgs VEV:

\begin{itemize}

\item[i.]{4D Hilbert Series and 6D Index}

We calculate the Hilbert Series (HS) \cite{Benvenuti:2006qr} on the mesonic Higgs branch of the circular-quiver theory. After deconstruction this reproduces the ``half-BPS'' limit of the $(2,0)_k$ superconformal index, Eqs.~\eqref{1/2BPSNonAbelian}, \eqref{eq:33}:
\begin{align}
\lim_{N \to\infty }{\rm HS}^N_k  = \prod_{m=1}^k \frac{1}{1-x^m} = \mathcal{I}^{(2,0)_k}_{\frac{1}{2}{\rm BPS}}\,.  \nonumber
\end{align}
In this manner, we compare and match local operators parametrising the 4D $\mathcal N=2$ mesonic Higgs branch with certain half-BPS local operators of the $(2,0)_k$ theory on $\mathbb R^6$, i.e. in the limit where the torus decompactifies.  In the process of doing so, we also find that the half-BPS index in the $(2,0)_k$ theory is the $k$-fold-symmetrised product of the Higgs-branch HS for the 4D abelian theory in the deconstruction limit.

\end{itemize}

The above calculation only matches a simple class of BPS operators. Moreover, it is insensitive to non-local operators in $\mathbb R^6$, which should also have a 4D origin.\footnote{These operators create the selfdual-string solitons of the (2,0) theory \cite{Howe:1997ue}.} We hence perform a second, more refined counting by writing down the full partition function for the 4D circular quiver on $S^4$, identifying the appropriate limits of parameters compatible with the deconstruction procedure of \cite{ArkaniHamed:2001ie}, and finally comparing the result with the full partition function of the (2,0) theory on $ S^4 \times T^2$ \cite{Haghighat:2013gba} to find precise agreement:

\begin{itemize}

\item[ii.]{Exact Partition Functions}

It is well known that both the 4D and 6D partition functions on $S^4$ and $S^4\times T^2$ can be obtained by glueing together two copies of a basic building block, $Z_{\rm 4D/6D}$. The building blocks are IR partition functions on the Coulomb/tensor branch of the 4D/6D theory on the $\Omega$ background, and the glueing is implemented by integrating over the set of Coulomb/tensor-branch parameters \cite{Pestun:2007rz,Lockhart:2012vp}. For our purposes, this implies that we can consider the deconstruction limit directly on these Coulomb/tensor-branch partition functions. We implement this by prescribing specific identifications of 4D/6D parameters in Eqs.~\eqref{H1} \eqref{H2}. Using these identifications we are able to explicitly show that 
\begin{align}
   \int [ da ] \left|Z_{\text{4D}}(\tau, a, m_{\rm bif};\epsilon_1, \epsilon_2)\right|^2 \stackrel{\rm Deconstruction}{\Longrightarrow} \int [ d\tilde{a} ]  \left|Z_{\text{6D}}(\tilde{\tau},\tilde{a}, \tilde{t}_m;\tilde{\epsilon}_1, \tilde{\epsilon}_2)\right|^2\;. \nonumber
\end{align}

\end{itemize}

The quantitative agreement found here provides strong motivation for using the circular quiver---combined with deconstruction---as a tool towards computing observables for the (2,0) theory on $T^2$. In particular, by reproducing the full 6D partition function from 4D we establish a precise dictionary that could be extended to a number of closely-related exact calculations, including e.g. chiral correlators \cite{Baggio:2014sna, Baggio:2015vxa, Gerchkovitz:2016gxx,Baggio:2016skg,Rodriguez-Gomez:2016ijh,Pini:2017ouj}, Wilson loop expectation values \cite{Pestun:2007rz,Mitev:2014yba,Fraser:2015xha}, the radiation emitted by a heavy quark \cite{Fiol:2015spa,Mitev:2015oty}, or other protected subsectors of operators \cite{Bourdier:2015sga}. It would be very interesting to import these results to 6D, and compare where possible with other approaches to the (2,0) theory. Ultimately, one would hope to be able to extend this procedure to non-protected sectors \cite{Pomoni:2013poa}. We leave these possibilities open as directions for future research.

The rest of this paper is organised as follows: We begin in Sec.~\ref{Sec2} with a brief review of the $(2,0)_k$-theory deconstruction from \cite{ArkaniHamed:2001ie}. We then proceed to calculate the Higgs-branch HS and reproduce the half-BPS limit of the $(2,0)_k$ superconformal index in Sec.~\ref{Sec3}. Finally, in Sec.~\ref{Sec4} we present the full partition function for the 4D circular-quiver theory on $S^4$ and give a prescription for how to take the deconstruction limit on the Coulomb-branch IR partition functions. Using this, we recover the full $(2,0)_k$ partition function on $S^4 \times T^2$. We include various derivations and background material in the Appendices.

\section{Review of  Dimensional Deconstruction for the (2,0)  Theory}\label{Sec2}

We commence with a short summary of the results of \cite{ArkaniHamed:2001ie} in order to establish conventions and notation. The starting point is an $N$-noded 4D circular-quiver theory, with $\SU(k)$ gauge groups. The nodes are connected by bifundamental chiral superfields, the scalar components of which are denoted by ${(X_{\alpha+1, \alpha})^{i_{(\alpha+1)}}}_{j_{(\alpha)}}$, ${(X_{\alpha,\alpha+1})^{j_{(\alpha)}}}_{i_{(\alpha+1)}}$; here $\alpha = \lfloor -N/2\rfloor + 1, \ldots , \lfloor N/2\rfloor$ labels the quiver nodes and the (down) up $i_{(\alpha)}$s are (anti)fundamental gauge indices associated with the $\alpha$-th gauge group. The minimal coupling to the gauge fields occurs via
\begin{align}
  \label{eq:30}
D_\mu  {(X_{\alpha+1, \alpha})^{i_{(\alpha+1)}}}_{j_{(\alpha)}} & = \partial_\mu  {(X_{\alpha+1, \alpha})^{i_{(\alpha+1)}}}_{j_{(\alpha)}} - i {(A_\mu^{(\alpha +1)})^{i_{(\alpha+1)}}}_{k_{(\alpha+1)}} {(X_{\alpha+1, \alpha})^{k_{(\alpha+1)}}}_{j_{(\alpha)}}\cr 
& \qquad + i {(X_{\alpha+1, \alpha})^{i_{(\alpha+1)}}}_{l_{(\alpha)}}{(A_\mu^{(\alpha )})^{l_{(\alpha)}}}_{j_{(\alpha)}}\;.
\end{align}
This theory is conformal and enjoys $\mathcal N =2$ supersymmetry. 

The first step towards implementing the deconstruction prescription of \cite{ArkaniHamed:2001ie} is to give a VEV to the scalar fields
\begin{align}
  \label{eq:25}
 \langle {(X_{\alpha,\alpha+1})^{j_{(\alpha)}}}_{i_{(\alpha+1)}} \rangle = \mathrm v \; {\delta ^{j_{(\alpha)}}}_{i_{(\alpha+1)}} \qquad \forall \;\alpha\;.  
\end{align}
This takes the theory onto the Higgs branch and has the effect of breaking the gauge group down to the diagonal subgroup $\SU(k)^N\to\SU(k)$. Consequently, the previously-independent $N-1$ gauge couplings $g^{(\alpha)}$ are replaced by a single coupling parameter denoted by $G$. 

The second step is to consider the  limit
\begin{align}
  \label{eq:1}
  N\to \infty\;,\qquad G\to\infty\;,\qquad {\rm v}\to\infty\;,
\end{align}
in a fashion that keeps 
\begin{align}
  \label{eq:2}
 g_{5}^2 := \frac{G}{{\rm v}}\to \textrm{fixed}\;, \qquad 2 \pi R_5:=\frac{N}{G {\rm v}}\to \textrm{fixed}\;.
\end{align}
For energies small compared to the scale $1/g_{5}^2$, the resultant
theory can be explicitly seen to reproduce 5D MSYM on a continuous
circle of radius $R_5$, with bare gauge coupling $g_5$
\cite{Lambert:2012qy}. Note that supersymmetry is enhanced in the
limit, with 16 preserved supercharges. One
straightforwardly recovers two towers of massive states:
\begin{align}
  \label{eq:3}
  M_{n_1}^2 =  \left(\frac{2 \pi n_1}{R_5}\right)^2\;,\qquad   \widetilde M_{n_{2}}^2 =  \left(\frac{4 \pi^2 n_2}{g_5^2}\right)^2\;.
\end{align}
The first is a tower of KK modes associated with $S^1_{R_5}$, while
the second with the BPS spectrum of $n_2$-instanton-soliton
states. The latter can also be identified with another KK tower when
the bare 5D coupling is related to the radius of an additional circle,
\begin{align}
  \label{eq:5}
g_5^2 = 2 \pi R_6\;.  
\end{align}
This identification is implied by Type IIA/M-theory duality, whence 5D
MSYM is interpreted as the low-energy effective description for the 6D
(2,0) theory on $S^1_{R_6}$. Even though this 5D picture is expected
to break down at high energies, the 4D description is UV complete and valid
for all values of parameters \eqref{eq:2}, therefore bypassing the
issue of non-completeness of 5D MSYM. The 4D $\mathcal N=2$
circular-quiver theory is thus claimed to be deconstructing the (2,0)
theory on a torus $T^2 = S^1_{R_5} \times S^1_{R_6}$ of any
size. It is interesting to observe that the S-duality action on the 4D
theory, which sends $G\leftrightarrow \frac{N}{G}$, also exchanges the
two circles of the $T^2$.

\subsection{Brane Engineering I}\label{Eng1}

\begin{table}[t]
\centering
\begin{tabular}{|c|c|c|c|c|c|c|c|c|c|c|c|}
\hline
&$x^0$ & $x^1$ &$x^2$ &$x^3$ &$x^4$ &$x^5$ &$x^6$ &$x^7$ &$x^8$ &$x^9$ \\
\hline
$k$ D3 branes &$-$&$-$&$-$&$-$&$\cdot$&$\cdot$&$\cdot$&$\cdot$&$\cdot$&$\cdot$\\
\hline
$A_{N-1}$ ALE  &$\cdot$&$\cdot$&$\cdot$&$\cdot$&$\cdot$&$\cdot$&$-$&$-$&$-$&$-$\\
\hline
\end{tabular}
\caption{Brane configuration in type IIB string theory. The ALE space
  extends in the directions $x^6,x^7,x^8,x^9$. The direction
  $x^6$ is compact with periodicity $x^6\to x^6 + 2
  \pi r$.}
\label{configB}
\end{table}

The above picture is reinforced using brane engineering and a chain of dualities. The circular-quiver theory can be obtained at low energies on a stack of $k$ D3 branes probing a $\mathbb C \times\mathbb C^2/\mathbb Z_N$ orbifold singularity. This system is parametrised by the orbifold rank, $N$, the string length, $l_s$, and string coupling, $g_s$. Turning on the chiral-multiplet VEVs corresponds to taking the $k$ D3 branes off the orbifold singularity and into the orbifolded transverse space by a distance $d$. The limit \eqref{eq:2} translates into taking $l_s\to 0$ and $N\to\infty$, while keeping
\begin{align}
  \label{eq:7}
g_s\to\textrm{fixed}\;,\qquad  \frac{d}{N l_s^2} = \frac{1}{R_5}\to\textrm{fixed}\;,\qquad
  \frac{d}{N l_s^2 g_s} = \frac{1}{R_6}\to\textrm{fixed}\;.
\end{align}
One can use the above to straightforwardly deduce the following
relation between the string and gauge-theory parameters
\begin{align}
  \label{eq:6}
  \frac{d}{2 \pi l_s^2} = G {\rm v}\;,\qquad \sqrt{g_s N} = G\;.
\end{align}
In the limit \eqref{eq:7} the geometry probed by the $k$ D3s can be locally approximated by $\mathbb R^5 \times S^1_r$, where $r = d/N = l_s^2/R_5$. The D3s can be T-dualised along this circle to obtain $k$ D4s wrapping $S^1_{R_5}$, with string coupling $g_s' = g_s R_5/l_s$. This becomes strong as $l_s\to 0$ upon which one has $k$ M5 branes on $S^1_{R_5}$, with the M-theory circle being $R_6= g_s' l_s = g_s R_5$. Moreover, in the deconstruction limit the 11D Planck length $l_p =l_s {g'_s}^{1/3}\to 0$ and one recovers $(2,0)_k$, the $A_{k-1}$ theory plus a free-tensor multiplet on $T^2 = S^1_{R_5}\times S^1_{R_6}$ \cite{ArkaniHamed:2001ie}.

Note that the low-energy description for this D-brane system is in terms of $\U(k)$ as opposed to $\SU(k)$ gauge groups. On the one hand, this distinction is unimportant for the procurement of the mass spectrum \eqref{eq:3} as well as for the deconstruction argument reviewed above (since, rather than providing any strong-coupling scale, confinement or any sort of exotic IR phenomena, the $\U(1)$ part is IR free and thus decouples). On the other, there are computations which are sensitive to it. For instance, the IIB brane picture clearly suggests that the IR theory on the Higgs branch should be associated with the untwisted-sector string degrees of freedom, and these can be straightforwardly isolated from the twisted-sector by using the $\U(k)$ version of the theory \cite{Douglas:1996sw}.\footnote{For $\SU(k)$ the blow-up modes resolving the orbifold singularity are moduli, while for the $\U(k)$ these are lifted by the $\U(1)$ D-terms and become parameters.} This observation will be important later on and we will therefore explicitly use the $\U(k)^N$ circular-quiver theory in the calculations that follow \cite{ArkaniHamed:2001ie}.

\subsection{Brane Engineering II}\label{Sec:BEII}

One can also use an alternative description of this brane system, where the T-duality transformation is implemented before taking the branes off the orbifold singularity. This will be more appropriate for our purposes. For the D3-brane probes, the space $\mathbb C^2/\mathbb Z_N$ resolves to the $A_{N-1}$ ALE metric, as summarised in Table~\ref{configB}. This can be T-dualised along the compact direction to give rise to a configuration of $N$ NS5s on the dual circle, with $k$ D4 branes stretched between them, as summarised in Table~\ref{configA} \cite{Ooguri:1995wj,Gregory:1997te,Tong:2002rq,Witten:2009xu}. There is a $\U(k)^N$ gauge theory associated with open strings that end on the D4 segment between two adjacent NS5s (each of which have their own coupling), massive modes coming from open strings stretching between adjacent D4 segments (across an NS5), as well as an overall centre of mass. At low energies the massive modes freeze out (they become parameters) and the remaining degrees of freedom comprise an $\SU(k)^N\times\U(1)$ theory. Rotations along $x^{7,8,9}$ correspond to the $\SU(2)_R$ of the $\mathcal{N}=2$ theory while rotations along the $x^{4,5}$ plane to the $\U(1)_r$ symmetry \cite{Witten:1997sc,Giveon:1997sn}.

\begin{table}[t]
\centering
\begin{tabular}{|c|c|c|c|c|c|c|c|c|c|c|c|c|}
\hline
&$x^0$ & $x^1$ &$x^2$ &$x^3$ &$x^4$ &$x^5$ &$x^6$ &$x^7$ &$x^8$ &$x^9$ &($x^{10}$)\\
\hline
$k$ D4 branes &$-$&$-$&$-$&$-$&$\cdot$&$\cdot$&$-$&$\cdot$&$\cdot$&$\cdot$&$-$\\
\hline
$N$ NS5 branes &$-$&$-$&$-$&$-$&$-$&$-$&$\cdot$&$\cdot$&$\cdot$&$\cdot$&$\cdot$\\
\hline
\end{tabular}
\caption{Brane configuration in type IIA string theory. The direction
  $x^6$ is compact with periodicity $x^6\to x^6 + 2
  \pi R_5$. The coordinate $x^{10}$ parametrises the M-theory circle at
  strong coupling with periodicity $x^{10} \to x^{10} + 2 \pi R_{6}$.}
\label{configA}
\end{table}

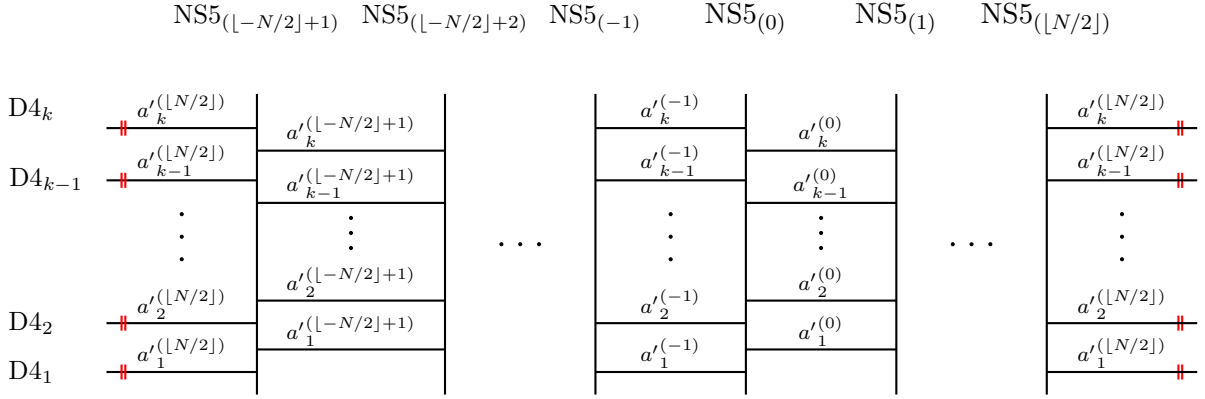
\begin{figure}[t]
\begin{center}
\begin{tikzpicture}
\node at (-3,-3.7) {\footnotesize $\text{D4}_1$};
\node at (-3,-3.05) {\footnotesize $\text{D4}_2$};
\node at (-2.8,-1.15) {\footnotesize $\text{D4}_{k-1}$};
\node at (-3,-0.2) {\footnotesize $\text{D4}_{k}$};
\draw [thick] (-2,-3.7) -- (0,-3.7);
\draw [thick,red] (-1.8,-3.8) -- (-1.8,-3.6);
\draw [thick,red] (-1.75,-3.8) -- (-1.75,-3.6);
\draw [thick,red] (12.3,-3.8) -- (12.3,-3.6);
\draw [thick,red] (12.25,-3.8) -- (12.25,-3.6);
\node at (-1,-3.45) {\scriptsize ${a'}_1^{(\lfloor N/2\rfloor)}$};
\draw [thick] (-2,-3.05) -- (0,-3.05);
\draw [thick,red] (-1.8,-2.95) -- (-1.8,-3.15);
\draw [thick,red] (-1.75,-2.95) -- (-1.75,-3.15);
\draw [thick,red] (12.3,-2.95) -- (12.3,-3.15);
\draw [thick,red] (12.25,-2.95) -- (12.25,-3.15);
\node at (-1,-2.8) {\scriptsize ${a'}_2^{(\lfloor N/2\rfloor)}$};
\draw [fill=black] (-1,-2.2) circle (0.2mm); 
\draw [fill=black] (-1,-1.9) circle (0.2mm); 
\draw [fill=black] (-1,-1.6) circle (0.2mm); 
\draw [thick] (-2,-1.15) -- (0,-1.15);
\draw [thick,red] (-1.8,-1.05) -- (-1.8,-1.25);
\draw [thick,red] (-1.75,-1.05) -- (-1.75,-1.25);
\draw [thick,red] (12.3,-1.05) -- (12.3,-1.25);
\draw [thick,red] (12.25,-1.05) -- (12.25,-1.25);
\node at (-1,-0.9) {\scriptsize ${a'}_{k-1}^{(\lfloor N/2\rfloor)}$};
\draw [thick] (-2,-0.455) -- (0,-0.455);
\draw [thick,red] (-1.8,-0.355) -- (-1.8,-0.555);
\draw [thick,red] (-1.75,-0.355) -- (-1.75,-0.555);
\draw [thick,red] (12.3,-0.355) -- (12.3,-0.555);
\draw [thick,red] (12.25,-0.355) -- (12.25,-0.555);
\node at (-1,-0.2) {\scriptsize ${a'}_{k}^{(\lfloor N/2\rfloor)}$};
\node at (0,1) {\footnotesize $\text{NS5}_{(\lfloor -N/2\rfloor+1)}$};
\draw [thick] (0,0) -- (0,-4);
\draw [thick] (0,-3.4) -- (2.5,-3.4);
\node at (1.25,-3.15) {\scriptsize ${a'}_1^{(\lfloor -N/2\rfloor+1)}$};
\draw [thick] (0,-2.75) -- (2.5,-2.75);
\node at (1.25,-2.5) {\scriptsize ${a'}_2^{(\lfloor -N/2\rfloor+1)}$};
\draw [fill=black] (1.25,-2.05) circle (0.2mm); 
\draw [fill=black] (1.25,-1.85) circle (0.2mm); 
\draw [fill=black] (1.25,-1.65) circle (0.2mm); 
\draw [thick] (0,-1.45) -- (2.5,-1.45);
\node at (1.25,-1.2) {\scriptsize ${a'}_{k-1}^{(\lfloor -N/2\rfloor+1)}$};
\draw [thick] (0,-0.755) -- (2.5,-0.755);
\node at (1.25,-0.5) {\scriptsize ${a'}_{k}^{(\lfloor -N/2\rfloor+1)}$};
\node at (2.5,1) {\footnotesize $\text{NS5}_{(\lfloor -N/2\rfloor+2)}$};
\draw [thick] (2.5,0) -- (2.5,-4);
\draw [fill=black] (3.25,-2) circle (0.2mm); 
\draw [fill=black] (3.5,-2) circle (0.2mm); 
\draw [fill=black] (3.75,-2) circle (0.2mm); 
\node at (4.5,1) {\footnotesize $\text{NS5}_{(-1)}$};
\draw [thick] (4.5,0) -- (4.5,-4);
\draw [thick] (4.5,-3.7) -- (6.5,-3.7);
\node at (5.5,-3.45) {\scriptsize ${a'}_1^{(-1)}$};
\draw [thick] (4.5,-3.05) -- (6.5,-3.05);
\node at (5.5,-2.8) {\scriptsize ${a'}_2^{(-1)}$};
\draw [fill=black] (5.5,-2.2) circle (0.2mm); 
\draw [fill=black] (5.5,-1.9) circle (0.2mm); 
\draw [fill=black] (5.5,-1.6) circle (0.2mm); 
\draw [thick] (4.5,-1.15) -- (6.5,-1.15);
\node at (5.5,-0.9) {\scriptsize ${a'}_{k-1}^{(-1)}$};
\draw [thick] (4.5,-0.455) -- (6.5,-0.455);
\node at (5.5,-0.2) {\scriptsize ${a'}_{k}^{(-1)}$};
\node at (6.5,1) {\small $\text{NS5}_{(0)}$};
\draw [thick] (6.5,0) -- (6.5,-4);
\draw [thick] (6.5,-3.4) -- (8.5,-3.4);
\node at (7.5,-3.15) {\scriptsize ${a'}_1^{(0)}$};
\draw [thick] (6.5,-2.75) -- (8.5,-2.75);
\node at (7.5,-2.5) {\scriptsize ${a'}_2^{(0)}$};
\draw [fill=black] (7.5,-2.05) circle (0.2mm); 
\draw [fill=black] (7.5,-1.85) circle (0.2mm); 
\draw [fill=black] (7.5,-1.65) circle (0.2mm); 
\draw [thick] (6.5,-1.45) -- (8.5,-1.45);
\node at (7.5,-1.2) {\scriptsize ${a'}_{k-1}^{(0)}$};
\draw [thick] (6.5,-0.755) -- (8.5,-0.755);
\node at (7.5,-0.5) {\scriptsize ${a'}_{k}^{(0)}$};
\draw [thick] (8.5,0) -- (8.5,-4);
\node at (8.5,1) {\small $\text{NS5}_{(1)}$};
\draw [fill=black] (9.25,-2) circle (0.2mm); 
\draw [fill=black] (9.5,-2) circle (0.2mm); 
\draw [fill=black] (9.75,-2) circle (0.2mm); 
\node at (10.5,1) {\small $\text{NS5}_{(\lfloor N/2\rfloor)}$};
\draw [thick] (10.5,0) -- (10.5,-4);
\draw [thick] (10.5,-3.7) -- (12.5,-3.7);
\node at (11.5,-3.45) {\scriptsize ${a'}_1^{(\lfloor N/2\rfloor)}$};
\draw [thick] (10.5,-3.05) -- (12.5,-3.05);
\node at (11.5,-2.8) {\scriptsize ${a'}_2^{(\lfloor N/2\rfloor)}$};
\draw [fill=black] (11.5,-2.2) circle (0.2mm); 
\draw [fill=black] (11.5,-1.9) circle (0.2mm); 
\draw [fill=black] (11.5,-1.6) circle (0.2mm); 
\draw [thick] (10.5,-1.15) -- (12.5,-1.15);
\node at (11.5,-0.9) {\scriptsize ${a'}_{k-1}^{(\lfloor N/2\rfloor)}$};
\draw [thick] (10.5,-0.455) -- (12.5,-0.455);
\node at (11.5,-0.2) {\scriptsize ${a'}_{k}^{(\lfloor N/2\rfloor)}$};
\end{tikzpicture}
\end{center}
\caption{The brane set up for the $\U(k)^N$ circular-quiver
  theory. The vertical lines represent NS5 branes while  the horizontal lines represent D4 branes. The cyclic identification of the end points is indicated by the double red line on the D4 branes.}\label{fig:branediagsu(k)N}
\end{figure}

The gauge theory possesses several deformations, which can be
appropriately encoded in terms of distances in the geometric
interpretation. In order to parametrise the D4/NS5 brane setup it is
useful to define the complex coordinates $v$ and $s$ as follows:
\begin{align}
  v = x^4 + i x^5 \quad {\mbox{and}} \quad s = x^6 + i x^{10} \, .
\end{align}
Using these variables, the separation of D4 branes can be measured
along $v$. Distances in this coordinate, associated with the $\U(1)_r$
symmetry, translate into Coulomb parameters $a_b^{(\alpha)}$, as well as
bifundamental masses $m^{(\alpha)}_{\rm bif}$. The Coulomb-branch moduli are straightforwardly
encoded as the distances between a colour-D4-brane position and the centre-of-mass position of the colour branes within a single gauge-group factor; see Fig.~\ref{fig:branediagsu(k)N}:
\begin{align}
  \label{eq:8}
a^{(\alpha)}_b  := \frac{1}{2  \pi l_s^2}\left({a'}^{(\alpha)}_b- \frac{1}{k}\sum_{c=1}^k {a'}^{(\alpha)}_c \right)\, .  
\end{align}
The only mass parameters are those of the bifundamental hypermultiplets that capture the relative centre-of-mass positions of two consecutive D4 stacks.\footnote{In $\U(k)^N$ gauge-theory language this translates into the relative $\U(1)$ of the two consecutive colour groups \cite{Witten:1997sc}.} For bifundamental fields---fundamental under the gauge group on the left ($(\alpha-1)$-th gauge group) and anti-fundamental under the gauge group on the right ($\alpha$-th gauge group)---the bifundamental mass is given by

\begin{align}
\label{biff-def}
{m}^{(\alpha)}_{\rm bif} :=  \frac{1}{2  \pi l_s^2}\left( \frac{1}{k}\sum_{b=1}^k {a'}^{(\alpha+1)}_b  -  \frac{1}{k}\sum_{b=1}^k {a'}^{(\alpha)}_b\right) \, .
\end{align}

In a similar manner, the coordinate $s$ measures the distance between NS5
branes.  This encodes the couplings for the $\alpha$-th D4
segment
\begin{align}
  \label{eq:9}
 \tau^{(\alpha)}=\frac{4 \pi i}{g^{(\alpha)2}}+\frac{\theta^{(\alpha)}}{8\pi^2} \;.
\end{align}
Due to the
periodic nature of the M-theory circle, $x^{10}$, it is also natural to
introduce the exponentiated coordinate $t = e^{-\frac{s}{R_{6}}}$,
where $R_{6}$ is the radius of M-theory circle
\cite{Witten:1997sc}. In the $t$ coordinate the distance between two
NS5 branes is given by
\begin{align}
  \label{eq:4}
 \mathbf{q}_{(\alpha)}=e^{2 i \pi \tau^{(\alpha)}}  \;.
\end{align} 

For the purposes of deconstruction we are interested in the ``maximally-Higgsed'' phase, where all chiral multiplets acquire a VEV and the gauge group gets broken to $\U(k)^N \to\U(k)$. In the absence of bifundamental masses, the corresponding classical-brane picture is in terms of coincident endpoints for all adjacent D4 segments. At this point of moduli space, where the Coulomb and Higgs branches meet, the brane segments can reconnect to form a single collection of $k$ D4s wrapping the circle $S^1_{R_5}$, with a single coupling $G$. These can then be moved off the NS5s in the $x^{7,8,9}$ directions. In the limit \eqref{eq:7} the resultant configuration leads once again to the $(2,0)_k$ theory on the same $T^2 = S^1_{R_5}\times S^1_{R_6}$ as in Sec.~\ref{Eng1}. We will use this picture of deconstruction as our starting point for the calculation of the partition functions in Sec.~\ref{Sec4}.

\section{4D/6D Matching: Higgs-Branch Hilbert Series and the (2,0) Index}\label{Sec3}

Having established the proposal of \cite{ArkaniHamed:2001ie}, we now set out to test it using exact methods. We have already discussed in Sec.~\ref{Sec2} how the 4D $\mathcal N=2$ circular-quiver theory deconstructs the (2,0) theory on a torus of fixed (but arbitrary) size. Nevertheless, a special limit of parameters in Eq.~\eqref{eq:2} can be considered, such that the radii $R_5, R_6\to\infty$ and the torus decompactifies. The deconstruction limit then relates two fixed-point theories in the UV (4D) and the IR (6D) (at $\mathrm v \to\infty$) and hence their local operator spectra, up to sectors of the 4D theory decoupling at low energies and large $N$. A quantitative diagnostic of this claim can be performed by setting up a counting problem for local operators on both sides. For instance, one could try and quantify this relationship by comparing appropriate limits of superconformal indices following the procedure of \cite{Gaiotto:2012xa}, but generalising that approach to the maximally-Higgsed phase of the circular-quiver theory proves difficult.

For this reason, we will focus on very special classes of BPS operators using the superconformal algebra (SCA) as a guide and analysing the embedding of the $\mathcal N =2$ 4D SCA into the 6D (2,0) SCA. This will lead in an identification between a set of simple half-BPS primary operators parametrising the 4D mesonic Higgs branch in the deconstruction limit, and operators in a 6D half-BPS ring. As we will explain in detail below, we will perform the 4D counting using the Higgs-branch Hilbert Series for the $\U(k)^N$ theory, while the 6D counting via the ``half-BPS'' limit of the superconformal index \cite{Bhattacharyya:2007sa,Kim:2012tr}.

\subsection{SCA Analysis}

The (2,0) 6D superconformal algebra is denoted by $\mathfrak{osp}(8^*|4)$. Primaries in the corresponding modules are in one-to-one correspondence with highest weights in irreducible representations of the maximal compact subalgebra $\mathfrak{so}(6)\oplus\mathfrak{so}(2)\oplus\mathfrak{so}(5)_R \subset\mathfrak{osp}(8^*|4)$. They are thus labelled by the eigenvalues of the respective Cartan generators: the conformal dimension $\Delta$, $\mathfrak{so(6)}$ Lorentz quantum numbers in the orthogonal basis $h_i$, and $R$-symmetry quantum numbers in the orthogonal basis $J_i$. There are also fermionic generators: sixteen Poincar\'e and superconformal supercharges, denoted by $\QQ_{\mathbf{A} a}$ and $\QS_{\mathbf{A} \dot a}$, where $\dot a ,a=1,\ldots,4$ are (anti)fundamental indices of $\mathfrak{su}(4)$ and $\bf{A} =1,\ldots,4$ a spinor index of $\mathfrak{so}(5)_R$; see e.g. \cite{Minwalla:1997ka,Beem:2014kka,Buican:2016hpb} for conventions and notation.

Similarly, the $\mathcal N=2$ 4D superconformal algebra is denoted by $\mathfrak{su}(2,2|2)$. Primaries in the corresponding modules are in one-to-one correspondence with highest weights in irreducible representations of the maximal compact subalgebra $\mathfrak{so}(4)\oplus\mathfrak{so}(2)\oplus \mathfrak{su}(2)_R\oplus \mathfrak{u}(1)_r \subset\mathfrak{su}(2,2|2)$. They are labelled by the eigenvalues of the respective Cartans: their conformal dimension $\Delta$, $\mathfrak{su}(2)\oplus\mathfrak{su}(2)\simeq\mathfrak{so(4)}$ Lorentz quantum numbers $m_i$, and $R$-symmetry quantum numbers $R,r$. There are also fermionic generators: eight Poincar\'e and superconformal supercharges,  $Q^{I}_{\alpha}$, $\widetilde{Q}_{I \dot\alpha }$ and $S^{\dot \alpha}_I$, $\widetilde{S}^{I\dot\alpha}$, where $\dot \alpha ,\alpha=\pm$ and $I=1,2$ are fundamental indices of $\mathfrak{su}(2)\oplus\mathfrak{su}(2)$ and $\mathfrak{su}(2)_R$ respectively. The 4D $\mathcal N = 2$ SCA is a subalgebra of the (2,0) 6D SCA. The 4D Cartans are related to the 6D ones through
\begin{align}
R=J_1,&\qquad r=J_2\,,\cr
m_1=\frac{1}{2}(h_2+h_3),&\qquad m_2=\frac{1}{2}(h_2-h_3)\, .
\end{align}
We choose an embedding of the 4D SCA into the 6D SCA, which relates
the supercharges as in Table~\ref{Qs} \cite{Beem:2014kka}.

\begin{table}[t]
\centering
\begin{tabular}{|c | c |c |c| c|| c|}
\hline 6D Supercharge & $(h_1,h_2,h_3)$ & $(m_1,m_2)$ & $R$ & $r$ & 4D Supercharge \\ \hline \hline 
$\mathcal{Q}_{\mathbf{1}1}$ & $(+,+,+)$ & $(+,0)$ & $+$ & $+$ &{\small $Q_+^1$ }\\ \hline
$\mathcal{Q}_{\mathbf{2}1}$ & $(+,+,+)$ & $(+,0)$ & $+$ & $-$ &  \\ \hline
$\mathcal{Q}_{\mathbf{3}1}$ & $(+,+,+)$ & $(+,0)$ & $-$ & $+$ &{\small $Q_+^2$} \\ \hline
$\mathcal{Q}_{\mathbf{4}1}$ & $(+,+,+)$ & $(+,0)$ & $-$ & $-$ &  \\ \hline \hline
$\mathcal{Q}_{\mathbf{1}2}$ & $(+,-,-)$ & $(-,0)$ & $+$ & $+$ &{\small $Q_-^1$} \\ \hline
$\mathcal{Q}_{\mathbf{2}2}$ & $(+,-,-)$ & $(-,0)$ & $+$ & $-$ &  \\ \hline
$\mathcal{Q}_{\mathbf{3}2}$ & $(+,-,-)$ & $(-,0)$ & $-$ & $+$ & {\small $Q_-^2$ }\\ \hline
$\mathcal{Q}_{\mathbf{4}2}$ & $(+,-,-)$ & $(-,0)$ & $-$ & $-$ &  \\ \hline \hline
$\mathcal{Q}_{\mathbf{1}3}$ & $(-,+,-)$ & $(0,+)$ & $+$ & $+$ & \\ \hline
$\mathcal{Q}_{\mathbf{2}3}$ & $(-,+,-)$ & $(0,+)$ & $+$ & $-$ & {\small $\widetilde{Q}_{2\dot{+}}$} \\ \hline
$\mathcal{Q}_{\mathbf{3}3}$ & $(-,+,-)$ & $(0,+)$ & $-$ & $+$ & \\ \hline
$\mathcal{Q}_{\mathbf{4}3}$ & $(-,+,-)$ & $(0,+)$ & $-$ & $-$ & {\small$\widetilde{Q}_{1\dot{+}}$ } \\ \hline \hline
$\mathcal{Q}_{\mathbf{1}4}$ & $(-,-,+)$ & $(0,-)$ & $+$ & $+$ & \\ \hline
$\mathcal{Q}_{\mathbf{2}4}$ & $(-,-,+)$ & $(0,-)$ & $+$ & $-$ &{\small $\widetilde{Q}_{2\dot{-}}$}  \\ \hline
$\mathcal{Q}_{\mathbf{3}4}$ & $(-,-,+)$ & $(0,-)$ & $-$ & $+$ & \\ \hline
$\mathcal{Q}_{\mathbf{4}4}$ & $(-,-,+)$ & $(0,-)$ & $-$ & $-$ & {\small $\widetilde{Q}_{1\dot{-}}$} \\ \hline 
\end{tabular}
\caption{Summary of supercharge quantum numbers in 6D and the 4D
  embedding. All orthogonal-basis quantum numbers have magnitude one half. The four-dimensional subalgebra acts on the $h_2$ and $h_3$ planes.}
\label{Qs}
\end{table}

The (2,0) SCA contains a special class of half-BPS short multiplets, denoted by $\mathcal D[0,0,0;J_1-J_2,0]$, for which $\Delta = 2 (J_1-J_2)$ and where the superconformal primary is annihilated by the set of Poincaré supercharges $\mathcal{Q}_{\mathbf{1}a}, \mathcal{Q}_{\mathbf{2}a}$ in addition to all the superconformal $\QS_{{\bf A} \dot a}$ \cite{Minwalla:1997ka,Beem:2014kka,Buican:2016hpb}. By inspecting Table~\ref{Qs}, one sees that $\mathcal{Q}_{\mathbf{1}1}, \mathcal{Q}_{\mathbf{1}2}$ and $\mathcal{Q}_{\mathbf{2}3}, \mathcal{Q}_{\mathbf{2}4}$ can be identified with certain supercharges $Q^1_\alpha,\widetilde{Q}_{2\dot{\alpha}}$ for a 4D $\mathcal N=2$ subalgebra, which can in turn be interpreted as the SCA for the 4D quiver.

Recall now that in four dimensions, operators $\mathcal{H}_{\mathcal I}$ satisfying
\begin{align}
[Q^1_{\alpha},\mathcal{H}_{\mathcal I}]  =0\;,\qquad[\widetilde{Q}_{2\dot{\alpha}},\mathcal{H}_{\mathcal I}]  =0\,
\end{align}
form a (non-freely-generated) ring that parametrises the Higgs branch of the theory; see e.g. \cite{Gerchkovitz:2016gxx}. In the notation of \cite{Dolan:2002zh} these are in the $\hat{\mathcal B}_R$ multiplet with shortening condition $\Delta = 2R$. Therefore, the 4D Higgs-branch operators $\mathcal H_{\mathcal I}$ are good candidates for reproducing the 6D half-BPS superconformal primaries in the deconstruction limit. It is these two classes of operators that we will be counting.

\subsection{The Half-BPS Index of the  6D (2,0) Theory}

For a choice of defining supercharge, $\QQ_{{\bf 4}4}$, the 6D (2,0)
superconformal index is given by the quantity \cite{Bhattacharya:2008zy,Kim:2012tr}
\begin{align}
  I &= \Tr(-1)^F e^{- \beta \delta}x^{\Delta + J_1}y_1^{h_1-h_2}
  y_2^{h_2+h_3} q^{h_1 + h_2 -h_3 -3 J_2}\cr
\delta & = \{\QQ_{{\bf 4}4}, \QS_{{\bf 1}\dot 4}\} = \Delta - h_1-h_2
         + h_3 + 2 J_1 + 2 J_2 \;,
\end{align}
where the trace is taken over the Hilbert space of the theory in
radial quantisation, the $x,y_1,y_2,q$ are a maximal set of fugacities
taking values inside the unit circle, such that e.g. $|x|<1$, and
their exponents are the already-defined Cartan generators of the (2,0)
SCA \cite{Bhattacharya:2008zy,Kim:2012tr}.  The superconformal index
receives contributions from operators in short representations of the
superconformal algebra with $\delta =0$, modulo combinations of short
representations that can pair up to form long representations.

Let us first consider the abelian case. The index of the free-tensor
multiplet in 6D---denoted in \cite{Beem:2014kka,Buican:2016hpb}
as $\mathcal D [0,0,0; 1,0]$---can be straightforwardly calculated
using letter counting and gives\footnote{Recall that the plethystic exponential is defined
  as
  $ {\rm PE}[f(\mathbf{x})]=\exp{\left[{\sum_{n=1}^{\infty}}
    \frac{f(\mathbf{x}^n)}{n}\right]}$, where $\mathbf{x}$ corresponds to a
  certain list of fugacities. The inverse operation is called the
  plethystic logarithm and given by
  ${\rm PL}[f(\mathbf{x})]=\sum_{n=1}^{\infty}
  \frac{\mu(n)}{n}\log{[f(\mathbf{x}^n)]}$, where $\mu(n)$ is the
  Möbius function.}
\begin{align} 
  \mathcal{I}^{(2,0)_1}={\rm
    PE}[f],\qquad f=\frac{x+x^2q^3-x^2q^2(y_1^{-1}+y_1y_2^{-1}+y_2)+x^3q^3}{(1-xqy_1)(1-xqy_1y_2^{-1})(1-xqy_2^{-1})}\,.
\end{align}

The index for the interacting (2,0) theory is known in closed form only in certain limits of fugacities; see e.g. \cite{Kim:2013nva} and also \cite{Beem:2014kka} for an alternative calculation using the associated $\mathcal W$ algebra. A particularly-simple such limit is obtained when $q\rightarrow 0$, whence the only operators counted are the half-BPS primaries
of the $\mathcal D [0,0,0; J_1-J_2,0]$ multiplets. The latter admit a free-field realisation in terms of $\mathcal D [0,0,0; 1,0]$ for which
$\lim_{q\to 0} f = x$.  One subsequently has for the ``half-BPS'' index in the (2,0) theory of $k$ M5 branes (including the c.o.m. free-tensor multiplet)
\cite{Bhattacharyya:2007sa,Kim:2012tr}
\begin{align}
\label{1/2BPSNonAbelian}
\mathcal{I}^{(2,0)_k}_{\frac{1}{2}{\rm BPS}}={\rm
    PE}\left[\sum_{m=1}^k x^{m} \right]= \prod_{m=1}^k \frac{1}{1-x^m}\,.  
\end{align}
Note that this is not the same as the Schur limit of the index, which also counts derivatives acting on the primaries.

Alternatively, the answer can be obtained as the coefficient of $\nu^k$ in the expansion of the generating function ${\rm PE}[\nu x]$, where $\nu$ an arbitrary parameter. A simple manipulation shows that
\begin{align}\label{kfold}
{\rm PE}[\mathcal{I}^{(2,0)_1}_{\frac{1}{2}{\rm BPS}}\nu  ] = \prod_{n=0}^{\infty} \frac{1}{1-x^n\nu}=\frac{1}{(\nu,x)}=\sum_{k=0}^{\infty} \frac{\nu^k}{(x,x)_k}\,
\end{align}
and the coefficient of $\nu^k$ is
\begin{align}\label{threeseven}
\frac{1}{(x,x)_k}=\prod_{n=1}^{k}\frac{1}{1-x^n}=\mathcal{I}^{(2,0)_k}_{\frac{1}{2}{\rm BPS}}\,.
\end{align}
This method reproduces the $k$-fold-symmetrised product of the
quantity multiplying $\nu$ on the left-hand-side of \eqref{kfold}. We
have thus recovered the half-BPS index of the interacting
theory from the symmetrised product of the half-BPS index for the
free-tensor theory.

\subsection{Higgs-Branch Hilbert Series for Circular Quivers}

We next proceed to the counting of the primary operators $\mathcal H _{\mathcal I}$ on the Higgs branch of the 4D circular quiver. These operators can be thought of as holomorphic functions on the Higgs-branch moduli space, and their counting is naturally accomplished by the Higgs-branch Hilbert Series (HS), which we next briefly review \cite{Benvenuti:2006qr,Benvenuti:2010pq}. To our knowledge, there is no limit of the 4D index for circular quivers that only receives contributions from such operators, although the Higgs-branch HS is closely related to the Hall-Littlewood limit of the 4D index. The Hall-Littlewood index counts the $\hat{\mathcal B}_R$, $\overline{\mathcal D}_{R(m_1,0)}$ (or $\mathcal D_{R(0,m_2)}$)-type multiplets, while the Higgs-branch HS only counts $\hat{\mathcal B}_R$ \cite{Gadde:2011uv,Gaiotto:2012uq}.

Typically, the classical moduli space of a (Lagrangian) $\mathcal{N}=2$ Quantum Field Theory contains a Coulomb branch (where vector multiplet scalars acquire VEVs), a Higgs branch (where hypermultiplet scalars acquire VEVs) and, possibly, a number of mixed Coulomb--Higgs branches. While on the Coulomb branch there is some residual gauge symmetry, on the Higgs branch the gauge group is usually completely broken. The Higgs branch $\mathscr{H}$ does not receive quantum corrections \cite{Argyres:1996eh} and can be described as a hyper-Kähler quotient by $D$ and $F$ terms.

Given the (holomorphic) manifold $\mathscr{H}$, we denote its
coordinate ring by
$\mathbb{C}[\mathscr{H}]=\bigoplus_i \mathcal{H}^{(0,i)}$, where the
elements of $\mathcal{H}^{(0,i)}$ are homogeneous, degree-$i$
holomorphic polynomials. Then, the Hilbert Series is defined as the
generating functional ${\rm HS}=\sum_i{\rm
  dim}(\mathcal{H}^{(0,i)})\,t^i$. Note that further refinements with respect to additional global symmetries can also be introduced. The construction of $\mathscr{H}$ through a hyper-Kähler quotient of the hypermultiplet space implies in particular that the holomorphic functions $\mathcal{H}_{\mathcal I}\in \bigcup_i \mathcal{H}^{(0,i)}$ are in one-to-one correspondence with gauge-invariant primary operators made out of the hypermultiplet fields. Therefore, the counting of the latter produces the HS of $\mathscr{H}$.

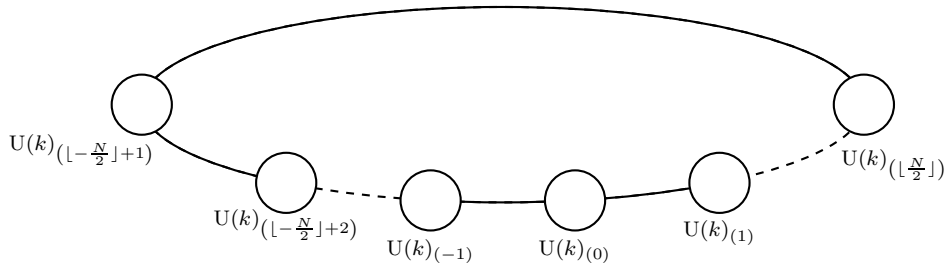
\begin{figure}[t]
\begin{center}
\begin{tikzpicture}
\draw [thick] (0,0) ellipse (4.8cm and 1.3cm);
\fill[white] (-2.88,-1.3) rectangle (-0.96,-1.04);
\fill[white] (2.88,0) rectangle (4.8,-1.04);
\draw  [thick,dashed] (0,0) ellipse (4.8cm and 1.3cm);
\draw  [thick, fill=white] (-4.8,0) circle (0.4cm); 
\node at (-5.6,-0.6) {\scriptsize $\U(k)_{\left(\lfloor-\frac{N}{2}\rfloor+1\right)}$}; 
\draw  [thick, fill=white] (-2.88,-1.04) circle (0.4cm); 
\node at (-2.88,-1.6) {\scriptsize $\U(k)_{\left(\lfloor-\frac{N}{2}\rfloor+2\right)}$}; 
\draw  [thick, fill=white] (-0.96,-1.27373) circle (0.4cm);
\node at (-0.96,-1.95) {\scriptsize $\U(k)_{(-1)}$};
\draw  [thick, fill=white] (0.96,-1.27373) circle (0.4cm);
\node at (0.96,-1.95) {\scriptsize $\U(k)_{(0)}$};
\draw  [thick, fill=white] (2.88,-1.04) circle (0.4cm);
\node at (2.88,-1.7) {\scriptsize $\U(k)_{(1)}$};
\draw  [thick, fill=white] (4.8,0) circle (0.4cm); 
\node at (5.2,-0.8) {\scriptsize $\U(k)_{\left(\lfloor \frac{N}{2}\rfloor\right)}$}; 
\end{tikzpicture}
\end{center}
\caption{The quiver diagram corresponding to the system of D4 and NS5 branes.}\label{fig: circquiv}
\end{figure}

With this in mind, we proceed to evaluate the Higgs-branch HS for the circular quiver theory with $\U(k)$ gauge symmetry at each of the $N$ nodes; c.f. Fig.~\ref{fig: circquiv}. In 4D $\mathcal{N}=1$ notation, each node $\alpha=\lfloor - N/2 \rfloor +1,\ldots, \lfloor N/2 \rfloor $ is associated with an adjoint scalar $X_{\alpha, \alpha}$. The hypermultiplets linking the $\alpha^{\rm th}$ and $(\alpha+1)^{\rm th}$ nodes contain bifundamental scalars $X_{\alpha, \alpha+1}$, $X_{\alpha+1, \alpha}$. The superpotential for this theory is given by
\begin{align}
  W=\sum_{\alpha=\lfloor - \frac{N}{2}\rfloor +1}^{\lfloor \frac{N}{2}\rfloor } \rm{tr}\Big(X_{\alpha+1, \alpha} X_{\alpha, \alpha}X_{\alpha, \alpha+1}-X_{\alpha, \alpha-1} X_{\alpha, \alpha}X_{\alpha-1, \alpha}\Big)\, , 
\end{align}
where $\rm{tr}$ denotes the trace over the gauge group. Note that each
chiral multiplet scalar is charged under a baryonic $\U(1)_{\rm B}$ symmetry,
such that $X_{\alpha, \alpha+1}$ has charge +1 and $X_{\alpha+1, \alpha}$ charge -1.

When the gauge group is completely broken, the Higgs-branch HS is typically computed through ``letter counting'' \cite{Benvenuti:2010pq}: On the Higgs branch all vector-multiplet scalars are set to zero and one is left with $F$ terms arising from the derivative of the superpotential $W$ with respect to Coulomb-branch scalars. The $\mathcal{H}_{\mathcal I}$ operators consist of all symmetrised, gauge-invariant combinations (words) made out of the hypermultiplet fields (letters) modulo these $F$ terms. To count them it is sufficient to consider a partition function over the plethystic exponential of the ``single-letter'' contribution. The $F$ terms are also taken into account but since they act as constraints they appear with opposite-sign coefficients. Taking the result, $g$, and projecting to gauge singlets through integration over the appropriate Haar measure $d\mu$ for a gauge group $G$ leads to the HS:
\begin{align} \label{letcount}
{\rm HS}=\int_G d\mu\,g\,.  
\end{align}

Let us now illustrate this procedure with the simple example of the abelian $k=1$ $\U(1)^N$ theory. Generically, the $k=1$ HS will depend on the following complex parameters: a fugacity $t$ keeping track of  operator scaling dimensions, fugacities for global symmetries (which we will set to 1 since the index \eqref{1/2BPSNonAbelian} only keeps track of scaling dimensions) and $k$ gauge fugacities $u_{i}$. The $t$ fugacities are complex parameters that take values inside the unit circle, that is $|t|<1$, while the gauge and possible flavour fugacities lie on the unit circle.

Let us take e.g.  $N=2$ where the superpotential is explicitly given by 
\begin{align}  
W=X_{2,1}^jX_{1,1}X_{1,2}^i\epsilon_{ij}+X_{1,2}^jX_{2,2}X_{2,1}^i\epsilon_{ij} \, .  
\end{align}
The relevant $F$ terms are $X_{1,2}^iX_{2,1}^j\epsilon_{ij}=0$ and
$X_{2,1}^iX_{1,2}^j\epsilon_{ij}=0$, explicitly
\begin{align}
X_{1,2}^1X_{2,1}^2=X_{1,2}^2X_{2,1}^1\;,\qquad X_{2,1}^1X_{1,2}^2=X_{2,1}^2X_{1,2}^2\, ,
\end{align}
and are obviously identical. This argument generalises to the $N$-noded case, where instead of $N$ $F$ terms, one only has $N-1$ conditions due to the circular nature of the quiver. Since the adjoint character for $k=1$ is equal to one, the $F$-term constraints can easily be accounted for.  Thus, in this case, we have that the letter counting of \eqref{letcount} should be modified into
\begin{align}
{\rm HS}_{k=1}^N={\rm PE}\left[-(N-1)t^2\right]\int \prod_{\alpha=\lfloor -\frac{N}{2}\rfloor+1}^{\lfloor \frac{N}{2}\rfloor}\frac{du_\alpha}{u_\alpha}{\rm PE}\left[ t \Big(\frac{u_\alpha}{u_{\alpha+1}}+\frac{u_{\alpha+1}}{u_{\alpha}}\Big)\right]\, ,
\end{align}
where $u_{\lfloor N/2\rfloor +1}\equiv u_{\lfloor -N/2\rfloor +1}$ from the cyclic identification, the factor outside the integral captures the $F$-term contributions, while the one under the integral the scalar-field contributions. After performing the gauge integrations one has
\begin{align}\label{anyN}
{\rm HS}_{k=1}^N={\rm PE}[t^2+2t^N-t^{2N}]\,.
\end{align}

The above is just the HS of the space $\mathbb{C}^2/\mathbb{Z}_N$ defined as the surface $\mathbb X \mathbb Y=\mathbb W^N$ embedded in $\mathbb{C}^3$. This is consistent with the fact that the Higgs branch of these quiver theories engineers the moduli space of instantons on ALE space \cite{Douglas:1996sw}. In order to show this explicitly, note that the generators $\mathbb X$, $\mathbb Y$ and $\mathbb W$ have respective weights $N$, $N$ and $2$, such that the defining equation is homogeneous of degree $2N$. From this it can also be deduced that the HS is ${\rm PE}[t^2+2t^N-t^{2N}]$ \cite{Benvenuti:2010pq}. In terms of operators, the $t^2$ term corresponds to the meson $\mathbb W=X_{\alpha, \alpha+1}X_{\alpha+1, \alpha}$---they are all equal due to the $F$ terms---while the $2t^N$ correspond to the two long mesons $\mathbb X=X_{1,2}X_{2,3}\ldots X_{N,1}$ and $\mathbb Y=X_{1,N}X_{N,N-1}\ldots X_{2,1}$.\footnote{The operators that
  parametrise the Higgs branch can be constructed out of these
  generators, such as a meson that ``connects'' the nodes 1 and 3,
  $\mathcal H_{13}=X_{1,2}X_{2,3}X_{3,2} X_{2,1}$ with contribution
  $t^4$, a meson that ``connects'' the nodes 1 and 4,
  $\mathcal H_{14}=X_{1,2}X_{2,3}X_{3,4}X_{4,3} X_{3,2} X_{2,1}$ with
  contribution $t^6$, and so on.} These clearly satisfy $\mathbb  X\mathbb Y=\mathbb W^N$.

This result is also to be expected on physical grounds. Upon considering adding $N_\alpha$ flavours to the $\alpha$-th node, the Higgs branch of the circular quiver with generic ranks $k_\alpha$ engineers the moduli space of $\U(\sum N_\alpha)$ instantons on $\mathbb{C}^2/\mathbb{Z}_N$ (see \textit{e.g.} \cite{Dey:2013fea} for the details of this identification). For $N_\alpha=0$, $k_\alpha=k=1$, the so-called ``rank-zero'' instanton only has position moduli and we thus recover the HS of the target-space algebraic variety $\mathbb{C}^2/\mathbb{Z}_N$. Moreover, we can borrow the computations from the appendix of \cite{Dey:2013fea} to provide a holographic interpretation of this result in terms of the quantisation of the phase space of dual giant gravitons in ${\rm AdS}_5\times S^5/\mathbb{Z}_N$.

Since $|t|<1$, it is now straightforward to take the large-$N$ limit of \eqref{anyN}. By relating $t^2=x$ we obtain
\begin{align}\label{threefourteen}
{\rm HS}_{k=1}^{\infty}=\frac{1}{1-x} = {\rm PE}\left[\lim_{q\to 0} f\right] = \mathcal{I}^{(2,0)_1}_{\frac{1}{2}{\rm BPS}}\,,
\end{align}
which is the half-BPS index for the 6D free-tensor theory. Note that a 6D scalar has scaling dimension two, while a 4D scalar has dimension one; this accounts for the redefinition $t^2=x$. 

Coming back to the general non-abelian case, it is clear that an obstruction to the direct use of letter counting will arise because of $F$ terms. However, extracting  modifications to the single-letter evaluation of higher-rank theories becomes increasingly complicated. Nevertheless, regarding the $\U(k)^N$ quiver as describing $k$ rank-zero instantons on $\mathbb{C}^2/\mathbb{Z}_N$ strongly suggests that  one can obtain the general formula for all $k$ and $N$ by simply considering the $k$-fold-symmetrised product of the $k=1$ case \cite{Benvenuti:2006qr,Dey:2013fea}. This is because rank-zero instantons do not have internal degrees of freedom and thus are akin to a gas of $k$ non-interacting particles on $\mathbb{C}^2/\mathbb{Z}_N$.\footnote{This expectation can be checked explicitly by  brute-force calculation for the first few values of $k,\,N$, as we show in App.~\ref{AppD}.} This can be implemented by considering the coefficient of $\nu^k$ in the expansion of
\begin{align}
{\rm PE}\left[{\rm HS}_{k=1}^{N}(t)\nu\right]={\rm PE}\left[\frac{(1-t^{2N})}{(1-t^2)(1-t^N)^2}\nu\right]\, .
\end{align}
Taking the large-$N$ limit  gives back the coefficient of $\nu^k$ in the expansion of 
\begin{align}
{\rm PE}\left[\frac{\nu}{(1-t^2)}\right]\, .
\end{align}
By setting $t^2 = x$, this expression is precisely \eqref{kfold} and one reproduces the superconformal-index result
\begin{align}
  \label{eq:33}
 \lim_{N \to\infty }{\rm HS}^N_k  = \prod_{m=1}^k \frac{1}{1-x^m} \;.
\end{align}
We have thus confirmed the expectation that, through the deconstruction procedure, the mesonic part of the 4D BPS-operator spectrum properly accounts for the appropriate piece of the 6D (2,0) spectrum.

\section{4D/6D Matching: Partition Functions}\label{Sec4}

While the matching that we have just performed is satisfying, it provides but a very simple test of the deconstruction proposal for a collection of protected states. Moreover, since it only involves 6D BPS local operators it is not sensitive to selfdual strings, which should be part of the spectrum and can e.g. wrap a torus of finite size. We will next rectify both of these drawbacks by performing a more sophisticated counting on both sides.

We thus switch gears and turn our attention to the deconstruction of the full partition function for the $(2,0)_k$ theory on $S^4\times T^2$. Putting the superconformal circular quiver on $S^4$ initially lifts the moduli space through the conformal coupling of the scalars to the curvature, thus naively posing a problem for deconstruction. However, this can be easily rectified by appropriately mass-deforming the theory \cite{Kim:2014kta}. With this in mind, our starting point will be the BPS partition function in the Coulomb-branch of the mass-deformed 4D $\mathcal{N}=2$ circular-quiver theory on $\mathbb{R}_{\epsilon_1, \epsilon_2}^4$. The latter can be used as a building block for the full 4D partition function on the ellipsoid $S^4_{\epsilon_1,\epsilon_2}$, upon taking two copies and integrating over the Coulomb-branch parameters with the appropriate Haar measure.\footnote{This IR building block counts BPS
  particles (monopoles, dyons, etc.) at low energies on the Coulomb
  branch and is the well-known 4D limit of the ``Nekrasov partition
  function'', in the definition of which we include the classical,
  one-loop and instanton contributions \cite{Nekrasov:2003rj,
    Nekrasov:2002qd}.} Such a construction is typical of
(Coulomb-branch) supersymmetric localisation
\cite{Pestun:2007rz,Pestun:2016zxk}; see also \cite{Alday:2009aq,
  Hama:2012bg}.\footnote{It is straightforward to
  convince oneself that the IR partition functions for the 
    mass-deformed $\SU(k)^N \times \U(1)$ circular quiver and the
   undeformed $\U(k)^N$ circular quiver on
  $\mathbb R^4_{\epsilon_1,\epsilon_2}$ are identical up to $\U(1)$
  vector-multiplet factors.}

We propose that the deconstruction limit of \cite{ArkaniHamed:2001ie} be taken directly on these building blocks. Appropriate implementation of this procedure leads to an analogous BPS building block for the 6D partition function on $\mathbb{R}^4_{\tilde{\epsilon}_1,\tilde{\epsilon}_2}\times T^2_{R_5, R_6}$ \cite{Haghighat:2013gba}. Taking two copies of this result and integrating them over the Coulomb-branch parameters with the appropriate Haar measure leads to the desired answer.

\subsection{The 4D Coulomb-Branch Partition Function}

We denote the IR partition function of the mass-deformed 4D $\mathcal N=2$ theory on $\mathbb R_{\epsilon_1,\epsilon_2}$ by $Z_{\rm 4D}(\tau, a,m_{\rm bif};\epsilon_1 , \epsilon_2)$. We use the symbol $a= \{a^{(\alpha)}_b\}$ for the set of Coulomb-branch parameters associated with the brane separations of Fig.~\ref{fig:branediagsu(k)N}, with the index $\alpha = \lfloor -N/2\rfloor + 1, ..., \lfloor N/2\rfloor$ counting the number of colour groups and $b = 1,...,k$ the Coulomb-branch parameters within a given colour group. Moreover, we denote the set of $N$ bifundamental-hypermultiplet masses as $m_{\rm bif} = \{m_{\rm bif}^{(\alpha)} \}$; we will keep these generic for the moment.  The partition function $Z_{\rm 4D}(\tau, a,m_{\rm bif};\epsilon_1 , \epsilon_2)$ is the holomorphic half of the integrand of the partition function on the ellipsoid $S^4_{\epsilon_1,\epsilon_2}$: 
\begin{align} 
  Z_{S^4_{\epsilon_1,\epsilon_2}}(\tau, \bar \tau, m_{\rm bif};\epsilon_1, \epsilon_2) = \int [ da ] \left| Z_{\rm 4D}(\tau, a,m_{\rm bif};\epsilon_1 , \epsilon_2) \right|^2\;, 
\end{align}
where complex conjugation is implemented by:
\begin{align}
  \label{eq:31}
\overline{ Z_{\rm 4D}}(\tau , a,m_{\rm bif};\epsilon_1 , \epsilon_2) := Z_{\rm 4D} (\bar \tau, -a,-m_{\rm bif};-\epsilon_1 , -\epsilon_2)\;.  
\end{align}

The Coulomb-branch partition function $Z_{\rm 4D}(a,m_{\rm bif} ;\epsilon_1 , \epsilon_2)$ comprises of a classical, a one-loop and a non-pertur\-bative piece, all of which can be constructed using existing results from the localisation literature. However, they are also known to arise from the circle reduction of the 5D Nekrasov partition function. In turn, the 5D one-loop and instanton pieces can be calculated using the refined topological vertex formalism \cite{Iqbal:2002we, Aganagic:2003db, Awata:2005fa, Iqbal:2007ii}, the technology of which we employ in the independent rederivation of our expressions in App.~\ref{AppC}.\footnote{The classical contribution in the calculation of the IR partition function cannot be recovered by dimensionally reducing the $\mathbb R^4_{\epsilon_1,\epsilon_2} \times S^1$ result obtained through the topological vertex formalism \cite{Iqbal:2012xm,Nieri:2013vba,Mitev:2014jza}. However, it is explicitly known from the localisation calculation of \cite{Pestun:2007rz}. In 5D it can be restored using symmetries, e.g. by imposing fibre-base duality \cite{Mitev:2014jza}.} Our conventions are given in App.~\ref{AppA}. 

Following \cite{Alday:2009aq,Hama:2012bg}, the classical piece of the Coulomb-branch partition function comprises of the factor
\begin{align}
  \label{eq:26}
  Z_{\text{4D,cl}}(\tau, a;\epsilon_1 , \epsilon_2) = \exp \left[ - \frac{2\pi i }{\epsilon_1 \epsilon_2} \sum_{\alpha =-\lfloor\frac{N}{2}\rfloor+1}^{\lfloor\frac{N}{2}\rfloor} \tau^{(\alpha)} \sum_{b=1}^k \left( a_b^{(\alpha)} \right)^2 \right]\;.
\end{align}
The one-loop part is obtained by appropriately assigning vector and hypermultiplet contributions for the quiver of Fig.~\ref{fig: circquiv}, 
\begin{align}
 Z_{\text{4D,1-loop}}( a,m_{\rm bif};\epsilon_1 , \epsilon_2) =\prod_{\alpha =\lfloor -\frac{N}{2}\rfloor+1}^{\lfloor \frac{N}{2}\rfloor }  Z_{\text{1-loop}}^{\text{vec}} (a ;\epsilon_1 , \epsilon_2)  
 Z_{\text{1-loop}}^{\text{bif}}(a,m_{\rm bif}\; ;\epsilon_1 , \epsilon_2)\;,
\end{align}
with the vector and  hypermultiplet one-loop determinant contributions given by\footnote{Note that, compared to \cite{Alday:2009aq,Hama:2012bg}, the vector multiplet piece includes additional factors arising from the Cartan contributions which we explicitly keep. Furthermore, by virtue of making contact with the topological vertex formalism we have shifted the one-loop contributions by various factors of $\epsilon_{1,2}$. See Appendix B.2 of \cite{Alday:2009aq} for details.}
\begin{align}
Z^{\text{bif}}_{\text{1-loop}}(a,m_{\rm bif};\epsilon_1 , \epsilon_2) &=\prod_{b,c=1}^k\Gamma_2\left(a_b^{(\alpha)}-a_c^{(\alpha+1)} -m_{\rm bif}^{(\alpha)}+\frac{\epsilon_+}{2}\big|\epsilon_1,\epsilon_2\right)\;,\cr
Z^{\text{vec}}_{\text{1-loop}}(a;\epsilon_1 , \epsilon_2)&=\prod_{b,c=1}^k \Gamma_2\left(a_b^{(\alpha)}-a_c^{(\alpha)}\big|\epsilon_1,\epsilon_2\right)^{-1}\;,
\end{align}
where $\epsilon_+ = \epsilon_1 + \epsilon_2$ and $\Gamma_2(x|\epsilon_1,\epsilon_2) $ is the Barnes double-Gamma function, given in its product representation in \eqref{barnes doubGamma}.

Similarly, the instanton part is given by \cite{Alday:2009aq}
\begin{align}
\label{eq:instantonPF}
 Z_{\text{4D,inst}}(\tau, a,m_{\rm bif };\epsilon_1 , \epsilon_2) = \sum_{\nu} \prod_{\alpha =\lfloor -\frac{N}{2}\rfloor+1}^{\lfloor \frac{N}{2}\rfloor }   \mathbf{q}_{(\alpha)}^{\sum_{b=1}^k| \nu^{(\alpha)}_{b} |} 
   Z_{\text{inst}}^{\text{vec}} (a; \nu )  
  Z_{\text{inst}}^{\text{bif}} (a,m_{\rm bif}; \nu )
\end{align}
where $\mathbf{q}_{(\alpha)}$ is the fugacity keeping track of the instanton number, associated with the complexified coupling $\tau^{(\alpha)}$ for the $\alpha^{th}$ colour group as in \eqref{eq:4}. A  Young diagram $\nu^{(\alpha)}_{b}$ appears for each of the Coulomb moduli  $a^{(\alpha)}_{b} $, collectively denoted by $\nu =\{\nu_b^{(\alpha)}\}$, while the instanton number is given in terms of the total number of boxes of the Young diagram $| \nu_b^{(\alpha)} |$.
The vector and hypermultiplet instanton contributions respectively read
\begin{align}
  Z_{\text{inst}}^{\text{bif}} (\tau, a,m_{\rm bif}; \nu)&=\prod_{b,c=1}^k{\rm N}_{\nu_{c}^{(\alpha+1)}\nu_{b}^{(\alpha)}}\left(a_b^{(\alpha)} -a^{(\alpha+1)}_c -m_{\rm bif}^{(\alpha)}-\frac{\epsilon_+}{2}\right)\;,\cr
 Z_{\text{inst}}^{\text{vec}} (a,m_{\rm bif}; \nu )&=\prod_{b,c=1}^k{\rm N}_{\nu_{c}^{(\alpha)}\nu_{b}^{(\alpha)}}\left(a^{(\alpha)}_b -a^{(\alpha)}_c\right)^{-1}\;,
\end{align}
where the functions $\textrm{N}_{\lambda\mu}(a)$ involved above are defined as\footnote{See App.~\ref{sec:B.1} for our conventions on labelling partitions.}
\begin{align}\label{Nfunctions}
\textrm{N}_{\lambda\mu}(a)=\prod_{(i,j)\in \lambda}\left[a+\epsilon_1(\lambda_i-j+1)+\epsilon_2(i-\mu_j^t)\right]
 \prod_{(i,j)\in \mu}\left[a+\epsilon_1(j-\mu_i)+\epsilon_2(\lambda_j^t -i+1)\right]\;.
\end{align}
At this stage we will find it convenient to introduce the following change of variables
\begin{align}\label{eq:rels}
\mathbb{F}^{(\alpha)}_{bc}:= a^{(\alpha)}_{c} -a^{(\alpha)}_{b} \;,\qquad m_b^{(\alpha)} := a_b^{(\alpha+1)} - a_b^{(\alpha)} + m_{\rm bif}^{(\alpha)}\;.
\end{align}
We then show in App.~\ref{AppC} that the full Coulomb-branch partition function can be rewritten as
\begin{align}
\label{eq:4DQuiverPF}
Z_{\rm 4D}&=Z_{\text{4D,cl}}  Z_{\text{4D,1-loop}} Z_{\text{4D,inst}}\;
\end{align}
where 
\begin{align}\label{eq:4DQuiverPFpertinst}
&  Z_{\text{4D,cl}} = \exp \left[ - \frac{2\pi i }{\epsilon_1 \epsilon_2} \sum_{\alpha =-\lfloor\frac{N}{2}\rfloor+1}^{\lfloor\frac{N}{2}\rfloor} \tau^{(\alpha)} \sum_{b=1}^k \left( a_b^{(\alpha)} \right)^2 \right]\;,\cr
& Z_{\text{4D,1-loop}} =\prod_{i,j=1}^{\infty}\prod_{\alpha =\lfloor -\frac{N}{2}\rfloor+1}^{\lfloor \frac{N}{2}\rfloor }\prod_{1\leq b\leq c\leq k}\frac{\left[-\mathbb{F}_{bc}^{(\alpha)}-m_c^{(\alpha)} +\epsilon_1(j-1/2)-\epsilon_2(i-1/2)\right]}{\left[-\mathbb{F}_{bc}^{(\alpha+1)} +\epsilon_1 i-\epsilon_2 (j-1)\right]}\cr 
&\qquad \qquad\times\prod_{1\leq b<c\leq k}\frac{\left[-\mathbb{F}_{bc}^{(\alpha)} +m_b^{(\alpha)} +\epsilon_1(j-1/2)-\epsilon_2(i-1/2)\right]}{\left[-\mathbb{F}_{bc}^{(\alpha)} +\epsilon_1(i-1)-\epsilon_2 j\right]}\;,\cr
 & Z_{\text{4D,inst}} =\sum_{\nu}\prod_{\alpha =\lfloor -\frac{N}{2}\rfloor+1}^{\lfloor \frac{N}{2}\rfloor }  \mathbf{q}_{(\alpha)}^{\sum_{b=1}^k|\nu_{b}^{(\alpha)}|} \prod_{1\leq b\leq c\leq k}\frac{{\rm N}_{\nu_{c}^{(\alpha+1)}\nu_{b}^{(\alpha)}}\left(-\mathbb{F}_{bc}^{(\alpha)} - m^{(\alpha)}_c -\frac{\epsilon_+}{2}\right)}{{\rm N}_{\nu_{c}^{(\alpha+1)}\nu_{b}^{(\alpha+1)}}\left(-\mathbb{F}_{bc}^{(\alpha+1)} \right)}\cr &\qquad \qquad \times  \prod_{1\leq b<c\leq k}\frac{{\rm N}_{\nu_{c}^{(\alpha)}\nu_{b}^{(\alpha+1)}}\left(-\mathbb{F}_{bc}^{(\alpha)} + m^{(\alpha)}_b -\frac{\epsilon_+}{2}\right)}{{\rm N}_{\nu_{c}^{(\alpha)}\nu_{b}^{(\alpha)}}\left(-\mathbb{F}_{bc}^{(\alpha)} -\epsilon_+\right)}\;.
\end{align}

This is the result for the Coulomb-branch partition function onto which we will perform the deconstruction limit. Before we proceed however, we will give the answer for the equivalent building block of the 6D partition function. 

\subsection{The 6D Tensor-Branch Partition Function}

The IR partition function for the $(2,0)_k$ theory on $\mathbb R^4_{\tilde{\epsilon}_1, \tilde{\epsilon}_2} \times T^2$ was given in \cite{Iqbal:2012xm,Haghighat:2013gba}. The key point for our purposes is that the BPS sector of the $(2,0)_k$ theory on the (twisted) torus $T_{R_5, R_6}^2 = S^1_{R_5}\times S^1_{R_6}$, can be captured by the (mass-deformed) 5D MSYM theory on $\mathbb R^4_{\tilde{\epsilon}_1, \tilde{\epsilon}_2} \times S^1_{R_5}$ \cite{Douglas:2010iu,Lambert:2010iw,Lockhart:2012vp,Kim:2012ava,Kim:2012qf}. The partition function associated with these BPS states is precisely given by the original calculation of \cite{Nekrasov:2002qd, Nekrasov:2003rj} and, as already mentioned, can be straightforwardly reproduced by the refined topological vertex formalism.\footnote{The 6D (2,0) theory has no Coulomb branch, since the role of the vector multiplet is played by a tensor multiplet. However, since the 6D result is recovered by a purely 5D calculation, both the Coulomb- and tensor-branch nomenclatures are appropriate.} As in 4D, taking two copies of this building block and integrating over the Coulomb/tensor-branch parameters $\tilde{a} = \{\tilde{a}_b\}$ leads to the full partition function of the 6D theory on $S^4_{\tilde{\epsilon}_1,\tilde{ \epsilon}_2} \times T_{R_5, R_6}^2$, that is
\begin{align}
  \label{eq:11}
Z_{S^4_{\tilde{\epsilon}_1,\tilde{\epsilon}_2}\times T_{R_5, R_6}^2}(\tilde{\tau}, \bar {\tilde{\tau}}, \tilde{t}_m;\tilde{\epsilon}_1,\tilde{\epsilon}_2) = \int [ d\tilde{a} ] \left| Z_{\rm 6D}(\widetilde{Q}_\tau, \widetilde{\mathbb Q}_{bc},\widetilde{Q}_m;\tilde{\mathfrak q},\tilde{ \mathfrak t}) \right|^2\;,   
\end{align}
where
\begin{align}
  \label{eq:12}
\overline{ Z_{\rm 6D}}(\widetilde{Q}_\tau, \widetilde{\mathbb Q}_{bc},\widetilde{Q}_m;\tilde{\mathfrak q},\tilde{ \mathfrak t}) := Z_{\rm 6D}(\overline{\widetilde{Q}}_\tau, \widetilde{\mathbb Q}_{bc}^{-1},\widetilde{Q}_m^{-1};\tilde{\mathfrak q}^{-1},\tilde{ \mathfrak t}^{-1})\;.
\end{align}
The fugacities appearing in the above expressions are related to chemical potentials through 
\begin{align}\label{chempot}
\tilde{\mathfrak{q}}= e^{-i R_5 \tilde{\epsilon}_1}\;,\quad
\tilde{\mathfrak{t}}= e^{ iR_5\tilde{ \epsilon}_2}\;,\quad
\widetilde{\mathbb{Q}}_{bc}= e^{iR_5 \tilde{t}_{bc}}\;,\quad
  \widetilde{Q}_m= e^{iR_5 \tilde{t}_m}\;,\quad \widetilde{Q}_\tau=
  e^{2\pi i\tilde{\tau}} \;,
\end{align}
which can in turn be readily identified with 6D geometric parameters as follows \cite{Haghighat:2013gba}:
\begin{align}
  \tilde{t}_m= \frac{\tilde{\tau} \tilde{m}}{R_5}\;,\quad \tilde{t}_{bc}= \tilde{a}_c-\tilde{a}_b= \frac{1}{2 \pi l_p^3} i R_6 \delta_{bc}\;.
\end{align}
Here $R_5, R_6$ are the radii of the circles, $\tilde{\tau}$ the modulus of $T^2$, the $\delta_{bc}$ denote the distance between the $b^{th}$ and $c^{th}$ separated M5 branes in the broken phase ($b,c = 1,...,k$), $\tilde{m}$ and $\tilde{\epsilon}_1,\tilde{\epsilon}_2$ are twists of the torus along $R_5$ and $R_6$ respectively, and $l_p$ is the 11D Planck length.

The 6D holomorphic block comprises of a classical, a one-loop and a non-perturbative piece
\begin{align}\label{6Dfull}
Z_{\rm 6D} = Z_{\text{6D,cl}} Z_{\text{6D, 1-loop}}Z_{\text{6D,inst}}\;,
\end{align}
with \cite{Kim:2012qf, Haghighat:2013gba}
\begin{align}\label{6Dexplicit}
& Z_{\text{6D,cl}} =\exp \left[ - \frac{2\pi i \tilde{\tau} }{\tilde \epsilon_1 \tilde \epsilon_2}  \left( \sum_{b=1}^k   \tilde{a}_b^2  \right) \right] \;,\cr
& Z_{\text{6D,1-loop}} =\prod_{i,j=1}^{\infty}\prod_{1\leq b\leq c\leq k} \frac{\left(1-\widetilde{\mathbb{Q}}_{b c}\widetilde{Q}_{m}\tilde{\mathfrak{q}}^{j-\frac{1}{2}}\tilde{\mathfrak{t}}^{i-\frac{1}{2}}\right)}{\left(1-\widetilde{\mathbb{Q}}_{b c}\tilde{\mathfrak{q}}^{i}\tilde{\mathfrak{t}}^{j-1}\right)}\prod_{1\leq b<c\leq k}\frac{\left(1-\widetilde{Q}_{m}^{-1}\widetilde{ \mathbb{Q}}_{b c}\tilde{\mathfrak{q}}^{j-\frac{1}{2}}\tilde{\mathfrak{t}}^{i-\frac{1}{2}}\right)}{\left(1-\widetilde{\mathbb{Q}}_{b c}\tilde{\mathfrak{q}}^{i-1}\tilde{\mathfrak{t}}^{j}\right)}\;,\cr 
& Z_{\text{6D,inst}}=\sum_{\nu}\widetilde{Q}_\tau^{\sum_{b=1}^k|\nu_b|}\left(\widetilde{Q}_m\sqrt{\frac{\tilde{\mathfrak{t}}}{\tilde{\mathfrak{q}}}}\right)^{-\sum_{b=1}^k|\nu_b|} \prod_{1\leq b\leq c \leq k} \frac{\mathcal{N}_{\nu_c \nu_b}\left(\widetilde{\mathbb{Q}}_{bc}\widetilde{Q}_m \sqrt{\frac{\tilde{\mathfrak{t}}}{\tilde{\mathfrak{q}}}};\tilde{\mathfrak{t}},\tilde{\mathfrak{q}}\right)}{\mathcal{N}_{\nu_c \nu_b}\left(\widetilde{\mathbb{Q}}_{bc};\tilde{\mathfrak{t}},\tilde{\mathfrak{q}}\right)}\cr 
 &\qquad\qquad\times\prod_{1\leq b<c \leq k}\frac{\mathcal{N}_{\nu_c \nu_b}\left(\widetilde{\mathbb{Q}}_{bc}\widetilde{Q}_m^{-1} \sqrt{\frac{\tilde{\mathfrak{t}}}{\tilde{\mathfrak{q}}}};\tilde{\mathfrak{t}},\tilde{\mathfrak{q}}\right)}{\mathcal{N}_{\nu_c \nu_b}\left(\widetilde{\mathbb{Q}}_{bc}\frac{\tilde{\mathfrak{t}}}{\tilde{\mathfrak{q}}};\tilde{\mathfrak{t}},\tilde{\mathfrak{q}}\right)}\;.
\end{align}
The functions $\mathcal{N}_{\lambda\mu}(Q;\mathfrak{t},\mathfrak{q})$ are defined as 
\begin{align}\label{mathcalN}
\mathcal{N}_{\lambda \mu}(Q;\mathfrak{t},\mathfrak{q}) = \left(Q\sqrt{\frac{\mathfrak{q}}{\mathfrak{t}}}\right)^{\frac{|\lambda|+|\mu|}{2}} \mathfrak{t}^{\frac{\|\mu^t\|^2-\|\lambda^t\|^2}{4}}\mathfrak{q}^{\frac{\|\lambda\|^2-\|\mu\|^2}{4}} \rm{N}^\beta_{\lambda\mu} (-\ell;\tilde{\epsilon}_1,\tilde{\epsilon}_2)\;,
\end{align}
for some generic $Q = e^{\beta \ell}$, with the ``Nekrasov functions'' $\rm N^\beta_{\lambda\mu}$ given by 
\begin{align}
\textrm{N}^{\beta}_{\lambda\mu}(-\ell;\epsilon_1,\epsilon_2)=\prod_{(i,j)\in \lambda}2 \sinh \frac{\beta}{2}\left[-\ell+\epsilon_1(\lambda_i-j+1)+\epsilon_2(i-\mu_j^t)\right]
\cr\times \prod_{(i,j)\in \mu}2 \sinh \frac{\beta}{2}\left[-\ell+\epsilon_1(j-\mu_i)+\epsilon_2(\lambda_j^t -i+1)\right]\;.
\end{align} 
We provide a complete derivation of the one-loop and instanton contributions in Eq.~\eqref{6Dfull} using the refined topological vertex formalism in App.~\ref{AppB}.

\subsection{Deconstructing the 6D Partition Function}

Equipped with both the 4D and 6D expressions, we can finally proceed with our prescription for implementing the dimensional-deconstruction limit at the level of IR partition functions.

\subsubsection{The One-Loop Piece}

Let us begin with the one-loop piece of Eq.~\eqref{eq:4DQuiverPFpertinst}:
\begin{align}
Z_{\text{4D,1-loop}} =&\prod_{i,j=1}^{\infty}\prod_{\alpha =\lfloor -\frac{N}{2}\rfloor+1}^{\lfloor \frac{N}{2}\rfloor }\prod_{1\leq b\leq c\leq k}\frac{\left[-m_c^{(\alpha)}-\mathbb{F}_{bc}^{(\alpha)} +\epsilon_1(j-1/2)-\epsilon_2(i-1/2)\right]}{\left[-\mathbb{F}_{bc}^{(\alpha+1)} +\epsilon_1 i-\epsilon_2 (j-1)\right]}\cr &\times\prod_{1\leq b<c\leq k}\frac{\left[-\mathbb{F}_{bc}^{(\alpha)} +m_b^{(\alpha)} +\epsilon_1(j-1/2)-\epsilon_2(i-1/2)\right]}{\left[-\mathbb{F}_{bc}^{(\alpha)} +\epsilon_1(i-1)-\epsilon_2 j\right]}\;.
\end{align}

We will now make the ``deconstruction'' identifications:\footnote{We will justify these choices a posteriori.}
\begin{align}
  \label{HiggsingId}
  m_c^{(\alpha)}+\mathbb{F}_{bc}^{(\alpha)} = \tilde{t}_m + \tilde{t}_{bc} +
  \frac{2 \pi  s_{1,bc}^{(\alpha)}}{ R_5}\;,\qquad & \mathbb{F}_{bc}^{(\alpha+1)} = \tilde{t}_{bc} + \frac{2 \pi  s_{3,bc}^{(\alpha)}}{ R_5}\cr
  -m_b^{(\alpha)}+\mathbb{F}_{bc}^{(\alpha)} = -\tilde{t}_m + \tilde{t}_{bc}
                                               +   \frac{2 \pi  s_{2,bc}^{(\alpha)}}{ R_5}\;,\qquad &
\mathbb{F}_{bc}^{(\alpha)} = \tilde{t}_{bc} + \frac{2 \pi  s_{4,bc}^{(\alpha)}}{ R_5}\;,
\end{align}
where the $s^{(\alpha)}_{n,bc}$ are $N$ unordered and distinct integers for fixed $b,c$ and each given $n=1,...,4$. With this definition, the $s^{(\alpha)}_{n,bc}$ are in one-to-one correspondence with all the integers in the limit $N\to\infty$. Using the definitions \eqref{eq:rels} it is straightforward to see that there exists a 2-parameter space of solutions to \eqref{HiggsingId} (given by $s_{1,bc}^{(\alpha)} = -s_{2,bc}^{(\alpha)} $ and $s_{3,bc}^{(\alpha)} = s_{4,bc}^{(\alpha+1)}$) any choice of which allows us to write
\begin{align}
\lim_{N\to\infty}Z_{\text{4D,1-loop}}^{\rm Higgs} =&\prod_{i,j=1}^{\infty}\prod_{\alpha = -\infty}^{\infty}\prod_{1\leq b\leq c\leq k}\frac{\left[-\frac{2 \pi s_{1,bc}^{(\alpha)}}{R_5}- \tilde{t}_m -\tilde{t}_{bc} +\epsilon_1(j-1/2)-\epsilon_2(i-1/2)\right]}{\left[-\frac{2 \pi s_{3,bc}^{(\alpha)}}{R_5} -\tilde{t}_{bc} +\epsilon_1 i-\epsilon_2 (j-1)\right]}\cr &\times\prod_{1\leq b<c\leq k}\frac{\left[-\frac{2 \pi s_{2,bc}^{(\alpha)}}{R_5} + \tilde{t}_m - \tilde{t}_{bc} +\epsilon_1(j-1/2)-\epsilon_2(i-1/2)\right]}{\left[-\frac{2 \pi s_{4,bc}^{(\alpha)}}{R_5} -\tilde{t}_{bc} +\epsilon_1(i-1)-\epsilon_2 j\right]}\;.
\end{align}
The following redefinitions significantly simplify the notation
\begin{align}
  \label{eq:10}
  X^1_{b,c,i,j} & = - \tilde{t}_m -\tilde{t}_{bc} +\epsilon_1(j-1/2)-\epsilon_2(i-1/2) \cr
  X^2_{b,c,i,j} & = \tilde{t}_m -\tilde{t}_{bc} +\epsilon_1(j-1/2)-\epsilon_2(i-1/2)\cr
  X^3_{b,c,i,j} & = -\tilde{t}_{bc} +\epsilon_1 i-\epsilon_2 (j-1) \cr
  X^4_{b,c,i,j} & = -\tilde{t}_{bc} +\epsilon_1(i-1)-\epsilon_2 j\;,
\end{align}
in terms of which we have that  
\begin{align}
\lim_{N\to\infty}Z_{\text{4D,1-loop}}^{\rm Higgs} =\prod_{i,j=1}^{\infty}\prod_{1\leq b\leq c\leq k}\frac{\prod_{\alpha = -\infty}^{\infty}\left[-\frac{2 \pi s_{1,bc}^{(\alpha)}}{R_5} + X^1_{b,c,i,j}\right]}{\prod_{\alpha = -\infty}^{\infty}\left[-\frac{2 \pi s_{3,bc}^{(\alpha)}}{R_5} + X^3_{b,c,i,j}\right]}\prod_{1\leq b<c\leq k}\frac{\prod_{\alpha = -\infty}^{\infty}\left[-\frac{2 \pi s_{2,bc}^{(\alpha)}}{R_5} + X^2_{b,c,i,j}\right]}{\prod_{\alpha = -\infty}^{\infty}\left[-\frac{2 \pi s_{4,bc}^{(\alpha)}}{R_5} + X^4_{b,c,i,j}\right]}\;.
\end{align}
Since the $s_{n,bc}^{(\alpha)}$ range over all integers, we can choose to order the factors in the products as
\begin{align}
\lim_{N\to\infty}Z_{\text{4D,1-loop}}^{\rm Higgs} =&\prod_{i,j=1}^{\infty}\prod_{1\leq b\leq c\leq k}\frac{\prod_{\alpha = -\infty}^{\infty}\left[-\frac{2 \pi \alpha}{R_5}+ X^1_{b,c,i,j}\right]}{\prod_{\alpha = -\infty}^{\infty}\left[-\frac{2 \pi \alpha}{R_5}+ X^3_{b,c,i,j}\right]}\prod_{1\leq b<c\leq k}\frac{\prod_{\alpha = -\infty}^{\infty}\left[-\frac{2 \pi \alpha}{R_5}+ X^2_{b,c,i,j}\right]}{\prod_{\alpha = -\infty}^{\infty}\left[-\frac{2 \pi \alpha}{R_5} + X^4_{b,c,i,j}\right]}\;.
\end{align}
We next rewrite the terms that appear in the numerators and denominators as 
\begin{align}\label{Prod2sin}
\prod_{\alpha = -\infty}^{\infty}\left(-\frac{2 \pi \alpha}{R_5} + X^n_{b,c,i,j}\right) = -X^n_{b,c,i,j} \prod_{\alpha = 1}^\infty \frac{4 \pi^2}{R_5^2  \alpha^2}\left(1 -\frac{ R_5^2(X^n_{b,c,i,j})^2}{4 \pi^2 \alpha^2}\right)
\end{align}
and use the product representation of the sine function such that the one-loop piece of the 4D partition function becomes
\begin{align}
\lim_{N\rightarrow\infty}Z_{\text{4D,1-loop}}^{\rm Higgs}=&\prod_{i,j=1}^{\infty}\prod_{1\leq b\leq c\leq k}\frac{\sin{ \frac{1}{2} R_5 X^1_{b,c,i,j}}}{\sin{\frac{1}{2} R_5 X^3_{b,c,i,j}}}\prod_{1\leq b<c\leq k}\frac{\sin{\frac{1}{2} R_5  X^2_{b,c,i,j}}}{\sin{\frac{1}{2} R_5  X^4_{b,c,i,j}}}\;.
\end{align}
Converting these trigonometric functions into their exponential form leads to cancellations between numerator and denominator prefactors resulting in
\begin{align}
\lim_{N\rightarrow\infty}Z_{\text{4D,1-loop}}^{\rm Higgs}
  =\prod_{i,j=1}^{\infty}e^{- i R_5 k \frac{\tilde{t}_m}{2} -  i R_5 k\frac{\epsilon_+}{4}}\prod_{1\leq b\leq c\leq k}\frac{\left(1-e^{- i R_5 X_{b,c,i,j}^1}\right)}{\left(1-e^{- i R_5  X_{b,c,i,j}^3}\right)} \prod_{1\leq b<c\leq k}\frac{\left(1-e^{- i R_5  X_{b,c,i,j}^2}\right)}{\left(1-e^{- i R_5 X_{b,c,i,j}^4}\right)}\;.
\end{align}
Upon identifying $\epsilon_{1,2} = \tilde{\epsilon}_{1,2}$ and expressing the above in terms of the fugacities \eqref{chempot} we arrive at the answer
\begin{align}
\label{eq:6DPFpert}
\lim_{N\rightarrow\infty}Z_{\text{4D,1-loop}}^{\rm Higgs}
&=\prod_{i,j=1}^{\infty}
\Big(\widetilde{Q}_m^{-2}\frac{\tilde{\mathfrak q}}{\tilde{\mathfrak t}}\Big)^{\frac{k}{4}}\prod_{1\leq b\leq c\leq k}\frac{\left(1-\widetilde{\mathbb{Q}}_{b c}\widetilde{Q}_{m}\tilde{\mathfrak{q}}^{j-\frac{1}{2}}\tilde{\mathfrak{t}}^{i-\frac{1}{2}}\right)}{ \left(1-\widetilde{\mathbb{Q}}_{b c}\tilde{\mathfrak{q}}^{i}\tilde{\mathfrak{t}}^{j-1}\right)} \prod_{1\leq b<c\leq k}\frac{\left(1-\widetilde{\mathbb{Q}}_{b c}\widetilde{Q}_{m}^{-1}\tilde{\mathfrak{q}}^{j-\frac{1}{2}}\tilde{\mathfrak{t}}^{i-\frac{1}{2}}\right) }{\left(1-\widetilde{\mathbb{Q}}_{b c}\tilde{\mathfrak{q}}^{i-1}\tilde{\mathfrak{t}}^{j}\right)}\;.
\end{align}
Comparing with \eqref{6Dexplicit}, one readily recognises this as\footnote{The additional factor $\prod_{i,j=1}^{\infty} \Big(\widetilde{Q}_m^{-2}\frac{\tilde{\mathfrak q}}{\tilde{\mathfrak t}}\Big)^{\frac{k}{4}}$ can be interpreted as the gravitational contribution to the genus-one part of the topological string partition function \cite{Haghighat:2013gba}, and is  also present in the localisation calculation of \cite{Kim:2013nva}. It is interesting to observe that this is reproduced from 4D through our deconstruction prescription. }
\begin{align}
  \label{eq:13}
  \lim_{N\rightarrow\infty}Z_{\text{4D,1-loop}}^{\rm Higgs} = Z_{\text{6D,1-loop}}\prod_{i,j=1}^{\infty} \Big(\widetilde{Q}_m^{-2}\frac{\tilde{\mathfrak q}}{\tilde{\mathfrak t}}\Big)^{\frac{k}{4}}\;.
\end{align}

\subsubsection{The Non-Perturbative Piece}

The calculation for the non-perturbative piece proceeds in a similar manner. Once again, we begin with the instanton part of the 4D Coulomb-branch partition function from Eq.~\eqref{eq:4DQuiverPFpertinst}
\begin{align}\label{eq:4DQuiverPFinst}
Z_{\text{4D,inst}} &=\sum_{\nu}\prod_{\alpha =\lfloor - \frac{N}{2}\rfloor +1}^{\lfloor  \frac{N}{2}\rfloor} \mathbf{q}_{(\alpha)}^{\sum_{b=1}^k|\nu_{b}^{(\alpha)}|} \prod_{1\leq b\leq c\leq k}\frac{{\rm N}_{\nu_{c}^{(\alpha+1)}\nu_{b}^{(\alpha)}}\left(-\mathbb{F}_{bc}^{(\alpha)} - m^{(\alpha)}_c -\frac{\epsilon_+}{2}\right)
}{
{\rm N}_{\nu_{c}^{(\alpha+1)}\nu_{b}^{(\alpha+1)}}\left(-\mathbb{F}_{bc}^{(\alpha+1)} \right)}\cr &\times \prod_{1\leq b<c\leq k}\frac{{\rm N}_{\nu_{c}^{(\alpha)}\nu_{b}^{(\alpha+1)}}\left(-\mathbb{F}_{bc}^{(\alpha)} + m^{(\alpha)}_b -\frac{\epsilon_+}{2}\right)}{{\rm N}_{\nu_{c}^{(\alpha)}\nu_{b}^{(\alpha)}}\left(-\mathbb{F}_{bc}^{(\alpha)} -\epsilon_+\right)}\;,
\end{align}
and impose the ``deconstruction'' identifications \eqref{HiggsingId}, along with the conditions
\begin{align}\label{decoupl}
  g^{(\alpha)} \to G \;, \qquad \theta^{(\alpha)} \to \Theta \;,\qquad \nu^{(\alpha)}_b \to \nu_b \;,
\end{align}
such that 
\begin{align}\label{taucoupling}
\prod_{\alpha=-\infty}^\infty\mathbf{q}_{(\alpha)} =\exp{\left(-\frac{8\pi^2 N}{G^2}+i\frac{N \Theta}{4\pi}\right)} =: \widetilde Q_\tau\;.
\end{align}
We have anticipated setting all the couplings to be equal in \eqref{eq:4} using the brane picture, from which the identification of theta angles and partitions follows naturally.

We then have 
 \begin{align}\label{instprehiggs}
Z_{\text{4D,inst}}^{\text{Higgs}}
   &=\sum_{\nu}\prod_{\alpha =\lfloor - \frac{N}{2}\rfloor +1}^{\lfloor \frac{N}{2}\rfloor}\widetilde Q_\tau^{\sum_{b=1}^k|\nu_{b}|}
     \prod_{1\leq b\leq c\leq k}\frac{{\rm
     N}_{\nu_{c}\nu_{b}}\left(-\tilde{t}_m - \tilde{t}_{bc}-\frac{2 \pi s_{1,bc}^{(\alpha)}}{R_5}- \frac{\epsilon_+}{2}\right)
}{
{\rm N}_{\nu_{c}\nu_{b}}\left(-\tilde{t}_{bc}-\frac{2 \pi s_{3,bc}^{(\alpha)}}{R_5} \right)}\cr
   &\qquad\times \prod_{1\leq b<c\leq k}\frac{{\rm
     N}_{\nu_{c}\nu_{b}}\left(\tilde{t}_m- \tilde{t}_{bc}-\frac{2 \pi s_{2,bc}^{(\alpha)}}{R_5} -\frac{\epsilon_+}{2}\right)}{{\rm N}_{\nu_{c}\nu_{b}}\left(-\tilde{t}_{bc}-\frac{2 \pi s_{4,bc}^{(\alpha)}}{R_5} -\epsilon_+\right)}\;.
 \end{align}
 Let us focus on the products of $\rm N_{\nu_c \nu_b}$ functions appearing in the numerators and denominators when considering the $N\rightarrow \infty$ limit, using the identifications
 \begin{align}
  \label{Ys}
  Y^1_{b,c}  =  \tilde{t}_m +\tilde{t}_{bc} +\frac{\epsilon_+}{2}\;,\qquad &  Y^3_{b,c}  = \tilde{t}_{bc} \cr
  Y^2_{b,c}  = -\tilde{t}_m +\tilde{t}_{bc} +\frac{\epsilon_+}{2}\;,\qquad &
  Y^4_{b,c}  = \tilde{t}_{bc} +\epsilon_+\;.
\end{align}
 Once again, since the $s_{n,bc}^{(\alpha)}$ range over all the integers we can reorder each of these terms for all $b,c$ as
\begin{align}
  \prod_{\alpha=-\infty}^{\infty}{\rm
  N}_{\nu_{c}\nu_{b}}\left(-\frac{2 \pi s_{n,bc}^{(\alpha)}}{R_5}-Y^n_{b,c}\right) = \prod_{\alpha=-\infty}^{\infty}{\rm
  N}_{\nu_{c}\nu_{b}}\left(-\frac{2 \pi \alpha}{R_5}-Y^n_{b,c}\right)\;.
\end{align}
Then, using the definition of the functions $\rm N_{\nu_c \nu_b}$ from \eqref{Nfunctions} as well as the product representation of the sine function, one arrives at
\begin{align}\label{Nmod}
\prod_{\alpha=-\infty}^{\infty}{\rm
  N}_{\nu_{c}\nu_{b}}\left(-\frac{2 \pi \alpha}{R_5}-Y^n_{b,c}\right) = \prod_{(i,j)\in \nu_c}\widetilde{Y}^n_{b,c,i,j}\frac{4 \pi^2\alpha^2}{R_5^2}\sin\left(-\frac{R_5}{2}\widetilde{Y}^n_{b,c,i,j}\right)
\cr\times \prod_{(i,j)\in \nu_b}\widetilde{\widetilde{Y}}^n_{b,c,i,j}\frac{4 \pi^2\alpha^2}{R_5^2}\sin\left(-\frac{R_5}{2} \widetilde{\widetilde{Y}}^n_{b,c,i,j}\right)\;,
\end{align}
where we have abbreviated
\begin{align}
  \label{eq:14}
  \widetilde{Y}^n_{b,c,i,j} &= - Y^n_{b,c}+\epsilon_1(\nu_{c,i}-j+1)+ \epsilon_2(i-\nu_{b,j}^t)\cr
\widetilde{\widetilde{Y}}^n_{b,c,i,j} &= - Y^n_{b,c}+\epsilon_1(j-\nu_{b,i})+\epsilon_2(\nu_{c,j}^t -i+1)\;.
\end{align}
We note that the factors appearing outside the sines will cancel out when plugging the expressions \eqref{Nmod} back into the ratios in \eqref{instprehiggs}. By writing the result in terms of the fugacities \eqref{chempot}, using the definition of the $\mathcal N_{\nu_c \nu_b}$ functions \eqref{mathcalN}, identifying $\epsilon_{1,2} = \tilde{\epsilon}_{1,2}$ and taking into account cancellations between the numerator and denominator terms once we put everything together, we arrive at the simple-looking expression
\begin{align}
\lim_{N\rightarrow \infty}Z^{\text{Higgs}}_{\text{4D,inst}}=&\sum_{\nu} \widetilde Q_\tau^{\sum_{b=1}^k|\nu_b|}\left(\widetilde{Q}_m\sqrt{ \frac{\tilde{\mathfrak{t}}}{\tilde{\mathfrak{q}}}}\right)^{-\sum_{b=1}^k|\nu_b|}\prod_{1\leq b \leq c \leq k}\frac{\mathcal{N}_{\nu_c\nu_b}\left(\widetilde{\mathbb{Q}}_{bc}\widetilde{Q}_m \sqrt{\frac{\tilde{\mathfrak{t}}}{\tilde{\mathfrak{q}}}}\right)}{\mathcal{N}_{\nu_c\nu_b}\left(\widetilde{\mathbb{Q}}_{bc}\right)}\cr
&\times \prod_{1\leq b < c \leq k}\frac{\mathcal{N}_{\nu_c\nu_b}\left(\widetilde{\mathbb{Q}}_{bc}\widetilde{Q}^{- 1}_m \sqrt{\frac{\tilde{\mathfrak{q}}}{\tilde{\mathfrak{t}}}}\right)}{\mathcal{N}_{\nu_c\nu_b}\left(\widetilde{\mathbb{Q}}_{bc}\frac{\tilde{\mathfrak{t}}}{\tilde{\mathfrak{q}}}\right)}\;.
\end{align}
Comparing with \eqref{6Dexplicit}, we see that we have reached the result
\begin{align}
  \label{eq:15}
  \lim_{N\rightarrow \infty}Z^{\text{Higgs}}_{\text{4D,inst}}= Z_{\text{6D,inst}}\;.
\end{align}

\subsubsection{The Classical Piece}

Finally, for the classical piece we start with the expression \eqref{eq:26}
\begin{align}
  \label{eq:27}
Z_{\rm 4D, cl} =    \exp \left[ - \frac{2\pi i }{\epsilon_1 \epsilon_2} \sum_{\alpha =\lfloor -\frac{N}{2}\rfloor+1}^{\lfloor\frac{N}{2}\rfloor} \tau^{(\alpha)} \sum_{b=1}^k \left( a_b^{(\alpha)} \right)^2 \right]
\end{align}
and impose the deconstruction identifications that we have already encountered. With the help of \eqref{eq:rels},  it can be shown that \eqref{HiggsingId} reduces to
\begin{align}
  \label{eq:elementary}
  a^{(\alpha)}_b = \tilde{a}_b + \frac{2 \pi r^{(\alpha)}_b}{R_5} \;, \qquad m_{\rm
  bif}^{(\alpha)} = \tilde{t}_m\;,
\end{align}
where the $r^{(\alpha)}_b$ are distinct integers such that 
\begin{align}
  \label{intrels}
  r^{(\alpha+1)}_c - r^{(\alpha)}_b = s^{(\alpha)}_{1, bc} \;,\qquad   r^{(\alpha)}_c - r^{(\alpha)}_b = s^{(\alpha)}_{4, bc}
\end{align}
are also distinct integers. Then along with \eqref{decoupl} one has
\begin{align}
  \label{eq:28}
  Z_{\rm 4D, cl}^{\rm Higgs} =  
  \exp \left[ - \frac{ N }{\tilde{\epsilon_1} \tilde{\epsilon_2}} \Big(-\frac{8\pi^2}{G^2}+i\frac{\Theta}{4 \pi}\Big)  \left(  \sum_{b=1}^k   \tilde{a}_b^2  + \frac{1}{N}   \sum_{b=1}^k  \sum_{\alpha =  \lfloor -\frac{N}{2}\rfloor +1}^{\lfloor \frac{N}{2}\rfloor} \Big( 2\tilde{a}_b r_{b}^{(\alpha)} + (r_{b}^{(\alpha)})^2 \Big) \right  ) \right]\;.
\end{align}
Since in the large-$N$ limit the integers $r^{(\alpha)}_b$ span all of $\mathbb Z$, one can use $\zeta$-function regularisation to show that both the quadratic and linear $r_b^{(\alpha)}$ terms vanish.\footnote{The linear term involves
$\sum_{\alpha = -\infty}^{\infty} \alpha = \sum_{\alpha = 1}^{\infty} \alpha  + \left(- \sum_{\alpha = 1}^{\infty} \alpha\right) = \zeta(-1) - \zeta(-1)    =0$.
For the quadratic piece $ \sum_{\alpha = -\infty}^{\infty} \alpha^2 = 2 \sum_{\alpha = 1}^{\infty} \alpha^2 = 2 \zeta(-2) = 0$.} 
Then, along with  \eqref{taucoupling}, Eq.~\eqref{eq:28}  becomes
\begin{align}
  \label{eq:29}
\lim_{N\to\infty}  Z_{\rm 4D, cl}^{\rm Higgs} =  \exp\left[- \frac{2\pi i \tau}{\tilde{\epsilon}_1 \tilde{\epsilon}_2}  \sum_{b=1}^k   \tilde{a}_b^2\right]
\;.
\end{align}

\subsection{Comments on the Exact  Deconstruction Procedure}

We summarise the results of this section and conclude with some comments. We have provided a prescription for implementing the deconstruction limit of \cite{ArkaniHamed:2001ie} at the level of exact partition functions. This requires making the following identifications between 4D and 6D parameters:
\begin{align}
  \label{H1}
  a^{(\alpha)}_b = \tilde{a}_b + \frac{2 \pi r^{(\alpha)}_b}{R_5} \;, \qquad m_{\rm
  bif}^{(\alpha)} = \tilde{t}_m\;,\qquad \epsilon_{1,2} = \tilde{\epsilon}_{1,2}\;,
\end{align}
along with
\begin{align}\label{H2}
  g^{(\alpha)} \to G \;, \qquad \theta^{(\alpha)}\to \Theta \;,\qquad \nu^{(\alpha)}_b \to \nu_b\;.
\end{align}
The $r^{(\alpha)}_b$ are distinct integers, which need to satisfy the conditions \eqref{intrels}, while $G, \Theta$ are related to the modulus $\tilde{\tau}$ of the $T^2$ through \eqref{taucoupling}.

With these identifications at hand, we have reached the following result:
\begin{align}
  \label{eq:17}
  \lim_{N\rightarrow\infty}Z_{\text{4D}}^{\rm Higgs} =  \prod_{i,j=1}^{\infty} \Big(\widetilde{Q}_m^{-2}\frac{\tilde{\mathfrak q}}{\tilde{\mathfrak t}}\Big)^{\frac{k}{4}}Z_{\text{6D}}\;.
\end{align}
Our  method reproduces the 6D IR partition function directly from 4D, up to an overall multiplicative piece, which encodes the  genus-one contribution to the topological-string partition function \cite{Haghighat:2013gba}. Nevertheless, given the complex-conjugation prescription \eqref{eq:12}, this piece cancels out once one takes two copies of the result to reproduce the $S^4\times T^2$ partition function. 

Upon integrating over the Coulomb/tensor-branch parameters one immediately gets the full partition function for the $(2,0)_k$ theory on $S^4_{\tilde{\epsilon}_1, \tilde{\epsilon}_2}\times T^2_{R_5, R_6}$ as per Eq.~\eqref{eq:11}
\begin{align}
  \label{eq:18}
 \int [ d\tilde{a} ] \left|\lim_{N\rightarrow\infty}Z_{\text{4D}}^{\rm Higgs}\right|^2  = \int [ d\tilde{a} ]  \left|Z_{\text{6D}}\right|^2  = Z_{S^4_{\tilde{\epsilon}_1, \tilde{\epsilon}_2}\times T^2_{R_5, R_6}}\;.
\end{align}
Note that if the complex structure of the torus wrapped by the M5 branes is taken to be purely imaginary $\tau = i \frac{R_6}{R_5}$, the torus is rectangular and twisted by turning on the $\Omega$-deformation parameters $\tilde{\epsilon}_{1,2}$ as well as the mass deformation $\tilde{m}$ \cite{Haghighat:2013gba}. However, as is evident from the calculation, our procedure can also deconstruct a torus with nonzero theta angle.

We have reached this deconstruction prescription using both physical and mathematical intuition. First, the classical brane picture of Sec.~\ref{Sec:BEII} immediately implies the conditions \eqref{H2}. Second, the Weierstrass factorisation theorem indicates that two meromorphic functions are equal (up to a nonzero holomorphic factor) when they have the same zeros and poles. Eq.~\eqref{H1} is a simple consequence of equating the zeros and poles of the 4D and 6D Coulomb-branch partition functions. Note that the same relationship between zeros/poles emerges when one ``q deforms'' rational to trigonometric functions and it is in this fashion that deconstruction explicitly recovers the extra dimension.

Even though we have presented our calculation starting directly from 4D, we also provide a derivation of the 4D Coulomb-branch partition function using the refined topological vertex formalism in App.~\ref{AppC}. The latter yields an intermediate 5D Coulomb-branch partition function on $\mathbb R^4_{\epsilon_1, \epsilon_2}\times S^1_\beta$, which is then reduced to 4D in the limit $\beta\to 0$. This 5D partition function is associated with a toric diagram that is dual to a $(p,q)$ 5-brane web. It is then natural to ask what our deconstruction prescription translates to in this 5D picture and how it is related to similar Higgsings that can be found in the literature \cite{Dimofte:2010tz, Hayashi:2013qwa}.

To answer this we observe the following: When expressed in terms of the 5D fugacities of App.~\ref{AppC}, Eqs.~\eqref{H1}, \eqref{H2} become
\begin{align}
  \label{eq:19}
 Q^{(\alpha)}_{mb}:=  \exp{\left[\beta m^{(\alpha)}_b\right]} &\to \widetilde Q_m^{-\frac{i\beta}{R_5}}\cr
 Q^{(\alpha)}_{fb}:=   \exp{\left[\beta F^{(\alpha)}_{b}\right]}  &\to \widetilde{ Q}_{fb}^{-\frac{i\beta}{R_5}}\cr
Q^{(\alpha)}_{gb}:=   \exp{\left[2 \pi i \tau^{(\alpha)} -\frac{1}{2}(m^{(\alpha+1)}_b+m^{(\alpha)}_b)\right]}  &\to \widetilde{ Q}^{-\frac{i\beta}{R_5}}
\end{align}
and 
\begin{align}
  \label{eq:20}
  \mathfrak q \to \tilde{\mathfrak q}^{-\frac{i\beta}{R_5}}\;,\qquad   \mathfrak t \to \tilde{\mathfrak t}^{-\frac{i\beta}{R_5}}\;,
\end{align}
such that all edges in the dual-toric diagram become equal. Upon further specialising the above to $\widetilde Q_m^{-i\frac{\beta}{R_5}} = \mathfrak q^{\frac{1}{2}} \mathfrak t^{-\frac{1}{2}}$ as dictated in \cite{Dimofte:2010tz,Hayashi:2013qwa}, one recovers the 6D Coulomb-branch partition function for $\tilde{t}_m = -\frac{\tilde{\epsilon}_+}{2}$.\footnote{When $\tilde{t}_m = \pm\frac{\tilde{\epsilon}_+}{2}$, the partition function coincides with that of $\mathcal{N}=4$ SYM with $\U(k)$ gauge group \cite{Haghighat:2013gba}, although generically the two are not the same.} It would be interesting to investigate the scenario $\widetilde Q_m^{-i\frac{\beta}{R_5}} = \mathfrak q^{r-\frac{1}{2}} \mathfrak t^{\frac{1}{2}-s}$ for $s,r> 1$, as this should correspond to the introduction of surface operators in the 6D theory that do not wrap the $T^2$.

We finally note that, although we had to tune the parameters of the 4D theory such that we are at the origin of a mixed Coulomb-Higgs branch, our Higgsing seems to be completely insensitive to the exact value of the chiral multiplet VEVs. This is not surprising, since we obtained the partition function via a Coulomb-branch (instead of a Higgs-branch) calculation. Still, one does need to consider ${\rm v}\to\infty$ to obtain a 6D theory on a torus of fixed radii so our procedure is indirectly sensitive to this choice.


\ack{ \bigskip We would like to thank S.~Beheshti, A.~Hanany and A.~Royston for several insightful comments. We would also like to thank M.~Buican, T.~Nishinaka, D.~Shih and especially G.~Moore for many enjoyable discussions, collaboration at the initial stages of this project and comments on the manuscript. J.H. is supported by an STFC research studentship and
  would like to thank Humboldt University of Berlin and CERN for hospitality
  during various stages of this work. C.P. is supported by the Royal
  Society through a University Research Fellowship. E.P. is supported by the German Research Foundation (DFG) via the Emmy Noether program “Exact results in Gauge theories”. D.R.G. is partly
  supported by the Spanish Government grant
  MINECO-13-FPA2012-35043-C02-02 and the Ramón y Cajal grant
  RYC-2011-07593 as well as the EU CIG grant UE-14-GT5LD2013-618459. This work was completed at the Aspen Center for Physics, which is supported by National Science Foundation grant PHY-1066293.}


\begin{appendix}

\section{Results for the Higgs-Branch HS}\label{AppD}

As described in the main text, the single-letter evaluation procedure of the Higgs-branch HS for $k>1$ is not straightforward. In these cases one has to solve the $F$-term constraints and perform the hyper-Kähler quotient explicitly, which can be done e.g. with the help of mathematical software such as \verb+macaulay2+ \cite{Mac2}. The algorithm can be summarised as follows: For a given $\mathcal N=2$ theory one first identifies the ring of operators that parametrise the Higgs branch of the moduli space ($\mathcal R$), labelled by their scaling dimensions and their charges under the $\U(k)$ gauge groups. One also finds the $F$-term conditions that arise from the superpotential; the  latter form an ideal of the ring ($\mathcal I$). The moduli space is the quotient $\mathcal R/\mathcal I$ and  \verb+macaulay2+ calculates directly the HS on that space. The resultant expression still needs to get integrated over with the appropriate Haar measure to yield the answer for gauge-invariant contributions to the HS. Here we will exhibit this brute-force procedure and list the Higgs-branch HS for $k=2$ and $N=2,3$ (the case $k=1$ for arbitrary $N$ can be easily seen to yield the same result as the modified letter-counting in the main text). This will confirm the expectation that the general formula for all $k$ is given by the $k$-fold-symmetrised  product of the $k=1$ result.

\paragraph{$k=2, N=2$:} The superpotential is given by:
\begin{align}
  \label{eq:16}
  W={\rm tr}\Big(X_{2,1}^jX_{1,1}X_{1,2}^i\epsilon_{ij}+X_{1,2}^jX_{2,2}X_{2,1}^i\epsilon_{ij} \Big)\, .  
\end{align}
For this example, the ring $\mathcal R$ contains 16 components, 4 each for $X_{1,2}^{i} ,X_{2,1}^{i}$, $i = 1,2$. These all have scaling dimension 1, are further characterised by their Cartan charges under the $\U(2)_1\times\U(2)_2\simeq \SU(2)_1\times\U(1)_1\times \SU(2)_2\times\U(1)_2$ gauge groups, and consequently labelled by a string $\{1,\pm 1, \pm 1, \pm 1, \pm 1\}$, where the signs are uncorrelated. The $F$-term conditions lead to  the matrix equation
\begin{align}
  \label{eq:22}
  X_{1,2}^1X_{2,1}^2=X_{1,2}^2X_{2,1}^1  
\end{align}
plus its hermitian conjugate, giving rise to 8 relations between the elements of the ring $\mathcal R$, encoded in the ideal $\mathcal I$. The above data, presented in  precisely this form, can be directly fed into \verb+macaulay2+, which then immediately evaluates the HS on $\mathcal R/\mathcal I$. The result for this simplest of nonabelian examples is:
\begin{align}
  \label{eq:23}
  {\rm HS}_{k=2}^{N=2} & = \Big( \frac{b_2^2 t^{10}}{b_1^2}+\frac{b_1^2 t^{10}}{b_2^2}+3 t^{10}-\frac{2 b_2 u_1 u_2 t^9}{b_1}-\frac{2 b_1 u_1 u_2 t^9}{b_2}-\frac{2 b_2 u_2 t^9}{b_1 u_1}-\frac{2 b_1 u_2 t^9}{b_2 u_1}-\frac{2 b_2 u_1 t^9}{b_1 u_2}\cr
& -\frac{2 b_1 u_1 t^9}{b_2 u_2}-\frac{2 b_2 t^9}{b_1 u_1 u_2}-\frac{2 b_1 t^9}{b_2 u_1 u_2}+u_1^2 t^8+u_1^2 u_2^2 t^8+\frac{u_2^2 t^8}{u_1^2}+u_2^2 t^8-\frac{b_2^2 t^8}{b_1^2}-\frac{b_1^2 t^8}{b_2^2}+\frac{t^8}{u_1^2}\cr
& +\frac{u_1^2 t^8}{u_2^2}+\frac{t^8}{u_2^2}+\frac{t^8}{u_1^2 u_2^2}-t^8+\frac{4 b_2 u_1 u_2 t^7}{b_1}+\frac{4 b_1 u_1 u_2 t^7}{b_2}+\frac{4 b_2 u_2 t^7}{b_1 u_1}+\frac{4 b_1 u_2 t^7}{b_2 u_1}+\frac{4 b_2 u_1 t^7}{b_1 u_2}\cr
& +\frac{4 b_1 u_1 t^7}{b_2 u_2}+\frac{4 b_2 t^7}{b_1 u_1 u_2}+\frac{4 b_1 t^7}{b_2 u_1 u_2}-3 u_1^2 t^6-2 u_1^2 u_2^2 t^6-3 u_2^2 t^6-\frac{b_2^2 t^6}{b_1^2}-\frac{b_1^2 t^6}{b_2^2}-\frac{2 u_2^2 t^6}{u_1^2}-\frac{3 t^6}{u_1^2}\cr
& -\frac{2 u_1^2 t^6}{u_2^2}-\frac{3 t^6}{u_2^2}-\frac{2 t^6}{u_1^2 u_2^2}-8 t^6-\frac{2 b_2 u_1 u_2 t^5}{b_1}-\frac{2 b_1 u_1 u_2 t^5}{b_2}-\frac{2 b_2 u_2 t^5}{b_1 u_1}-\frac{2 b_1 u_2 t^5}{b_2 u_1}-\frac{2 b_2 u_1 t^5}{b_1 u_2}\cr
&-\frac{2 b_1 u_1 t^5}{b_2 u_2}-\frac{2 b_2 t^5}{b_1 u_1 u_2}-\frac{2 b_1 t^5}{b_2 u_1 u_2}+\frac{b_2^2 t^4}{b_1^2}+3 u_1^2 t^4+u_1^2 u_2^2 t^4+\frac{u_2^2 t^4}{u_1^2}+3 u_2^2 t^4+\frac{b_1^2 t^4}{b_2^2}+\frac{3 t^4}{u_1^2}\cr
& +\frac{u_1^2 t^4}{u_2^2}+\frac{3 t^4}{u_2^2}+\frac{t^4}{u_1^2 u_2^2}+8 t^4-u_1^2 t^2-u_2^2 t^2-\frac{t^2}{u_1^2}-\frac{t^2}{u_2^2}-3 t^2+1\Big)\cr
& \times \frac{1}{\left(1-\frac{b_1 t}{b_2 u_1 u_2}\right)^2 \left(1-\frac{b_2 t}{b_1 u_1 u_2}\right)^2 \left(1-\frac{b_1 t u_1}{b_2 u_2}\right)^2 \left(1-\frac{b_2 t u_1}{b_1 u_2}\right)^2}\cr
& \times\frac{1}{ \left(1-\frac{b_1 t u_2}{b_2 u_1}\right)^2 \left(1-\frac{b_2 t u_2}{b_1 u_1}\right)^2 \left(1-\frac{b_1 t u_1 u_2}{b_2}\right)^2 \left(1-\frac{b_2 t u_1 u_2}{b_1}\right)^2}\;,
\end{align}
where $u_{1,2}$ denote the $\SU(2)_{1,2}$ fugacities, $b_{1,2}$ are the $\U(1)_{1,2}$ fugacities and $t$ the fugacity keeping track of the scaling dimension. The above expression can be integrated over the $\U(2)^2$ gauge group with Haar measure
\begin{align}
  \label{eq:24}
  d\mu = du_1\, du_2\, db_1\, db_2\, \frac{1}{b_1}\,\frac{1}{b_2}\,\frac{1-u_1^2}{u_1}\,\frac{1-u_2^2}{u_2}
\end{align}
and gives rise to the simple-looking expression
\begin{align}
{\rm HS}_{k=2}^{N=2}=\frac{1+t^2+4t^4+t^6+t^8}{(1-t^2)^4(1+t^2)^2}\, .
\end{align}
This result can be re-written in terms of
\begin{align}
{\rm HS}_{k=2}^{N=2}(t)=\frac{1}{2}\left[ {\rm HS}_{k=1}^{N=2}(t^2)+({\rm HS}_{k=1}^{N=2}(t))^2\right]\,,
\end{align}
which in turn can be encoded more succinctly as the coefficient of $\nu^2$ in the expansion of
\begin{align}
{\rm PE}[{\rm HS}_{k=1}^{N=2}(t)\nu]\, .
\end{align}
As we have already mentioned, the $\nu^k$ term in the expansion of
this function computes the HS for the $k$-fold-symmetrised product of
the variety for which the HS is given by ${\rm HS}_{k=1}^{N=2}$. 

\paragraph{$k=2, N=3$:} We may proceed as before and find a function
$g_{k=2}^{N=3}$, which is then projected to a gauge-invariant
result. As the closed-form expressions are rather cumbersome, we will
choose to present the HS perturbatively in $t$, finding
\begin{align}
{\rm HS}_{k=2}^{N=3}=1+t^2+2t^3+2t^4+4t^5+7t^6+6t^7+11 t^8 +14 t^9+\mathcal{O}(t^{10})\, .
\end{align}
One recognises here the expansion of the function
\begin{align}
{\rm HS}_{k=2}^{N=3}=\frac{1-t+t^3+t^5-t^7+t^8}{(1-t)^4 (1+t)^2 \left(1-t+t^2\right) \left(1+t+t^2\right)^2}\, .
\end{align}
In turn, this is nothing but the coefficient of $\nu^2$ in
\begin{align}
{\rm PE}\left[{\rm HS}_{k=1}^{N=3}(t)\nu\right]\, .
\end{align}
Extending the calculation of the Higgs-branch HS to higher $k$ and $N$ is a computationally daunting task. Nevertheless, the above fit the expectation \cite{Benvenuti:2006qr} that the result for arbitrary $N$ and $k$ is the coefficient of $\nu^k$ in the expansion of 
\begin{align} {\rm PE}\left[{\rm HS}_{k=1}^{N}(t)\nu\right]={\rm PE}\left[\frac{(1-t^{2N})}{(1-t^2)(1-t^N)^2}\nu\right]\;.
\end{align}
%

\section{The Refined Topological Vertex}\label{AppA}

In this appendix we review the computation of the topological amplitude for the basic building block of our Coulomb-branch partition functions, the ``strip geometry''. 

\subsection{Introduction to the Refined Topological Vertex}\label{sec:B.1}

We take this opportunity to review the rules of the refined topological vertex formalism  \cite{Iqbal:2007ii}; for a pedagogical account we refer the reader to \cite{Bao:2013pwa}.

We define the $\Omega$-background parameters $\mathfrak{q}$ and $\mathfrak{t}$ by 
\begin{align}
\mathfrak{q}=e^{-\beta \epsilon_1}\;,\qquad \mathfrak{t} = e^{\beta \epsilon_2}\;.
\end{align}
We reserve Greek letters for partitions of natural numbers, e.g. $\lambda$ with $\lambda_1\geq \lambda_2 \geq \ldots \geq \lambda_{l(\lambda)}>0$, where $l(\lambda)$ denotes the length of the partition. Each partition can be represented as a Young diagram, with the coordinates $(i,j)$ identifying a box in a given diagram. We have that $(i,j) \in \{ (i,j) |i=1,\ldots,l(\lambda); j=1,\ldots,\lambda_i\}$, so the number of boxes in the $i^{\text{th}}$ column is $\lambda_i$. It is also useful to define the following quantities
\begin{align}
|\lambda | =\sum_{(i,j)\in \lambda}^{l(\lambda)} 1 = \sum_{i=1}^{l(\lambda)}\lambda_i\;,\qquad \| \lambda \|^2 = \sum_{(i,j)\in \lambda} \lambda_i = \sum_{i=1}^{l(\lambda)} \lambda_i^2\;,
\end{align}
as well as the transposed Young diagram $\lambda^t$. These are to be used as data that label the refined topological vertex. 

Each refined topological vertex consists of three directed edges
emanating from the same point. The edges either all point outwards or
inwards, forming two-vectors $\vec v_{1,2,3}$ which satisfy $\sum_{i =
  1}^3\vec v_{i} = 0$ and $v_i \wedge v_{i+1}:= v_i^1 v_{i+1}^2 -
v_i^2 v_{i+1}^1 = -1$. Upon picking a ``preferred
direction''---this will be indicated in our diagrams by a
  double red line---the basic vertices can be glued together in a
unique fashion (outgoing to incoming edges and vice-versa) to form a
dual-toric diagram. A chain of dualities relates this geometry to a IIB 5-brane web represented by the same diagram. To every connected edge one associates a partition: $\lambda$ when the arrow points out of a vertex and $\lambda^t$ when the arrow points into a vertex. External edges are assigned an empty partition, $\emptyset$. 

The dual-toric diagram can be converted into a closed topological string amplitude (which in turn yields a 5D partition function on $\mathbb R_{\epsilon_1, \epsilon_2} \times S^1_{\beta}$ in the field-theory limit) based on the following rules: To each vertex with outgoing edges we assign the vertex factor
\begin{align}\label{vertex}
C_{\lambda \mu \nu}(\mathfrak{t},\mathfrak{q})=\mathfrak{q}^{\frac{\| \mu \|^2+\| \nu\|^2}{2}} \mathfrak{t}^{-\frac{\|\mu^t\|}{2}}\tilde{Z}_\nu (\mathfrak{t},\mathfrak{q}) \sum_{\eta}\left(\frac{\mathfrak{q}}{\mathfrak{t}}\right)^{\frac{|\eta | + | \lambda | - |\mu |}{2}}S_{\lambda^t / \eta}(\mathfrak{t}^{-\rho}\mathfrak{q}^{-\nu})S_{\mu / \eta}(\mathfrak{q}^{-\rho}\mathfrak{t}^{-\nu^t})\;,
\end{align}
such that $\nu$ is the partition associated with the edge vector aligned with the ``preferred direction''. If the edges are incoming then the partitions are simply replaced by their transposes. The $\widetilde{Z}$ functions are a specialisation of the Macdonald polynomials $P_\nu (\bf{x};\mathfrak{t},\mathfrak{q})$ given by 
\begin{align}
\widetilde{Z}_\nu (\mathfrak{t},\mathfrak{q}) = \mathfrak{t}^{-\frac{\|\nu^t\|^2}{2}}P_\nu (\mathfrak{t}^{-\rho};\mathfrak{q},\mathfrak{t})= \prod_{(i,j)\in\nu}\left(1-\mathfrak{t}^{\nu_j^t - i +1}\mathfrak{q}^{\nu_i-j}\right)^{-1}\;,
\end{align}
while the $S_{\lambda/\mu}(\mathbf{x})$ are the skew-Schur functions for the vector $\mathbf{x}=(x_1,\ldots)$. For a partition $\nu$, the vector $\mathfrak{t}^{-\rho}\mathfrak{q}^{-\nu}$ is
\begin{align}
\mathfrak{t}^{-\rho}\mathfrak{q}^{-\nu}= ( \mathfrak{t}^{\frac{1}{2}}\mathfrak{q}^{-\nu_1},\mathfrak{t}^{\frac{3}{2}}\mathfrak{q}^{-\nu_2},\mathfrak{t}^{\frac{5}{2}}\mathfrak{q}^{-\nu_3},\ldots).
\end{align}

Internal edges in the dual-toric diagram correspond to K\"ahler moduli in
the geometry, generically denoted by $Q$. More precisely, each
internal edge is assigned an ``edge factor''  given by
\begin{align}
\text{edge factor} = (-Q)^{|\lambda|}\times \text{framing factor}\;.
\end{align} 
The ``framing factor'' is determined as follows: After glueing two vertices together, one can assign to each connected edge vector $\vec v$ an incoming  and outgoing external vector $\vec v_{in,out}$, such that $\vec v_{in}\cdot \vec v_{out}> 0$. External-edge vectors with $\vec v_{in}\wedge \vec v_{out}\neq 0$ will have non-trivial framing factors; however all examples that we will consider in this appendix have $\vec v_{in}\wedge \vec v_{out} =0$, so our framing factors will all be equal to one. For a comprehensive discussion on this matter we refer the reader to \cite{Iqbal:2012mt, Bao:2013pwa}.

Equipped with the above definitions, we can finally write the topological string partition function with $M$ internal edges as
\begin{align}
Z = \sum_{\lambda^1,\ldots,\lambda^M} \prod_{\text{edges}}\text{edge factor}\prod_{\text{vertices}}\text{vertex factor}\;.
\end{align}

\subsection{Derivation of the Strip-Geometry Amplitude}

The partition functions that we will be deriving in this appendix can all be constructed from a single building block, the ``strip geometry''. In view of using this for the $\U(k)^N$ circular-quiver gauge theory, we assign Kähler moduli to each edge: For each vertical edge a $Q_{g}^{(\alpha)}$, each diagonal edge a $Q_{mb}^{(\alpha)}$ and each horizontal edge a $Q_{fb}^{(\alpha)}$, as well as associated partitions $\nu_{b}^{(\alpha)}$, $\mu_{b}^{(\alpha)}$ and $\lambda_b^{(\alpha)}$ respectively. The indices are $\alpha = \lfloor -\frac{N}{2}\rfloor +1,\ldots,\lfloor \frac{N}{2}\rfloor$ with $N$ the number of nodes for the quiver and $b=1,\ldots,k$ where $k$ is the rank of $\U(k)$.

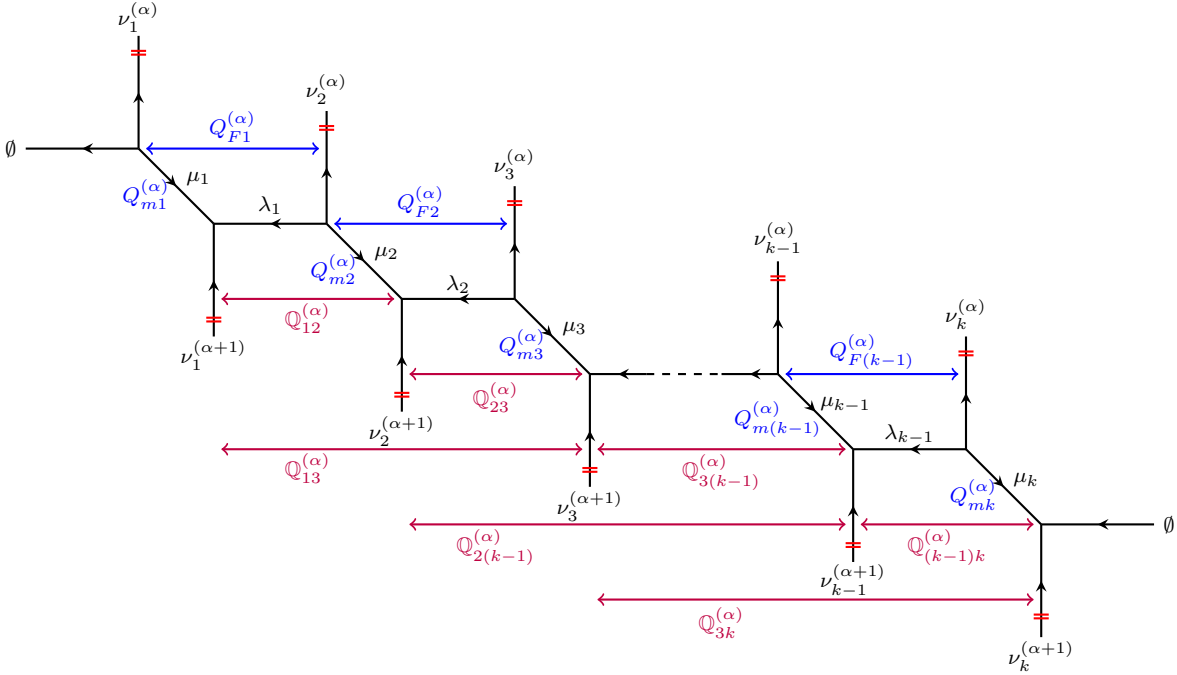
\begin{figure}[t]
\begin{center}
\begin{tikzpicture}
\tikzset{
  ->-/.style={decoration={markings, mark=at position 0.5 with {\arrow{stealth}}},
              postaction={decorate}},
}
\draw  [thick, ->-] (1.5,0) -- (0,0);
\node at (-0.2,0){\scriptsize $\emptyset$};
\draw  [thick, ->-] (1.5,0) -- (1.5,1.5);
\node at (1.5,1.75) {\scriptsize $\nu_{1}^{(\alpha)}$};
\draw  [thick, red] (1.4,1.3) -- (1.6,1.3);
\draw  [thick, red] (1.4,1.25) -- (1.6,1.25);
\draw  [thick, ->-] (1.5,0) -- (2.5,-1);
\node at  (2.3,-0.4) {\scriptsize $\mu_1$};
\node [blue] at  (1.6,-0.6) {\scriptsize $Q_{m1}^{(\alpha)}$};
\draw  [thick, ->-] (2.5,-2.5) -- (2.5,-1);
\node at (2.5,-2.75) {\scriptsize $\nu_{1}^{(\alpha+1)}$};
\draw  [thick, red] (2.4,-2.3) -- (2.6,-2.3);
\draw  [thick, red] (2.4,-2.25) -- (2.6,-2.25);
\draw  [thick, blue,<->]  (1.6,0)--(3.9,0);
\node [blue] at (2.75,0.3) { \scriptsize $Q_{F1}^{(\alpha)}$};
\draw  [thick, purple,<->]  (2.6,-2)--(4.9,-2);
\node [purple] at (3.75,-2.25) { \scriptsize $\mathbb{Q}^{(\alpha)}_{12}$};
\draw  [thick, purple,<->]  (2.6,-4)--(7.4,-4);
\node [purple] at (3.75,-4.25) { \scriptsize $\mathbb{Q}^{(\alpha)}_{13}$};
\draw  [thick, ->-] (4,-1) -- (2.5,-1);
\node at  (3.25,-0.75) {\scriptsize $\lambda_1$};
\draw  [thick, ->-] (4,-1) -- (4,0.5);
\node at (4,0.8) {\scriptsize $\nu_{2}^{(\alpha)}$};
\draw  [thick, red] (3.9,0.3) -- (4.1,0.3);
\draw  [thick, red] (3.9,0.25) -- (4.1,0.25);
\draw  [thick, ->-] (4,-1) -- (5,-2);
\node at  (4.8,-1.4) {\scriptsize $\mu_2$};
\node [blue] at  (4.1,-1.6) {\scriptsize $Q_{m2}^{(\alpha)}$};
\draw  [thick, ->-] (5,-3.5) -- (5,-2);
\node at (5,-3.75) {\scriptsize $\nu_{2}^{(\alpha+1)}$};
\draw  [thick, red] (4.9,-3.3) -- (5.1,-3.3);
\draw  [thick, red] (4.9,-3.25) -- (5.1,-3.25);
\draw  [thick, ->-] (6.5,-2) -- (5,-2);
\node at  (5.75,-1.8) {\scriptsize $\lambda_2$};
\draw  [thick, blue,<->]  (4.1,-1)--(6.4,-1);
\node [blue] at (5.25,-0.7) { \scriptsize $Q_{F2}^{(\alpha)}$};
\draw  [thick, purple,<->]  (5.1,-3)--(7.4,-3);
\node [purple] at (6.25,-3.3) { \scriptsize $\mathbb{Q}^{(\alpha)}_{23}$};
\draw  [thick, purple,<->]  (5.1,-5)--(10.9,-5);
\node [purple] at (6.25,-5.3) { \scriptsize $\mathbb{Q}^{(\alpha)}_{2(k-1)}$};
\draw  [thick, ->-] (6.5,-2) -- (6.5,-0.5);
\node at (6.5,-0.2) {\scriptsize $\nu_{3}^{(\alpha)}$};
\draw  [thick, red] (6.4,-0.75) -- (6.6,-0.75);
\draw  [thick, red] (6.4,-0.7) -- (6.6,-0.7);
\draw  [thick, ->-] (6.5,-2) -- (7.5,-3);
\node at  (7.3,-2.4) {\scriptsize $\mu_3$};
\node [blue] at  (6.6,-2.6) {\scriptsize $Q_{m3}^{(\alpha)}$};
\draw  [thick, ->-] (7.5,-4.5) -- (7.5,-3);
\node at (7.5,-4.75) {\scriptsize $\nu_{3}^{(\alpha+1)}$};
\draw  [thick, red] (7.4,-4.3) -- (7.6,-4.3);
\draw  [thick, red] (7.4,-4.25) -- (7.6,-4.25);
\draw  [thick, ->-] (8.25,-3) -- (7.5,-3);
\draw  [thick, purple,<->]  (7.6,-4)--(10.9,-4);
\node [purple] at (9.25,-4.3) { \scriptsize $\mathbb{Q}^{(\alpha)}_{3(k-1)}$};
\draw  [thick, purple,<->]  (7.6,-6)--(13.4,-6);
\node [purple] at (9.25,-6.3) { \scriptsize $\mathbb{Q}^{(\alpha)}_{3k}$};
\draw  [thick, dashed]  (8.25,-3) -- (9.25,-3);
\draw  [thick, ->-] (10,-3) -- (9.25,-3);
\draw  [thick, ->-] (10,-3) -- (10,-1.5);
\node at (10,-1.2) {\scriptsize $\nu_{k-1}^{(\alpha)}$};
\draw  [thick, red] (9.9,-1.7) -- (10.1,-1.7);
\draw  [thick, red] (9.9,-1.75) -- (10.1,-1.75);
\draw  [thick, ->-] (10,-3) -- (11,-4);
\node at  (10.9,-3.4) {\scriptsize $\mu_{k-1}$};
\node [blue] at  (10,-3.6) {\scriptsize $Q_{m(k-1)}^{(\alpha)}$};
\draw  [thick, ->-] (12.5,-4) -- (11,-4);
\draw  [thick, ->-] (11,-5.5) -- (11,-4);
\node at (11,-5.75) {\scriptsize $\nu_{k-1}^{(\alpha+1)}$};
\draw  [thick, red] (10.9,-5.25) -- (11.1,-5.25);
\draw  [thick, red] (10.9,-5.3) -- (11.1,-5.3);
\node at  (11.75,-3.8) {\scriptsize $\lambda_{k-1}$};
\draw  [thick, blue,<->]  (10.1,-3)--(12.4,-3);
\node [blue] at (11.25,-2.7) { \scriptsize $Q_{F(k-1)}^{(\alpha)}$};
\draw  [thick, purple,<->]  (11.1,-5)--(13.4,-5);
\node [purple] at (12.25,-5.3) { \scriptsize $\mathbb{Q}^{(\alpha)}_{(k-1)k}$};
\draw  [thick, ->-] (12.5,-4) -- (12.5,-2.5);
\node at (12.5,-2.2) {\scriptsize $\nu_{k}^{(\alpha)}$};
\draw  [thick, red] (12.4,-2.75) -- (12.6,-2.75);
\draw  [thick, red] (12.4,-2.7) -- (12.6,-2.7);
\draw  [thick, ->-] (12.5,-4) -- (13.5,-5);
\node at  (13.3,-4.4) {\scriptsize $\mu_k$};
\node [blue] at  (12.6,-4.6) {\scriptsize $Q_{mk}^{(\alpha)}$};
\draw  [thick, ->-] (15,-5) -- (13.5,-5);
\node at (15.2,-5) {\scriptsize $\emptyset$};
\draw  [thick, ->-] (13.5,-6.5) -- (13.5,-5);
\node at (13.5,-6.75) {\scriptsize $\nu_{k}^{(\alpha+1)}$};
\draw  [thick, red] (13.4,-6.2) -- (13.6,-6.2);
\draw  [thick, red] (13.4,-6.25) -- (13.6,-6.25);
\end{tikzpicture}
\caption{The ``strip geometry'' for the $\U(k)^N$ theory. The preferred direction is highlighted in red. There are no non-trivial framing factors.}\label{fig: U(k)^N building}
\end{center}
\end{figure}

The contribution to the partition function from this building block is given by
\begin{align}
\mathcal{W}^{(\alpha)}_k\left(\nu^{(\alpha)},\nu^{(\alpha+1)}\right)=\sum_{\vec{\mu},\vec{\lambda}}\prod_{b=1}^k \left(-Q_{mb}^{(\alpha)}\right)^{|\mu_b|}\left(-Q_{fb}^{(\alpha)}\right)^{|\lambda_b|}C_{\mu_b\lambda_{b-1}\nu_{(b,\alpha)}}(\mathfrak{t},\mathfrak{q})C_{\mu_b^\mathfrak{t} \lambda_b^t \nu_{(b,\alpha+1)}^t}(\mathfrak{q},\mathfrak{t})\;,
\end{align}
where we have denoted $\nu^{(\alpha)} = \{ \nu^{(\alpha)}_b\}$ for a fixed $\alpha$ and $\lambda_0 = \lambda_k =\emptyset$. The corresponding dual-toric diagram is given in Fig.~\ref{fig: U(k)^N building}. Using the definitions \eqref{vertex}, as well as the identity $Q^{|\mu|-|\nu|}S_{\mu/\nu}(x) = S_{\mu/\nu}(Q x)$, we then have that 
\begin{align}
  &\mathcal{W}^{(\alpha)}_k(\nu^{(\alpha)},\nu^{(\alpha+1)})=\prod_{b=1}^k \mathfrak{q}^{\frac{\|\nu_{b}^{(\alpha)}\|^2}{2}}\widetilde{Z}_{\nu_{b}^{(\alpha)}}(\mathfrak{t},\mathfrak{q})\mathfrak{t}^{\frac{\|\nu_{b}^{(\alpha+1)t}\|^2}{2}}\widetilde{Z}_{\nu_{b}^{(\alpha+1)t}}(\mathfrak{q},\mathfrak{t})\cr
  &\times\sum_{\substack{\vec{\mu},\vec{\lambda}\\ \vec{\eta},\vec{\xi}}}S_{\mu_b^t / \eta_b}\left(-\frac{Q_{mb}^{(\alpha)}}{\widetilde{Q}^{(\alpha)}_{b} }\mathfrak{t}^{-\rho}\mathfrak{q}^{-\nu_{b}^{(\alpha)}}\right) S_{\lambda_{b-1}/\eta_b}\left(-\frac{\widetilde{Q}_{b}^{(\alpha)}}{ Q_{mb}^{(\alpha)}}\mathfrak{q}^{-\rho-\frac{1}{2}}\mathfrak{t}^{-\nu_{b}^{(\alpha)t}+\frac{1}{2}}\right)\cr
 &\times S_{\mu_b / \xi_b}\left(-\widetilde{Q}_{b}^{(\alpha)} \mathfrak{q}^{-\rho}\mathfrak{t}^{-\nu_{b}^{(\alpha+1)}t}\right)S_{\lambda^t_b / \xi_b}\left(-(\widetilde{Q}_{b}^{(\alpha)})^{-1}\mathfrak{t}^{-\rho-\frac{1}{2}}\mathfrak{q}^{-\nu_{b}^{(\alpha+1)}+\frac{1}{2}}\right)\;,
\end{align}
with $\widetilde{Q}_{b}^{(\alpha)} := \prod_{k=1}^{b-1}Q_{mb}^{(\alpha)}Q_{fk}
^{(\alpha)}$. The sum can be computed exactly with the help of the Cauchy relations
\begin{align}
\sum_{\lambda}Q^{|\lambda |} S_{\lambda / \mu_1}(x)S_{\lambda^t / \mu_2}(y)&= \prod_{i,j=1}^{\infty}(1+Q x_i y_j) \sum_\lambda Q^{|\mu_1|+|\mu_2|-|\lambda|} S_{\mu_2^t / \lambda }(x)S_{\mu_1^t / \lambda^t }(y)\;,\cr
\sum_{\lambda}Q^{|\lambda |} S_{\lambda / \mu_1}(x)S_{\lambda / \mu_2}(y)&= \prod_{i,j=1}^{\infty}(1-Q x_i y_j)^{-1} \sum_\lambda Q^{|\mu_1|+|\mu_2|-|\lambda|} S_{\mu_2 / \lambda }(x)S_{\mu_1 / \lambda }(y)\;,
\end{align}
 to obtain
\begin{align}
\mathcal{W}^{(\alpha)}_k\left(\nu^{(\alpha)},\nu^{(\alpha+1)}\right)&=\prod_{b=1}^k \mathfrak{q}^{\frac{\|\nu_{b}^{(\alpha)}\|^2}{2}}\widetilde{Z}_{\nu_{b}^{(\alpha)}}(\mathfrak{t},\mathfrak{q})\mathfrak{t}^{\frac{\|\nu_{b}^{(\alpha+1)t}\|^2}{2}}\widetilde{Z}_{\nu_{b}^{(\alpha+1)t}}(\mathfrak{q},\mathfrak{t})\cr
&\times \prod_{i,j=1}^{\infty}\frac{\prod_{1\leq b\leq c\leq k}\left(1-\mathbb{Q}^{(\alpha)}_{b c}Q_{mc}^{(\alpha)}\mathfrak{q}^{j-\frac{1}{2}-\nu_{b,i}^{(\alpha)}}\mathfrak{t}^{i-\frac{1}{2}-\nu_{c,j}^{(\alpha+1)t}}\right)}{\prod_{1\leq b<c\leq k}\left(1-\mathbb{Q}^{(\alpha)}_{b c}\mathfrak{q}^{i-1-\nu_{b,j}^{(\alpha)}}\mathfrak{t}^{j- \nu_{c,i}^{(\alpha)t}}\right)}\cr 
&\times \prod_{i,j=1}^{\infty}\frac{\prod_{1\leq b<c\leq k}\left(1-(Q_{mb}^{(\alpha)})^{-1}\mathbb{Q}^{(\alpha)}_{b c}\mathfrak{q}^{j-\frac{1}{2}-\nu_{b,i}^{(\alpha+1)}}\mathfrak{t}^{i-\frac{1}{2}-\nu_{c,j}^{(\alpha)t}}\right)}{\prod_{1\leq b<c\leq k}\left(1-(Q_{mb}^{(\alpha)})^{-1}\mathbb{Q}^{(\alpha)}_{b c}Q_{mc}^{(\alpha)}\mathfrak{q}^{i-\nu_{b,j}^{(\alpha+1)}}\mathfrak{t}^{j-1- \nu_{c,i}^{(\alpha+1)t}}\right)}\;,
\end{align}
where we have defined 
\begin{align}
\mathbb{Q}^{(\alpha)}_{bc} := \prod_{l=b}^{c-1}Q_{ml}^{(\alpha)}Q_{fl}^{(\alpha)}=:\prod_{l=b}^{c-1}Q_{Fl}^{ (\alpha)}\;.
\end{align}
This is the refined version of the strip geometry found in \cite{Taki:2007dh, Bao:2011rc}. 

It is helpful to split the building block into one-loop and  non-perturbative pieces. We define the former as $\mathcal{W}^{(\alpha)}_k(\emptyset,\emptyset)$. The latter is defined as 
\begin{align}
\mathcal{D}^{(\alpha)}_k\left(\nu^{(\alpha)},\nu^{(\alpha+1)}\right) = \frac{\mathcal{W}^{(\alpha)}_k\left(\nu^{(\alpha)},\nu^{(\alpha+1)}\right)}{\mathcal{W}^{(\alpha)}_k(\emptyset,\emptyset)}\;.
\end{align}
Moreover, we  normalise the one-loop piece using the MacMahon function
\begin{align}
\mathcal{M}(Q;\mathfrak{t},\mathfrak{q}):=\prod_{i,j=1}^{\infty}(1-Q\mathfrak{t}^{i-1}\mathfrak{q}^j)^{-1}\;,\qquad M(\mathfrak{t},\mathfrak{q}) = \mathcal{M}(1;\mathfrak{t},\mathfrak{q})\;,
\end{align}
to give
\begin{align} \widehat{\mathcal{W}}^{(\alpha)}_k(\emptyset,\emptyset)=\mathcal{M}(1;\mathfrak{t},\mathfrak{q})^k\mathcal{W}^{(\alpha)}_k(\emptyset,\emptyset)\;.
\end{align}

An additional simplification is afforded to us, since in both examples of interest we will be identifying the top and bottom vertical edges of the dual-toric diagram. For instance, for an $N$-noded circular quiver we identify the indices $\alpha =\lfloor -\frac{N}{2} \rfloor + 1$ with $\alpha = \lfloor \frac{N}{2}\rfloor$. We can use this cyclicity alongside the identity   
\begin{align}
\tilde{Z}_\nu (\mathfrak{t},\mathfrak{q})\tilde{Z}_{\nu^t}(\mathfrak{q},\mathfrak{t})&=\left(-\sqrt{\frac{\mathfrak{q}}{\mathfrak{t}}}\right)^{|\nu|}\mathfrak{t}^{-\frac{\| \nu^t\|^2}{2}}\mathfrak{q}^{-\frac{\| \nu\|^2}{2}}\mathcal{N}_{\nu\nu}(1;\mathfrak{t},\mathfrak{q})^{-1}\;,
\end{align}
to write the non-perturbative piece of the building block as 
\begin{align}
\mathcal{D}^{(\alpha)}_k\left(\nu^{(\alpha)},\nu^{(\alpha+1)}\right)&=\left(-\sqrt{\frac{\mathfrak{q}}{\mathfrak{t}}}\right)^{\sum_{b=1}^k|\nu_b^{(\alpha)}|}\prod_{1\leq b\leq c\leq k}\frac{\mathcal{N}_{\nu_{c}^{(\alpha+1)}\nu_{b}^{(\alpha)}} \left(\mathbb{Q}_{bc}^{(\alpha)} Q_{mc}^{(\alpha)}\sqrt{\frac{\mathfrak{t}}{\mathfrak{q}}};\mathfrak{t},\mathfrak{q}\right)}{\mathcal{N}_{\nu_{c}^{(\alpha+1)}\nu_{b}^{(\alpha+1)}} \left((Q_{mb}^{(\alpha)})^{-1}\mathbb{Q}_{bc}^{(\alpha)}Q_{mc}^{(\alpha)};\mathfrak{t},\mathfrak{q}\right)}\cr 
 &\qquad\times \prod_{1\leq b<c\leq k}\frac{\mathcal{N}_{\nu_{c}^{(\alpha)}\nu_{b}^{(\alpha+1)}} \left( (Q_{mb}^{(\alpha)})^{-1}\mathbb{Q}_{bc}^{(\alpha)}\sqrt{\frac{\mathfrak{t}}{\mathfrak{q}}};\mathfrak{t},\mathfrak{q}\right)}{\mathcal{N}_{\nu_{c}^{(\alpha)}\nu_{b}^{(\alpha)}} \left(\mathbb{Q}_{bc}^{(\alpha)}\frac{\mathfrak{t}}{\mathfrak{q}};\mathfrak{t},\mathfrak{q}\right)}\;.
\end{align}
In the above we have defined the  function
\begin{align}
\mathcal{N}_{\lambda\mu}(Q;\mathfrak{t},\mathfrak{q})&=\prod_{i,j=1}^{\infty}\frac{1-Q \mathfrak{t}^{i-1-\lambda_j^t}\mathfrak{q}^{j-\mu_i}}{1-Q \mathfrak t^{i-1}\mathfrak{q}^j}\cr&=\prod_{(i,j)\in \lambda}(1-Q \mathfrak{q}^{\lambda_i -j+1}\mathfrak{t}^{\mu_j^t -i})\prod_{(i,j)\in \mu}(1-Q \mathfrak{q}^{-\mu_i +j}\mathfrak{t}^{-\lambda_j^t +i-1})\;.
\end{align}
The normalised one-loop piece is simply 
\begin{align}
 \widehat{\mathcal{W}}^{(\alpha)}_k(\emptyset,\emptyset)&=\prod_{i,j=1}^{\infty} \prod_{1\leq b\leq c\leq k}\frac{\left(1-\mathbb{Q}^{(\alpha)}_{b c}Q_{mc}^{(\alpha)}\mathfrak{q}^{j-\frac{1}{2}}\mathfrak{t}^{i-\frac{1}{2}}\right)}{\left(1-(Q_{mb}^{(\alpha)})^{-1}\mathbb{Q}^{(\alpha)}_{b c}Q_{mc}^{(\alpha)}\mathfrak{q}^{i}\mathfrak{t}^{j-1}\right)}\prod_{1\leq b<c\leq k}\frac{\left(1-(Q_{mb}^{(\alpha)})^{-1}\mathbb{Q}^{(\alpha)}_{b c}\mathfrak{q}^{j-\frac{1}{2}}\mathfrak{t}^{i-\frac{1}{2}}\right)}{\left(1-\mathbb{Q}^{(\alpha)}_{b c}\mathfrak{q}^{i-1}\mathfrak{t}^{j}\right)}\;.
\end{align}

\section{Derivation of the 4D Coulomb-Branch Partition Function}\label{AppC}

We will next use the results of App.~\ref{AppA} to calculate the Coulomb-branch partition function for a certain 5D theory on $\mathbb{R}^4_{\epsilon_1, \epsilon_2}\times S_\beta^1$. Upon dimensionally reducing on the circle, this will lead to a derivation of the 4D $\U(k)^N$ circular-quiver Coulomb-branch partition function on $\mathbb{R}^4_{\epsilon_1, \epsilon_2}$.

\subsection{The 5D Coulomb-Branch Partition Function on $\mathbb{R}_{\epsilon_1, \epsilon_2}^4\times S^1_\beta$}\label{AppC.1}

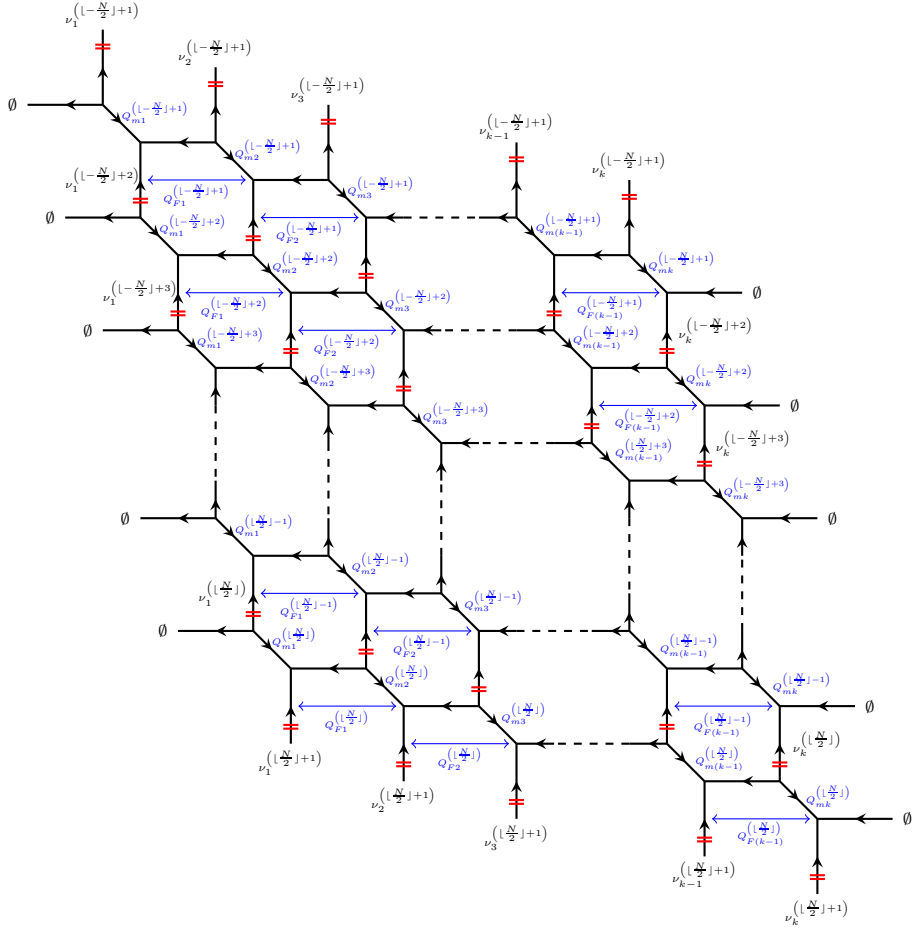
\begin{figure}[t]
\begin{center}
\begin{tikzpicture}
  \tikzset{
  ->-/.style={decoration={markings, mark=at position 0.5 with {\arrow{stealth}}},
              postaction={decorate}},
}
\draw  [thick,thick,->-] (1,0) -- (0,0);
\node at (-0.2,0){\tiny $\emptyset$};
\draw  [thick,thick,->-] (1,0) -- (1,1);
\node [scale=0.7] at (1,1.2) {\tiny $\nu_{1}^{\left(\lfloor -\frac{N}{2}\rfloor+1\right)}$};
\draw  [thick,thick,red] (0.9,0.8) -- (1.1,0.8);
\draw  [thick,thick,red] (0.9,0.75) -- (1.1,0.75);
\draw  [thick,thick,->-] (1,0) -- (1.5,-0.5);
\draw  [thick,thick,->-] (1.5,-1.5) -- (1.5,-0.5);
\draw  [thick,red] (1.4,-1.3) -- (1.6,-1.3);
\draw  [thick,red] (1.4,-1.25) -- (1.6,-1.25);
\node at (0.3,-1.5){\tiny $\emptyset$};
\node at (0.8,-3){\tiny $\emptyset$};
\node at (1.3,-5.5){\tiny $\emptyset$};
\node at (1.8,-7){\tiny $\emptyset$};
\node   [scale=0.7] at (1,-1){\tiny $\nu_{1}^{\left(\lfloor -\frac{N}{2}\rfloor+2\right)}$};
\node  [scale=0.7] at (1.5,-2.5){\tiny $\nu_{1}^{\left(\lfloor -\frac{N}{2}\rfloor+3\right)}$};
\node  [scale=0.7] at (2.6,-6.5){\tiny $\nu_{1}^{\left(\lfloor\frac{N}{2}\rfloor\right)}$};
\draw  [thick,->-] (1.5,-1.5) -- (0.5,-1.5);
\draw  [thick,->-] (1.5,-1.5) -- (2,-2);
\draw  [thick,->-] (3,-2) -- (2,-2);
\draw  [thick,->-] (2,-3) -- (2,-2);
\draw  [thick,red] (1.9,-2.8)-- (2.1,-2.8);
\draw  [thick,red] (1.9,-2.75)-- (2.1,-2.75);
\draw  [thick,->-] (2,-3) -- (1,-3);
\draw  [thick,->-] (2,-3) -- (2.5,-3.5);
\draw  [thick,->-] (3.5,-3.5) -- (2.5,-3.5);
\draw  [thick,->-] (2.5,-4) -- (2.5,-3.5);
\draw  [thick,dashed] (2.5,-4) -- (2.5,-5);
\draw  [thick,->-] (2.5,-5.5) -- (2.5,-5);
\draw  [thick,->-] (2.5,-5.5) -- (1.5,-5.5);
\draw  [thick,->-] (2.5,-5.5) -- (3,-6);
\draw  [thick,->-] (4,-6) -- (3,-6);
\draw  [thick,->-] (3,-7) -- (3,-6);
\draw  [thick,red] (2.9,-6.75) -- (3.1,-6.75);
\draw  [thick,red] (2.9,-6.8) -- (3.1,-6.8);
\draw  [thick,->-] (3,-7) -- (2,-7);
\draw  [thick,->-] (3,-7) -- (3.5,-7.5);
\draw  [thick,->-] (4.5,-7.5) -- (3.5,-7.5);
\draw  [thick,->-] (3.5,-8.5) -- (3.5,-7.5);
\draw  [thick,red] (3.4,-8.25) -- (3.6,-8.25);
\draw  [thick,red] (3.4,-8.3) -- (3.6,-8.3);
\node  [scale=0.7] at (3.5,-8.75) {\tiny $\nu_{1}^{\left(\lfloor\frac{N}{2}\rfloor+1\right)}$};
\draw  [thick,blue,thin,<->]  (1.6,-1)--(2.9,-1);
\node [blue,scale =0.6] at (2.25,-1.2) { {\tiny $Q_{F1}^{\left(\lfloor -\frac{N}{2}\rfloor+1\right)}$}};
\draw  [thick,blue,thin,<->]  (2.1,-2.5)--(3.4,-2.5);
\node [blue,scale =0.6] at (2.75,-2.7) { {\tiny $Q_{F1}^{\left(\lfloor -\frac{N}{2}\rfloor+2\right)}$}};
\draw  [thick,blue,thin,<->]  (3.1,-6.5)--(4.4,-6.5);
\node [blue,scale =0.6] at (3.75,-6.7) { {\tiny $Q_{F1}^{\left(\lfloor\frac{N}{2}\rfloor-1\right)}$}};
\draw  [thick,blue,thin,<->]  (3.6,-8)--(4.9,-8);
\node [blue,scale =0.6] at (4.25,-8.2) { {\tiny $Q_{F1}^{\left(\lfloor\frac{N}{2}\rfloor\right)}$}};
\node [blue,scale = 0.6] at (1.7,-0.1) {\tiny $Q_{m1}^{\left(\lfloor -\frac{N}{2}\rfloor+1\right)}$};
\node [blue,scale = 0.6] at (2.2,-1.6) {\tiny $Q_{m1}^{\left(\lfloor -\frac{N}{2}\rfloor+2\right)}$};
\node [blue,scale = 0.6] at (2.7,-3.1) {\tiny $Q_{m1}^{\left(\lfloor -\frac{N}{2}\rfloor+3\right)}$};
\node [blue,scale = 0.6] at (3.15,-5.6) {\tiny $Q_{m1}^{\left(\lfloor\frac{N}{2}\rfloor-1\right)}$};
\node [blue,scale = 0.6] at (3.55,-7.1) {\tiny $Q_{m1}^{\left(\lfloor\frac{N}{2}\rfloor\right)}$};
\draw  [thick,->-] (2.5,-0.5) -- (1.5,-0.5);
\draw  [thick,->-] (2.5,-0.5) -- (2.5,0.5);
\node [scale=0.7] at (2.5,0.7){\tiny $\nu_{2}^{\left(\lfloor -\frac{N}{2}\rfloor+1\right)}$};
\draw  [thick,red] (2.4,0.3) -- (2.6,0.3);
\draw  [thick,red] (2.4,0.25) -- (2.6,0.25);
\draw  [thick,->-] (2.5,-0.5) -- (3,-1);
\draw  [thick,->-] (3,-2) -- (3,-1);
\draw  [thick,red] (2.9,-1.8) -- (3.1,-1.8);
\draw  [thick,red] (2.9,-1.75) -- (3.1,-1.75);
\draw  [thick,->-] (3,-2) -- (3.5,-2.5);
\draw  [thick,->-] (4.5,-2.5) -- (3.5,-2.5);
\draw  [thick,->-] (3.5,-3.5) -- (3.5,-2.5);
\draw  [thick,red] (3.4,-3.25) -- (3.6,-3.25);
\draw  [thick,red] (3.4,-3.3) -- (3.6,-3.3);
\draw  [thick,->-] (3.5,-3.5) -- (4,-4);
\draw  [thick,->-] (5,-4) -- (4,-4);
\draw  [thick,->-] (4,-4.5) -- (4,-4);
\draw  [thick,dashed] (4,-4.5) -- (4,-5.5);
\draw  [thick,->-] (4,-6) -- (4,-5.5);
\draw  [thick,->-] (4,-6) -- (4.5,-6.5);
\draw  [thick,->-] (5.5,-6.5) -- (4.5,-6.5);
\draw  [thick,->-] (4.5,-7.5) -- (4.5,-6.5);
\draw  [thick,red] (4.4,-7.25) -- (4.6,-7.25);
\draw  [thick,red] (4.4,-7.3) -- (4.6,-7.3);
\draw  [thick,->-] (4.5,-7.5) -- (5,-8);
\draw  [thick,->-] (6,-8) -- (5,-8);
\draw  [thick,->-] (5,-9) -- (5,-8);
\draw  [thick,red] (4.9,-8.75) -- (5.1,-8.75);
\draw  [thick,red] (4.9,-8.8) -- (5.1,-8.8);
\node  [scale=0.7] at (5,-9.25) {\tiny $\nu_{2}^{\left(\lfloor\frac{N}{2}\rfloor+1\right)}$};
\draw  [thick,blue,thin,<->]  (3.1,-1.5)--(4.4,-1.5);
\node [blue,scale =0.6] at (3.75,-1.7) { {\tiny $Q_{F2}^{\left(\lfloor -\frac{N}{2}\rfloor+1\right)}$}};
\draw  [thick,blue,thin,<->]  (3.6,-3)--(4.9,-3);
\node [blue,scale =0.6] at (4.25,-3.2) { {\tiny $Q_{F2}^{\left(\lfloor -\frac{N}{2}\rfloor+2\right)}$}};
\draw  [thick,blue,thin,<->]  (4.6,-7)--(5.9,-7);
\node [blue,scale =0.6] at (5.25,-7.2) { {\tiny $Q_{F2}^{\left(\lfloor\frac{N}{2}\rfloor-1\right)}$}};
\draw  [thick,blue,thin,<->]  (5.1,-8.5)--(6.4,-8.5);
\node [blue,scale =0.6] at (5.75,-8.7) { {\tiny $Q_{F2}^{\left(\lfloor\frac{N}{2}\rfloor\right)}$}};
\node [blue,scale = 0.6] at (3.2,-0.6) {\tiny $Q_{m2}^{\left(\lfloor -\frac{N}{2}\rfloor+1\right)}$};
\node [blue,scale = 0.6] at (3.7,-2.1) {\tiny $Q_{m2}^{\left(\lfloor -\frac{N}{2}\rfloor+2\right)}$};
\node [blue,scale = 0.6] at (4.2,-3.6) {\tiny $Q_{m2}^{\left(\lfloor -\frac{N}{2}\rfloor+3\right)}$};
\node [blue,scale = 0.6] at (4.7,-6.1) {\tiny $Q_{m2}^{\left(\lfloor\frac{N}{2}\rfloor-1\right)}$};
\node [blue,scale = 0.6] at (5.05,-7.6) {\tiny $Q_{m2}^{\left(\lfloor\frac{N}{2}\rfloor\right)}$};
\draw  [thick,->-] (4,-1) -- (3,-1);
\draw  [thick,->-] (4,-1) -- (4,0);
\node  [scale=0.7] at (4,0.2) {\tiny $\nu_{3}^{\left(\lfloor -\frac{N}{2}\rfloor+1\right)}$};
\draw  [thick,red] (3.9,-0.2) -- (4.1,-0.2);
\draw  [thick,red] (3.9,-0.25) -- (4.1,-0.25);
\draw  [thick,->-] (4,-1) -- (4.5,-1.5);
\draw  [thick,->-] (5,-1.5) -- (4.5,-1.5);
\draw  [thick,->-] (4.5,-2.5) -- (4.5,-1.5);
\draw  [thick,red] (4.4,-2.3) -- (4.6,-2.3);
\draw  [thick,red] (4.4,-2.25) -- (4.6,-2.25);
\draw  [thick,->-] (4.5,-2.5) -- (5,-3);
\draw  [thick,->-] (5,-4) -- (5,-3); 
\draw  [thick,red] (4.9,-3.75) -- (5.1,-3.75);
\draw  [thick,red] (4.9,-3.8) -- (5.1,-3.8);
\draw  [thick,->-] (5.5,-3) -- (5,-3);
\draw  [thick,dashed] (6.5,-3)--(5.5,-3);
\draw  [thick,->-] (7,-3)--(6.5,-3);
\draw  [thick,->-] (5,-4) -- (5.5,-4.5);
\draw  [thick,->-] (6,-4.5) --(5.5,-4.5);
\draw  [thick,dashed] (7,-4.5) -- (6,-4.5);
\draw  [thick,->-] (7.5,-4.5) -- (7,-4.5);
\draw  [thick,->-] (5.5,-5) -- (5.5,-4.5);
\draw  [thick,dashed] (5.5,-6) -- (5.5,-5);
\draw  [thick,->-] (5.5,-6.5) -- (5.5,-6);
\draw  [thick,->-] (5.5,-6.5) -- (6,-7);
\draw  [thick,->-] (6.5,-7) -- (6,-7);
\draw  [thick,dashed] (6.5,-7)--(7.5,-7);
\draw  [thick,->-] (8,-7) -- (7.5,-7);
\draw  [thick,->-] (6,-8) -- (6,-7);
\draw  [thick,red] (5.9,-7.75) -- (6.1,-7.75);
\draw  [thick,red] (5.9,-7.8) -- (6.1,-7.8);
\draw  [thick,->-] (6,-8) -- (6.5,-8.5);
\draw  [thick,->-] (6.5,-9.5) -- (6.5,-8.5);
\draw  [thick,red] (6.4,-9.25) -- (6.6,-9.25);
\draw  [thick,red] (6.4,-9.3) -- (6.6,-9.3);
\draw  [thick,->-] (7,-8.5) -- (6.5,-8.5);
\draw  [thick,dashed] (8,-8.5) -- (7,-8.5);
\draw  [thick,->-] (8.5,-8.5) -- (8,-8.5);
\node  [scale=0.7] at (6.5,-9.75) {\tiny $\nu_{3}^{\left(\lfloor\frac{N}{2}\rfloor+1\right)}$};
\node [blue,scale = 0.6] at (4.7,-1.1) {\tiny $Q_{m3}^{\left(\lfloor -\frac{N}{2}\rfloor+1\right)}$};
\node [blue,scale = 0.6] at (5.2,-2.6) {\tiny $Q_{m3}^{\left(\lfloor -\frac{N}{2}\rfloor+2\right)}$};
\node [blue,scale = 0.6] at (5.7,-4.1) {\tiny $Q_{m3}^{\left(\lfloor -\frac{N}{2}\rfloor+3\right)}$};
\node [blue,scale = 0.6] at (6.2,-6.6) {\tiny $Q_{m3}^{\left(\lfloor\frac{N}{2}\rfloor-1\right)}$};
\node [blue,scale = 0.6] at (6.6,-8.1) {\tiny $Q_{m3}^{\left(\lfloor\frac{N}{2}\rfloor\right)}$};
\draw  [thick,dashed]  (5,-1.5) -- (6,-1.5);
\draw  [thick,->-] (6.5,-1.5) -- (6,-1.5);
\draw  [thick,->-] (6.5,-1.5) -- (6.5,-0.5);
\node  [scale=0.7] at (6.5,-0.3) {\tiny $\nu_{k-1}^{\left(\lfloor -\frac{N}{2}\rfloor+1\right)}$};
\draw  [thick,red] (6.4,-0.7) -- (6.6,-0.7);
\draw  [thick,red] (6.4,-0.75) -- (6.6,-0.75);
\draw  [thick,->-] (6.5,-1.5) -- (7,-2);
\draw  [thick,->-] (8,-2) -- (7,-2);
\draw  [thick,->-] (7,-3) -- (7,-2);
\draw  [thick,red] (6.9,-2.8) -- (7.1,-2.8);
\draw  [thick,red] (6.9,-2.75) -- (7.1,-2.75);
\draw  [thick,->-] (7,-3) -- (7.5,-3.5);
\draw  [thick,->-] (8.5,-3.5) -- (7.5,-3.5);
\draw  [thick,->-] (7.5,-4.5) -- (7.5,-3.5);
\draw  [thick,red] (7.4,-4.25) -- (7.6,-4.25);
\draw  [thick,red] (7.4,-4.3) -- (7.6,-4.3);
\draw  [thick,->-] (7.5,-4.5) -- (8,-5);
\draw  [thick,->-] (9,-5) -- (8,-5);
\draw  [thick,->-] (8,-5.5) -- (8,-5);
\draw  [thick,dashed] (8,-5.5)--(8,-6.5);
\draw  [thick,->-] (8,-7)--(8,-6.5);
\draw  [thick,->-] (8,-7) -- (8.5,-7.5);
\draw  [thick,->-] (9.5,-7.5) -- (8.5,-7.5);
\draw  [thick,->-] (8.5,-8.5) -- (8.5,-7.5);
\draw  [thick,red] (8.4,-8.25)-- (8.6,-8.25);
\draw  [thick,red] (8.4,-8.3) -- (8.6,-8.3);
\draw  [thick,->-] (8.5,-8.5) -- (9,-9);
\draw  [thick,->-] (9,-10)--(9,-9);
\draw  [thick,red] (8.9,-9.75) -- (9.1,-9.75);
\draw  [thick,red] (8.9,-9.8) -- (9.1,-9.8);
\node  [scale=0.7] at (9,-10.25) {\tiny $\nu_{k-1}^{\left(\lfloor\frac{N}{2}\rfloor+1\right)}$};
\draw  [thick,blue,thin,<->]  (7.1,-2.5)--(8.4,-2.5);
\node [blue,scale =0.6] at (7.75,-2.7) { {\tiny $Q_{F(k-1)}^{\left(\lfloor -\frac{N}{2}\rfloor+1\right)}$}};
\draw  [thick,blue,thin,<->]  (7.6,-4)--(8.9,-4);
\node [blue,scale =0.6] at (8.25,-4.2) { {\tiny $Q_{F(k-1)}^{\left(\lfloor -\frac{N}{2}\rfloor+2\right)}$}};
\draw  [thick,blue,thin,<->]  (8.6,-8)--(9.9,-8);
\node [blue,scale =0.6] at (9.25,-8.25) { {\tiny $Q_{F(k-1)}^{\left(\lfloor\frac{N}{2}\rfloor-1\right)}$}};
\draw  [thick,blue,thin,<->]  (9.1,-9.5)--(10.4,-9.5);
\node [blue,scale =0.6] at (9.75,-9.7) { {\tiny $Q_{F(k-1)}^{\left(\lfloor\frac{N}{2}\rfloor\right)}$}};
\node [blue,scale = 0.6] at (7.2,-1.6) {\tiny $Q_{m(k-1)}^{\left(\lfloor -\frac{N}{2}\rfloor+1\right)}$};
\node [blue,scale = 0.6] at (7.7,-3.1) {\tiny $Q_{m(k-1)}^{\left(\lfloor -\frac{N}{2}\rfloor+2\right)}$};
\node [blue,scale = 0.6] at (8.2,-4.6) {\tiny $Q_{m(k-1)}^{\left(\lfloor\frac{N}{2}\rfloor+3\right)}$};
\node [blue,scale = 0.6] at (8.8,-7.2) {\tiny $Q_{m(k-1)}^{\left(\lfloor\frac{N}{2}\rfloor-1\right)}$};
\node [blue,scale = 0.6] at (9.2,-8.7) {\tiny $Q_{m(k-1)}^{\left(\lfloor\frac{N}{2}\rfloor\right)}$};
\draw  [thick,->-] (10,-9) -- (9,-9);
\draw  [thick,->-] (8,-2) -- (8,-1);
\node  [scale=0.7] at (8,-0.8) {\tiny $\nu_{k}^{\left(\lfloor -\frac{N}{2}\rfloor+1\right)}$};
\draw  [thick,red] (7.9,-1.2) -- (8.1,-1.2);
\draw  [thick,red] (7.9,-1.25) -- (8.1,-1.25);
\draw  [thick,->-] (8,-2) -- (8.5,-2.5);
\draw  [thick,->-] (9.5,-2.5) -- (8.5,-2.5);
\node at (9.7,-2.5) {\tiny $\emptyset$};
\node at (10.2,-4){\tiny $\emptyset$};
\node at (10.7,-5.5){\tiny $\emptyset$};
\node at (11.2,-8){\tiny $\emptyset$};
\node at (11.7,-9.5){\tiny $\emptyset$};
\node [scale=0.7] at (9.15,-3){\tiny $\nu_{k}^{\left(\lfloor -\frac{N}{2}\rfloor+2\right)}$};
\node [scale=0.7] at (9.65,-4.5){\tiny $\nu_{k}^{\left(\lfloor -\frac{N}{2}\rfloor+3\right)}$};
\node  [scale=0.7] at (10.5,-8.5){\tiny $\nu_{k}^{\left(\lfloor\frac{N}{2}\rfloor\right)}$};
\draw  [thick,->-] (8.5,-3.5) -- (8.5,-2.5);
\draw  [thick,red] (8.4,-3.3) -- (8.6,-3.3);
\draw  [thick,red] (8.4,-3.25) -- (8.6,-3.25);
\draw  [thick,->-] (8.5,-3.5) -- (9,-4);
\draw  [thick,->-] (10,-4) -- (9,-4);
\draw  [thick,->-] (9,-5) -- (9,-4);
\draw  [thick,red] (8.9,-4.75) -- (9.1,-4.75);
\draw  [thick,red] (8.9,-4.8) -- (9.1,-4.8);
\draw  [thick,->-] (9,-5) -- (9.5,-5.5);
\draw  [thick,->-] (10.5,-5.5) -- (9.5,-5.5);
\draw  [thick,->-] (9.5,-6) -- (9.5,-5.5);
\draw  [thick,dashed] (9.5,-7) --(9.5,-6);
\draw  [thick,->-] (9.5,-7.5) -- (9.5,-7);
\draw  [thick,->-] (9.5,-7.5) -- (10,-8);
\draw  [thick,->-] (11,-8) -- (10,-8);
\draw  [thick,->-] (10,-9) -- (10,-8); 
\draw  [thick,red] (9.9,-8.75) -- (10.1,-8.75);
\draw  [thick,red] (9.9,-8.8) -- (10.1,-8.8);
\draw  [thick,->-] (10,-9) -- (10.5,-9.5);
\draw  [thick,->-] (11.5,-9.5) -- (10.5,-9.5);
\draw  [thick,->-] (10.5,-10.5) -- (10.5,-9.5);
\draw  [thick,red] (10.4,-10.25) -- (10.6,-10.25);
\draw  [thick,red] (10.4,-10.3) -- (10.6,-10.3);
\node  [scale=0.7] at (10.5,-10.75) {\tiny $\nu_{k}^{\left(\lfloor\frac{N}{2}\rfloor+1\right)}$};.\node [blue,scale = 0.6] at (8.7,-2.1) {\tiny $Q_{mk}^{\left(\lfloor -\frac{N}{2}\rfloor+1\right)}$};
\node [blue,scale = 0.6] at (9.2,-3.6) {\tiny $Q_{mk}^{\left(\lfloor -\frac{N}{2}\rfloor+2\right)}$};
\node [blue,scale = 0.6] at (9.7,-5.1) {\tiny $Q_{mk}^{\left(\lfloor -\frac{N}{2}\rfloor+3\right)}$};
\node [blue,scale = 0.6] at (10.3,-7.7) {\tiny $Q_{mk}^{\left(\lfloor\frac{N}{2}\rfloor-1\right)}$};
\node [blue,scale = 0.6] at (10.65,-9.2) {\tiny $Q_{mk}^{\left(\lfloor\frac{N}{2}\rfloor\right)}$};
\end{tikzpicture}
\caption{The dual-toric diagram for the $\U(k)^N$ theory. The preferred direction is along the compactified circle and is highlighted in red. There are no non-trivial framing factors. We close the quiver by identifying the partitions $\nu_b^{\left(\lfloor\frac{N}{2}\rfloor+1\right)}=\nu_b^{\left(\lfloor-\frac{N}{2}\rfloor+1\right)}$.}\label{fig: U(k)^N diag}
\end{center}
\end{figure}

The Coulomb-branch partition function for the 5D theory depicted in Fig.~\ref{fig: U(k)^N diag} can be obtained by fusing the building blocks $\widehat{\mathcal{W}}_k^{(\alpha)}$ and $\mathcal D_k^{(\alpha)}$ with the help of the glueing parameters $\sum_{\nu^{(\alpha)}}(-Q_g^{(\alpha)})^{\sum_{b = 1}^k |\nu_b^{(\alpha)}|}$ and the cyclic identification of  partitions $\nu_{b}^{(\lfloor N/2 \rfloor +1)} = \nu_{b}^{(\lfloor -N/2\rfloor] +1)}$. 
This results in the partition function
\begin{align}
Z_{\rm 5D}^{\U(k)^N} = \sum_{\nu}\prod_{\alpha=\lfloor - \frac{N}{2}\rfloor +1}^{\lfloor \frac{N}{2}\rfloor} \widehat{\mathcal{W}}^{(\alpha)}_k\left(\emptyset,\emptyset\right) (-Q_{g}^{(\alpha)})^{\sum_{b = 1}^k |\nu_b^{(\alpha)}|}\mathcal{D}^{(\alpha)}_k\left(\nu^{(\alpha)},\nu^{(\alpha+1)}\right)\;,
\end{align}
where $\nu = \{ \nu_b^{(\alpha)}\}$. At this stage we can define the ``physical parameters'' (where $\beta$ is the radius of the $S^1$)
\begin{align}
\mathbf{q}_{(\alpha)} = Q_{gb}^{(\alpha)}
  \sqrt{Q_{mb}^{(\alpha)}Q_{mb}^{(\alpha+1)}}\quad\forall\; b \;,\quad Q_{Fb}^{(\alpha)} = Q_{fb}^{(\alpha)} Q_{mb}^{(\alpha)} = e^{\beta F_{b}^{(\alpha)}}\;,\quad  \mathbb{Q}^{(\alpha)}_{bc} = \prod_{l=b}^{c-1}Q_{Fl}^{(\alpha)}\;,
\end{align}
alongside $Q_{mb}^{(\alpha)} = e^{\beta m_b^{(\alpha)}}$ to explicitly write:
\begin{align}
Z_{\rm 5D}^{\U(k)^N}=Z_{\text{5D,1-loop}}^{\U(k)^N}Z_{\text{5D,inst}}^{\U(k)^N}\;,
\end{align}
with
\begin{align}\label{circinst}
Z_{\text{5D,inst}}^{\U(k)^N} &=\sum_{\nu}\prod_{\alpha=\lfloor - \frac{N}{2}\rfloor +1}^{\lfloor \frac{N}{2}\rfloor}  (-Q_{g}^{(\alpha)})^{\sum_{b=1}^k|\nu_b^{(\alpha)}|}\mathcal{D}^{(\alpha)}_k \left(\nu^{(\alpha)},\nu^{(\alpha+1)}\right)\cr
& =\sum_{\nu} \prod_{\alpha=\lfloor - \frac{N}{2}\rfloor +1}^{\lfloor \frac{N}{2}\rfloor}\left(\mathbf{q}_{(\alpha)}\left(Q_{mb}^{(\alpha)}Q_{mb}^{(\alpha+1)}\right) ^{-\frac{1}{2}}\sqrt{\frac{\mathfrak{q}}{\mathfrak{t}}}\right)^{\sum_{b=1}^k|\nu_b^{(\alpha)}|}\cr 
\times\prod_{1\leq b\leq c\leq k}&\frac{\mathcal{N}_{\nu_{c}^{(\alpha+1)}\nu_{b}^{(\alpha)}} \left(\mathbb{Q}_{bc}^{(\alpha)} Q_{mc}^{(\alpha)}\sqrt{\frac{\mathfrak{t}}{\mathfrak{q}}};\mathfrak{t}, \mathfrak{q}\right)}{\mathcal{N}_{\nu_{c}^{(\alpha+1)}\nu_{b}^{(\alpha+1)}} \left((Q_{mb}^{(\alpha)})^{-1}\mathbb{Q}_{bc}^{(\alpha)}Q_{mc}^{(\alpha)};\mathfrak{t}, \mathfrak{q}\right)} \prod_{1\leq b<c\leq k}\frac{\mathcal{N}_{\nu_{c}^{(\alpha)}\nu_{b}^{(\alpha+1)}} \left( (Q_{mb}^{(\alpha)})^{-1}\mathbb{Q}_{bc}^{(\alpha)}\sqrt{\frac{\mathfrak{t}}{\mathfrak{q}}}; \mathfrak{t},\mathfrak{q}\right)}{\mathcal{N}_{\nu_{c}^{(\alpha)}\nu_{b}^{(\alpha)}}  \left(\mathbb{Q}_{bc}^{(\alpha)}\frac{\mathfrak{t}}{\mathfrak{q}};\mathfrak{t}, \mathfrak{q}\right)}\;\cr
\end{align}
and
\begin{align}\label{circpert}
Z_{\text{5D,1-loop}}^{\U(k)^N} &=\prod_{\alpha=\lfloor - \frac{N}{2}\rfloor +1}^{\lfloor \frac{N}{2}\rfloor}\widehat{\mathcal{W}}^{(\alpha)}_k(\emptyset,\emptyset)\cr
&=\prod_{\alpha=\lfloor - \frac{N}{2}\rfloor +1}^{\lfloor \frac{N}{2}\rfloor} \prod_{i,j=1}^{\infty} \prod_{1\leq b\leq c\leq k}\frac{\left(1-\mathbb{Q}^{(\alpha)}_{b c}Q_{mc}^{(\alpha)}\mathfrak{q}^{j-\frac{1}{2}} \mathfrak{t}^{i-\frac{1}{2}}\right)}{\left(1-(Q_{mb}^{(\alpha)})^{-1}\mathbb{Q}^{(\alpha)}_{b c}Q_{mc}^{(\alpha)}\mathfrak{q}^{i}\mathfrak{t}^{j-1}\right)}\cr 
&\qquad\qquad\times\prod_{1\leq b<c\leq k}\frac{\left(1-(Q_{mb}^{(\alpha)})^{-1}\mathbb{Q}^{(\alpha)}_{b c}\mathfrak{q}^{j-\frac{1}{2}}\mathfrak{t}^{i-\frac{1}{2}}\right)}{\left(1-\mathbb{Q}^{(\alpha)}_{b c}\mathfrak{q}^{i-1}\mathfrak{t}^{j}\right)}\;.
\end{align}

\subsection{The 4D Coulomb-Branch Partition Function on $\mathbb{R}^4_{\epsilon_1,\epsilon_2}$} \label{AppC.2}

The 4D $\mathcal{N}=2$ $\U(k)^N$ circular-quiver result is obtained by reducing the 5D answer along the circle.
\begin{align}
Z_{4D}^{\U(k)^N} = \lim_{\beta\rightarrow 0} Z_{5D}^{\U(k)^N}\;.
\end{align}
This operation can be straightforwardly performed, since having a Lagrangian description implies that the limit and sum over partitions commute \cite{Mitev:2014isa}. With respect to the non-perturbative piece, we will convert the functions $\mathcal{N}_{\lambda\nu}(Q;\mathfrak{t},\mathfrak{q})$ into the so-called ``Nekrasov functions'' $\rm{N}^{\beta}_{\lambda\nu}(m;\epsilon_1,\epsilon_2)$, the behaviour of which is very simple in the $\beta \rightarrow 0$ limit. We will reduce the two parts of the partition function separately, since they require slightly different approaches.

\subsubsection{The  One-Loop Piece}

First let us consider the normalised one-loop piece of the 5D $\U(k)^N$ Coulomb-branch partition function. In terms of chemical potentials we have that
\begin{align}
  \label{eq:21}
 \mathbb{Q}_{bc}^{(\alpha)} = \exp\left[\beta\sum_{l=b}^{c-1}F_{l}^{(\alpha)}\right]=\exp \left[\beta \mathbb{F}^{(\alpha)}_{bc}\right]\;.  
\end{align}
We can then write  
\begin{align}
&Z_{\text{5D, 1-loop}}^{\U(k)^N} = \prod_{\alpha=\lfloor - \frac{N}{2}\rfloor +1}^{\lfloor \frac{N}{2}\rfloor}\prod_{i,j=1}^{\infty}\prod_{1\leq b<c\leq k}\frac{\sinh\frac{-\beta}{2}\left[\mathbb{F}_{bc}^{(\alpha)} -m_b^{(\alpha)} -\epsilon_1(j-1/2)+\epsilon_2(i-1/2)\right]}{\sinh\frac{-\beta}{2}\left[\mathbb{F}_{bc}^{(\alpha)} -\epsilon_1(i-1)+\epsilon_2 j\right]} \cr 
&\qquad\qquad\prod_{1\leq b< c\leq k}\frac{\sinh\frac{-\beta}{2}\left[m_c^{(\alpha)}+\mathbb{F}_{bc}^{(\alpha)} -\epsilon_1(j-1/2)+\epsilon_2(i-1/2)\right]}{\sinh\frac{-\beta}{2}\left[m_c^{(\alpha)}+\mathbb{F}_{bc}^{(\alpha)}-m_b^{(\alpha)} -\epsilon_1 i+\epsilon_2 (j-1)\right]}\cr
&\qquad \qquad\prod_{1\leq b\leq k}\frac{\sinh\frac{-\beta}{2}\left[m_b^{(\alpha)} -\epsilon_1(j-1/2)+\epsilon_2(i-1/2)\right]}{\sinh\frac{-\beta}{2}\left[ -\epsilon_1 i+\epsilon_2 (j-1)\right]}\exp\frac{-\beta}{2}\left[m_b^{(\alpha)} + \frac{\epsilon_+}{2}\right]\;,
\end{align}
where we have defined $\epsilon_+=\epsilon_1+\epsilon_2$. Upon taking the $\beta\rightarrow 0$ limit we find that 
\begin{align}\label{pertred}
&Z_{\text{4D,1-loop}}^{\U(k)^N} =\prod_{i,j=1}^{\infty}\prod_{\alpha=\lfloor - \frac{N}{2}\rfloor +1}^{\lfloor \frac{N}{2}\rfloor}\prod_{1\leq b\leq c\leq k}\frac{\left[-\mathbb{F}_{bc}^{(\alpha)}-m_c^{(\alpha)} +\epsilon_1(j-1/2)-\epsilon_2(i-1/2)\right]}{\left[- \mathbb{F}_{bc}^{(\alpha+1)} +\epsilon_1 i-\epsilon_2 (j-1)\right]}\cr 
&\qquad \qquad\times\prod_{i,j=1}^{\infty}\prod_{\alpha=\lfloor - \frac{N}{2}\rfloor +1}^{\lfloor \frac{N}{2}\rfloor}\prod_{1\leq b<c\leq k}\frac{\left[-\mathbb{F}_{bc}^{(\alpha)} +m_b^{(\alpha)} +\epsilon_1(j-1/2)-\epsilon_2(i-1/2)\right]}{\left[-\mathbb{F}_{bc}^{(\alpha)} +\epsilon_1(i-1)-\epsilon_2 j\right]}\;.
\end{align}

\subsubsection{The Non-Perturbative Piece}

To bring the instanton partition function into a more suitable form, we  re-express the $\mathcal{N}_{\lambda\mu}(Q;\mathfrak{t},\mathfrak{q})$ functions as 
\begin{align}
\mathcal{N}_{\lambda \mu}(Q;\mathfrak{t},\mathfrak{q}) = \left(Q\sqrt{\frac{\mathfrak{q}}{\mathfrak{t}}}\right)^{\frac{|\lambda|+|\mu|}{2}}\mathfrak{t}^{\frac{\|\mu^t\|^2-\|\lambda^t\|^2}{4}}\mathfrak{q}^{\frac{\|\lambda\|^2-\|\mu\|^2}{4}}\rm{N}^\beta_{\lambda\mu}(-m;\epsilon_1,\epsilon_2)\;,
\end{align}
for some generic $Q = e^{\beta m}$. The ``Nekrasov functions'' appearing above are of the form 
\begin{align}
\textrm{N}^{\beta}_{\lambda\mu}(-m;\epsilon_1,\epsilon_2)=\prod_{(i,j)\in \lambda}2 \sinh \frac{\beta}{2}\left[-m+\epsilon_1(\lambda_i-j+1)+\epsilon_2(i-\mu_j^t)\right]
\cr\times \prod_{(i,j)\in \mu}2 \sinh \frac{\beta}{2}\left[-m+\epsilon_1(j-\mu_i)+\epsilon_2(\lambda_j^t -i+1)\right]\;,
\end{align} 
and their behaviour in the $\beta \rightarrow 0$ limit is simply $\rm{N}^{\beta}_{\lambda\mu} \xrightarrow{\beta\rightarrow 0} \beta^{|\lambda|+|\mu|}\rm{N}_{\lambda\mu}$ with 
\begin{align}
\textrm{N}_{\lambda\nu}(-m;\epsilon_1,\epsilon_2)=\prod_{(i,j)\in \lambda}\left[-m+\epsilon_1(\lambda_i-j+1)+\epsilon_2(i-\mu_j^t)\right]
\cr\times \prod_{(i,j)\in \mu}\left[-m+\epsilon_1(j-\mu_i)+\epsilon_2(\lambda_j^t -i+1)\right]\;.
\end{align}
We therefore obtain after the reduction
\begin{align}\label{instred}
Z_{\text{4D,inst}}^{\U(k)^N} &=\sum_{\nu}\prod_{\alpha=\lfloor - \frac{N}{2}\rfloor +1}^{\lfloor \frac{N}{2}\rfloor} \mathbf{q}_{(\alpha)}^{\sum_{b=1}^k|\nu_b^{(\alpha)}|} \prod_{1\leq b\leq c\leq k}\frac{{\rm N}_{\nu_{c}^{(\alpha+1)}\nu_{b}^{(\alpha)}}\left(-\mathbb{F}_{bc}^{(\alpha)} - m^{(\alpha)}_c -\frac{\epsilon_+}{2}\right)}{{\rm N}_{\nu_{c}^{(\alpha+1)}\nu_{b}^{(\alpha+1)}}\left(-\mathbb{F}_{bc}^{(\alpha+1)}\right)}\cr 
&\times \prod_{1\leq b<c\leq k}\frac{{\rm N}_{\nu_{c}^{(\alpha)}\nu_{b}^{(\alpha+1)}}\left(-\mathbb{F}_{bc}^{(\alpha)} + m^{(\alpha)}_b -\frac{\epsilon_+}{2}\right)}{{\rm N}_{\nu_{c}^{(\alpha)}\nu_{b}^{(\alpha)}}\left(-\mathbb{F}_{bc}^{(\alpha)} -\epsilon_+\right)}\;,
\end{align}
where we have used the shorthand $\rm{N}^\beta (-m;\epsilon_1,\epsilon_2) = \rm{N}^\beta (-m)$.

\subsection{Comparison with Localisation Results}

We will finally compare the results we obtained using the refined topological vertex with those from supersymmetric localisation \cite{Pestun:2007rz,Alday:2009aq}. We start by providing a quick map between the two formalisms and rewriting the various parameters in terms of the 4D Coulomb parameters $a_{b}^{(\alpha)}$ and $m_{\rm bif}^{(\alpha)}$, which from Eqs.~\eqref{eq:8}-\eqref{biff-def} read:
\begin{align}
\mathbb{F}^{(\alpha)}_{bc} = a^{(\alpha)}_{c} -a^{(\alpha)}_{b} \;,\qquad m_b^{(\alpha)} = a_b^{(\alpha+1)} - a_b^{(\alpha)}+m_{\text{bif}}^{(\alpha)}\;.
\end{align}

The non-perturbative piece of the partition function \eqref{instred} can be simplified by modifying the indices of its second line; swapping $b\leftrightarrow c$ and using the exchange relations \cite{Mitev:2014isa}
\begin{align}
\label{eq:propertiestN}
{\rm N}_{\lambda\mu}(m;-\epsilon_2,-\epsilon_1)&= {\rm N}_{\mu^t\lambda^t}(m-\epsilon_1-\epsilon_2;\epsilon_1,\epsilon_2),\nonumber\\
{\rm N}_{\lambda\mu}(-m;\epsilon_1,\epsilon_2)&=(-1)^{|\lambda|+|\mu|} {\rm N}_{\mu\lambda}(m-\epsilon_1-\epsilon_2;\epsilon_1,\epsilon_2),\\
 {\rm N}_{\lambda\mu}(m;\epsilon_2,\epsilon_1)&= {\rm N}_{\lambda^t\mu^t}(m;\epsilon_1,\epsilon_2),\nonumber
\end{align}
we can absorb it into the first line. By additionally using the cyclicity of the $\alpha$ index, we obtain 
\begin{align}\label{toptolocinst}
Z_{\text{4D,inst}}^{\U(k)^N} &=\sum_{\nu}\prod_{\alpha =\lfloor -\frac{N}{2}\rfloor+1}^{\lfloor \frac{N}{2}\rfloor } \mathbf{q}_\alpha^{\sum_{b=1}^k|\nu_{b}^{(\alpha)}|} \prod_{b,c=1}^k\frac{{\rm N}_{\nu_{c}^{(\alpha+1)}\nu_{b}^{(\alpha)}}\left(a_b^{(\alpha)} -a^{(\alpha+1)}_c -m_{\text{bif}}^{(\alpha)}-\frac{\epsilon_+}{2}\right)}{{\rm N}_{\nu_{c}^{(\alpha)}\nu_{b}^{(\alpha)}}\left(a^{(\alpha)}_b -a^{(\alpha)}_c\right)}\;.
\end{align}

We next move on to the one-loop piece. This can also be expressed in terms of the above 4D variables as
\begin{align}\label{4DPert Coulomb}
Z_{\text{4D,1-loop}}^{\U(k)^N} &=\prod_{i,j=1}^{\infty}\prod_{\alpha =\lfloor -\frac{N}{2}\rfloor+1}^{\lfloor \frac{N}{2}\rfloor } \prod_{1\leq b\leq c\leq k}\frac{\left[a_b^{(\alpha)}-a_c^{(\alpha+1)}-m_{ \text{bif}}^{(\alpha)} + \epsilon_1 (j-1) - i\epsilon_2\right]}{\left[a^{(\alpha)}_b-a^{(\alpha)}_c +i\epsilon_1 -\epsilon_2 (j-1)\right]}\cr 
&\qquad \times \prod_{1\leq b<c\leq k}\frac{\left[a_b^{(\alpha+1)}-a_c^{(\alpha)}+m_{ \text{bif}}^{(\alpha)} +j\epsilon_1-\epsilon_2(i-1)\right]}{\left[a^{(\alpha)}_b-a^{(\alpha)}_c + \epsilon_1(i-1)- j\epsilon_2\right]}\;.
\end{align}
and treated in a similar way. That is, one can swap the indices $b\leftrightarrow c$ for one part of the two products in \eqref{4DPert Coulomb}, use the product representation of Barnes' double Gamma function for $\epsilon_1>0 $, $\epsilon_2<0$ \cite{Alday:2009aq}\footnote{Recall that our fugacities are given by $\mathfrak{q}= e^{-\beta \epsilon_1}$ and $\mathfrak{t}= e^{\beta \epsilon_2}$. The choice $\epsilon_1 >0$ and $\epsilon_2 <0$ ensures that the 5D partition functions are well defined.}
\begin{align}\label{barnes doubGamma}
\Gamma_2(x|\epsilon_1,\epsilon_2) \propto \prod_{i,j\geq 1}(x +\epsilon_1(i-1) -j \epsilon_2)
\end{align}
and apply the identity 
\begin{align}\label{barnesid}
\Gamma_2(-x|\epsilon_1,\epsilon_2) = \Gamma_2(x+\epsilon_+|\epsilon_1,\epsilon_2)
\end{align}
to obtain 
\begin{align}\label{toptolocpert}
Z_{\text{4D,1-loop}}^{\U(k)^N} &=\prod_{\alpha =\lfloor -\frac{N}{2}\rfloor+1}^{\lfloor \frac{N}{2}\rfloor }\prod_{b,c=1}^k \frac{\Gamma_2\left(a_b^{(\alpha)}-a_c^{(\alpha+1)} -m_{ \text{bif}}^{(\alpha)}+\frac{\epsilon_+}{2}\big|\epsilon_1,\epsilon_2\right) }{\Gamma_2\left(a_b^{(\alpha)}-a_c^{(\alpha)}\big|\epsilon_1,\epsilon_2\right)}\;.
\end{align}

Combining \eqref{toptolocpert} with \eqref{toptolocinst} one gets a 4D partition function which matches the results of \cite{Alday:2009aq,Hama:2012bg,Nekrasov:2003rj}.

\section{Derivation of the 6D Tensor-Branch Partition Function}\label{AppB}

In this section we introduce the tensor-branch partition function of the 6D $\text{(2,0)}_k$ theory, which emerges in the low-energy limit of $k$ separated M5 branes on $\mathbb{R}^4_{\epsilon_1,\epsilon_2}\times T^2_{R_5, R_6}$---that is the $A_{k-1}$ theory plus a free tensor multiplet---following \cite{Haghighat:2013gba}. 

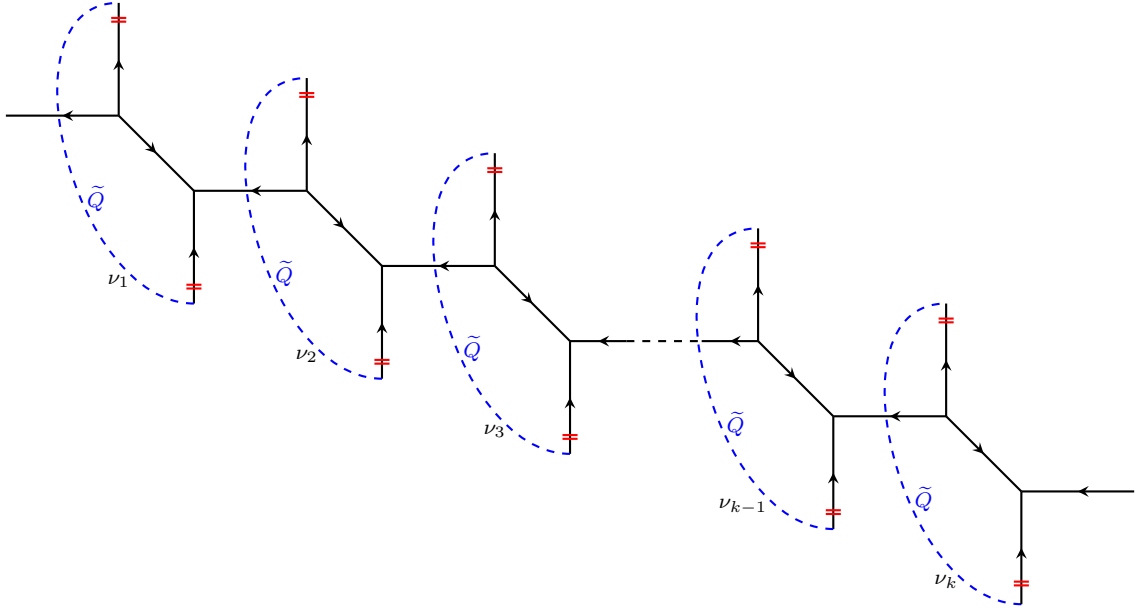
\begin{figure}[t]
\begin{center}
\begin{tikzpicture}
\tikzset{
  ->-/.style={decoration={markings, mark=at position 0.5 with {\arrow{stealth}}},
              postaction={decorate}},
}
\draw  [thick, ->-] (1.5,0) -- (0,0);
\draw  [thick, ->-] (1.5,0) -- (1.5,1.5);
\node  [blue] at (1.2,-1.1) {\scriptsize $\widetilde{Q}$};
\node at (1.5,-2.2) {\scriptsize $\nu_{1}$};
\draw  [thick, red] (1.4,1.3) -- (1.6,1.3);
\draw  [thick, red] (1.4,1.25) -- (1.6,1.25);
\draw  [thick, ->-] (1.5,0) -- (2.5,-1);
\draw  [thick, ->-] (2.5,-2.5) -- (2.5,-1);
\draw  [thick, blue, dashed] (2.5,-2.5) to [out=180,in=180] (1.5,1.5);
\draw  [thick, red] (2.4,-2.3) -- (2.6,-2.3);
\draw  [thick, red] (2.4,-2.25) -- (2.6,-2.25);
\draw  [thick, ->-] (4,-1) -- (2.5,-1);
\draw  [thick, ->-] (4,-1) -- (4,0.5);
\node  [blue] at (3.7,-2.1) {\scriptsize $\widetilde{Q}$};
\node at (4,-3.2) {\scriptsize $\nu_{2}$};
\draw  [thick, red] (3.9,0.3) -- (4.1,0.3);
\draw  [thick, red] (3.9,0.25) -- (4.1,0.25);
\draw  [thick, ->-] (4,-1) -- (5,-2);
\draw  [thick, ->-] (5,-3.5) -- (5,-2);
\draw  [thick, blue, dashed] (5,-3.5) to [out=180,in=180] (4,0.5);
\draw  [thick, red] (4.9,-3.3) -- (5.1,-3.3);
\draw  [thick, red] (4.9,-3.25) -- (5.1,-3.25);
\draw  [thick, ->-] (6.5,-2) -- (5,-2);
\draw  [thick, ->-] (6.5,-2) -- (6.5,-0.5);
\node  [blue] at (6.2,-3.1) {\scriptsize $\widetilde{Q}$};
\node at (6.5,-4.2) {\scriptsize $\nu_{3}$};
\draw  [thick, red] (6.4,-0.75) -- (6.6,-0.75);
\draw  [thick, red] (6.4,-0.7) -- (6.6,-0.7);
\draw  [thick, ->-] (6.5,-2) -- (7.5,-3);
\draw  [thick, ->-] (7.5,-4.5) -- (7.5,-3);
\draw  [thick, blue, dashed] (7.5,-4.5) to [out=180,in=180] (6.5,-0.5);
\draw  [thick, red] (7.4,-4.3) -- (7.6,-4.3);
\draw  [thick, red] (7.4,-4.25) -- (7.6,-4.25);
\draw  [thick, ->-] (8.25,-3) -- (7.5,-3);
\draw  [thick, dashed]  (8.25,-3) -- (9.25,-3);
\draw  [thick, ->-] (10,-3) -- (9.25,-3);
\draw  [thick, ->-] (10,-3) -- (10,-1.5);
\node  [blue] at (9.7,-4.1) {\scriptsize $\widetilde{Q}$};
\node at (9.8,-5.2) {\scriptsize $\nu_{k-1}$};
\draw  [thick, red] (9.9,-1.7) -- (10.1,-1.7);
\draw  [thick, red] (9.9,-1.75) -- (10.1,-1.75);
\draw  [thick, ->-] (10,-3) -- (11,-4);
\draw  [thick, ->-] (12.5,-4) -- (11,-4);
\draw  [thick, ->-] (11,-5.5) -- (11,-4);
\draw  [thick, blue, dashed] (11,-5.5) to [out=180,in=180] (10,-1.5);
\draw  [thick, red] (10.9,-5.25) -- (11.1,-5.25);
\draw  [thick, red] (10.9,-5.3) -- (11.1,-5.3);
\draw  [thick, ->-] (12.5,-4) -- (12.5,-2.5);
\node  [blue] at (12.2,-5.1) {\scriptsize $\widetilde{Q}$};
\node at (12.5,-6.2) {\scriptsize $\nu_{k}$};
\draw  [thick, red] (12.4,-2.75) -- (12.6,-2.75);
\draw  [thick, red] (12.4,-2.7) -- (12.6,-2.7);
\draw  [thick, ->-] (12.5,-4) -- (13.5,-5);
\draw  [thick, ->-] (15,-5) -- (13.5,-5);
\draw  [thick, blue, dashed] (13.5,-6.5) to [out=180,in=180] (12.5,-2.5);
\draw  [thick, ->-] (13.5,-6.5) -- (13.5,-5);
\draw  [thick, red] (13.4,-6.2) -- (13.6,-6.2);
\draw  [thick, red] (13.4,-6.25) -- (13.6,-6.25);
\end{tikzpicture}
\caption{The dual-toric diagram for the $(2,0)_k$ theory on $\mathbb R_{\epsilon_1,\epsilon_2}\times T^2$. The preferred direction is along the compactified circle and is highlighted in red, and the wrapping around the circle is indicated by blue, along with its partition $\nu_b$ and the associated fugacity $\widetilde{Q}$. This is the same strip geometry as in Fig.~\ref{fig: U(k)^N building} but corresponding to a theory with a single node. There are no non-trivial framing factors.}\label{fig: (2,0) A_k diag}
\end{center}
\end{figure}

When $S^1_{R_6}$ is considered as the M-theory circle, the  partition function of interest coincides with the partition function on a generic point of the Coulomb brach for 5D $\U(N)$ MSYM on $\mathbb{R}^4_{\epsilon_1,\epsilon_2}\times S^1_{R_5}$. The latter can also be readily calculated using the refined topological vertex formalism and specifically the strip geometry of App.~\ref{AppA}; one needs to glue the top and bottom vertical edges to account for the compact nature of $S^1_{R_5}$ by identifying $\nu_{b}^{(1)} =\nu_{b}^{(2)} = \nu_b$ and including an edge factor $\sum_{\nu_b}(-\widetilde{Q})^{|\nu_b|}$ as shown in Fig.~\ref{fig: (2,0) A_k diag}.

The fugacities appearing in the dual-toric diagram are now identified with the following quantities
\begin{equation}
  \widetilde{Q}_m = e^{ i R_5 \tilde{t}_m} 
  \;,\qquad
\widetilde{Q}_\tau = e^{\tilde{t}_e} = \widetilde{Q} \widetilde{Q}_m\;, \qquad \widetilde{Q}_{t_{f_b}} = \widetilde{Q}_{fb} \widetilde{Q}_m\;,\qquad \widetilde{\mathbb{Q}}_{bc}= e^{iR_5 \tilde{t}_{bc}}\;.
\end{equation}
As we are using the same building block for both 4D and 6D calculations, we distinguish the latter by using tildes. The partition function will moreover be a function of the $\Omega$-deformation parameters, which are related to the refinement parameters of the topological string as
\begin{align}
\tilde{\mathfrak{q}} = e^{-i  R_5 \tilde{\epsilon}_1}\;,\qquad \tilde{\mathfrak{t}}= e^{ i R_5 \tilde{\epsilon}_2}\;.
\end{align}

One can then directly import the results of  \eqref{circinst}, \eqref{circpert} for $N=1$ to obtain\footnote{In the language of App.~\ref{AppC} this configuration corresponds to a single node and all quantities would be accompanied by a superscript $(0)$, which we will however drop from the subsequent expressions.}
\begin{align}
Z_{\rm 6D} = Z_{\text{6D,1-loop}}Z_{\text{6D,inst}}\;,
\end{align}
with 
\begin{align}\label{eq:(2,0)_k PF}
Z_{\text{6D,inst}}&=\sum_{\nu}(-\widetilde{Q})^{\sum_{b=1}^k|\nu_b|} \mathcal{D}_k\left(\nu,\nu\right)\cr
&=\sum_{\nu}\widetilde{Q}_\tau^{\sum_{b=1}^k|\nu_b|}\left(\widetilde{Q}_m\sqrt{\frac{\tilde{\mathfrak{t}}}{\tilde{\mathfrak{q}}}}\right)^{-\sum_{b=1}^k|\nu_b|} \prod_{1\leq b\leq c \leq k} \frac{\mathcal{N}_{\nu_c \nu_b}\left(\widetilde{\mathbb{Q}}_{bc}\widetilde{Q}_m \sqrt{\frac{\tilde{\mathfrak{t}}}{\tilde{\mathfrak{q}}}};\tilde{\mathfrak{t}},\tilde{\mathfrak{q}}\right)}{\mathcal{N}_{\nu_c \nu_b}\left(\widetilde{\mathbb{Q}}_{bc};\tilde{\mathfrak{t}},\tilde{\mathfrak{q}}\right)}\cr
&\times  \prod_{1\leq b<c \leq k}\frac{\mathcal{N}_{\nu_c \nu_b}\left(\widetilde{\mathbb{Q}}_{bc}\widetilde{Q}_m^{-1} \sqrt{\frac{\tilde{\mathfrak{t}}}{\tilde{\mathfrak{q}}}};\tilde{\mathfrak{t}},\tilde{\mathfrak{q}}\right)}{\mathcal{N}_{\nu_c \nu_b}\left(\widetilde{\mathbb{Q}}_{bc}\frac{\tilde{\mathfrak{t}}}{\tilde{\mathfrak{q}}};\tilde{\mathfrak{t}},\tilde{\mathfrak{q}}\right)}\;,\cr
Z_{\text{6D,1-loop}}&=\widehat{\mathcal{W}}_k(\emptyset,\emptyset)\cr
 &=\prod_{i,j=1}^{\infty}\prod_{1\leq b\leq c\leq k} \frac{\left(1-\widetilde{\mathbb{Q}}_{b c}\widetilde{Q}_{m}\tilde{\mathfrak{q}}^{j-\frac{1}{2}}\tilde{\mathfrak{t}}^{i-\frac{1}{2}}\right)}{\left(1-\widetilde{\mathbb{Q}}_{b c}\tilde{\mathfrak{q}}^{i}\tilde{\mathfrak{t}}^{j-1}\right)}\prod_{1\leq b<c\leq k}\frac{\left(1-\widetilde{Q}_{m}^{-1}\widetilde{ \mathbb{Q}}_{b c}\tilde{\mathfrak{q}}^{j-\frac{1}{2}}\tilde{\mathfrak{t}}^{i-\frac{1}{2}}\right)}{\left(1-\widetilde{\mathbb{Q}}_{b c}\tilde{\mathfrak{q}}^{i-1}\tilde{\mathfrak{t}}^{j}\right)}\;.
\end{align}
This concludes the derivation of the 6D $(2,0)_k$ partition function.

\end{appendix}

\newpage



\bibliography{deconstruction}

\begin{thebibliography}{10}
\ifx\href\asklfhas\newcommand{\href}[2]{#2}\fi
\ifx\arxivref\asklfhas\newcommand{\arxivref}[2]{\href{http://arxiv.org/abs/#1}{#2}}\fi
\ifx\doiref\asklfhas\newcommand{\doiref}[2]{\href{http://dx.doi.org/#1}{#2}}\fi
\parskip 0pt
\normalsize

\bibitem{Romelsberger:2005eg}
C.~Romelsberger,
\textit{``{Counting chiral primaries in N = 1, d=4 superconformal field
  theories}''},
\doiref{10.1016/j.nuclphysb.2006.03.037}{Nucl.~Phys. \textbf{B747}, 329
  (2006)\ignorespaces}\ignorespaces,
\normalsize{\texttt{\arxivref{hep-th/0510060}{hep-th/0510060}}}\ignorespaces
\bibitem{Kinney:2005ej}
J.~Kinney, J.~M. Maldacena, S.~Minwalla \& S.~Raju,
\textit{``{An Index for 4 dimensional super conformal theories}''},
\doiref{10.1007/s00220-007-0258-7}{Commun.~Math.~Phys. \textbf{275}, 209
  (2007)\ignorespaces}\ignorespaces,
\normalsize{\texttt{\arxivref{hep-th/0510251}{hep-th/0510251}}}\ignorespaces
\bibitem{Bhattacharya:2008zy}
J.~Bhattacharya, S.~Bhattacharyya, S.~Minwalla \& S.~Raju,
\textit{``{Indices for Superconformal Field Theories in 3,5 and 6
  Dimensions}''},
\doiref{10.1088/1126-6708/2008/02/064}{JHEP \textbf{0802}, 064
  (2008)\ignorespaces}\ignorespaces,
\normalsize{\texttt{\arxivref{0801.1435}{arXiv:0801.1435}}}\ignorespaces
\bibitem{Pestun:2007rz}
V.~Pestun,
\textit{``{Localization of gauge theory on a four-sphere and supersymmetric
  Wilson loops}''},
\doiref{10.1007/s00220-012-1485-0}{Commun.~Math.~Phys. \textbf{313}, 71
  (2012)\ignorespaces}\ignorespaces,
\normalsize{\texttt{\arxivref{0712.2824}{arXiv:0712.2824}}}\ignorespaces
\bibitem{Pestun:2016zxk}
V.~Pestun et~al.,
\textit{``{Localization Techniques in Quantum Field Theories}''},
\normalsize{\texttt{\arxivref{1608.02952}{arXiv:1608.02952}}}\ignorespaces
\bibitem{Kallen:2012cs}
J.~Källén \& M.~Zabzine,
\textit{``{Twisted supersymmetric 5D Yang-Mills theory and contact
  geometry}''},
\doiref{10.1007/JHEP05(2012)125}{JHEP \textbf{1205}, 125
  (2012)\ignorespaces}\ignorespaces,
\normalsize{\texttt{\arxivref{1202.1956}{arXiv:1202.1956}}}\ignorespaces
\bibitem{Kallen:2012va}
J.~Källén, J.~Qiu \& M.~Zabzine,
\textit{``{The perturbative partition function of supersymmetric 5D Yang-Mills
  theory with matter on the five-sphere}''},
\doiref{10.1007/JHEP08(2012)157}{JHEP \textbf{1208}, 157
  (2012)\ignorespaces}\ignorespaces,
\normalsize{\texttt{\arxivref{1206.6008}{arXiv:1206.6008}}}\ignorespaces
\bibitem{Kim:2012ava}
H.-C. Kim \& S.~Kim,
\textit{``{M5-branes from gauge theories on the 5-sphere}''},
\doiref{10.1007/JHEP05(2013)144}{JHEP \textbf{1305}, 144
  (2013)\ignorespaces}\ignorespaces,
\normalsize{\texttt{\arxivref{1206.6339}{arXiv:1206.6339}}}\ignorespaces
\bibitem{Kim:2012qf}
H.-C. Kim, J.~Kim \& S.~Kim,
\textit{``{Instantons on the 5-sphere and M5-branes}''},
\normalsize{\texttt{\arxivref{1211.0144}{arXiv:1211.0144}}}\ignorespaces
\bibitem{Kallen:2012zn}
J.~Källén, J.~A. Minahan, A.~Nedelin \& M.~Zabzine,
\textit{``{$N^3$-behavior from 5D Yang-Mills theory}''},
\doiref{10.1007/JHEP10(2012)184}{JHEP \textbf{1210}, 184
  (2012)\ignorespaces}\ignorespaces,
\normalsize{\texttt{\arxivref{1207.3763}{arXiv:1207.3763}}}\ignorespaces
\bibitem{Jafferis:2012iv}
D.~L. Jafferis \& S.~S. Pufu,
\textit{``{Exact results for five-dimensional superconformal field theories
  with gravity duals}''},
\doiref{10.1007/JHEP05(2014)032}{JHEP \textbf{1405}, 032
  (2014)\ignorespaces}\ignorespaces,
\normalsize{\texttt{\arxivref{1207.4359}{arXiv:1207.4359}}}\ignorespaces
\bibitem{Kim:2013nva}
H.-C. Kim, S.~Kim, S.-S. Kim \& K.~Lee,
\textit{``{The general M5-brane superconformal index}''},
\normalsize{\texttt{\arxivref{1307.7660}{arXiv:1307.7660}}}\ignorespaces
\bibitem{Bern:2012di}
Z.~Bern, J.~J. Carrasco, L.~J. Dixon, M.~R. Douglas, M.~von~Hippel \&
  H.~Johansson,
\textit{``{D=5 maximally supersymmetric Yang-Mills theory diverges at six
  loops}''},
\doiref{10.1103/PhysRevD.87.025018}{Phys.~Rev. \textbf{D87}, 025018
  (2013)\ignorespaces}\ignorespaces,
\normalsize{\texttt{\arxivref{1210.7709}{arXiv:1210.7709}}}\ignorespaces
\bibitem{Douglas:2010iu}
M.~R. Douglas,
\textit{``{On D=5 super Yang-Mills theory and (2,0) theory}''},
\doiref{10.1007/JHEP02(2011)011}{JHEP \textbf{1102}, 011
  (2011)\ignorespaces}\ignorespaces,
\normalsize{\texttt{\arxivref{1012.2880}{arXiv:1012.2880}}}\ignorespaces
\bibitem{Lambert:2010iw}
N.~Lambert, C.~Papageorgakis \& M.~Schmidt-Sommerfeld,
\textit{``{M5-Branes, D4-Branes and Quantum 5D super-Yang-Mills}''},
\doiref{10.1007/JHEP01(2011)083}{JHEP \textbf{1101}, 083
  (2011)\ignorespaces}\ignorespaces,
\normalsize{\texttt{\arxivref{1012.2882}{arXiv:1012.2882}}}\ignorespaces
\bibitem{Papageorgakis:2014dma}
C.~Papageorgakis \& A.~B. Royston,
\textit{``{Revisiting Soliton Contributions to Perturbative Amplitudes}''},
\doiref{10.1007/JHEP09(2014)128}{JHEP \textbf{1409}, 128
  (2014)\ignorespaces}\ignorespaces,
\normalsize{\texttt{\arxivref{1404.0016}{arXiv:1404.0016}}}\ignorespaces
\bibitem{Kim:2016usy}
S.~Kim \& K.~Lee,
\textit{``{Indices for 6 dimensional superconformal field theories}''},
\normalsize{\texttt{\arxivref{1608.02969}{arXiv:1608.02969}}}\ignorespaces,
in \textit{``{Localization Techniques in Quantum Field Theories}''}
\bibitem{Lockhart:2012vp}
G.~Lockhart \& C.~Vafa,
\textit{``{Superconformal Partition Functions and Non-perturbative Topological
  Strings}''},
\normalsize{\texttt{\arxivref{1210.5909}{arXiv:1210.5909}}}\ignorespaces
\bibitem{Haghighat:2013gba}
B.~Haghighat, A.~Iqbal, C.~Kozçaz, G.~Lockhart \& C.~Vafa,
\textit{``{M-Strings}''},
\doiref{10.1007/s00220-014-2139-1}{Commun.~Math.~Phys. \textbf{334}, 779
  (2015)\ignorespaces}\ignorespaces,
\normalsize{\texttt{\arxivref{1305.6322}{arXiv:1305.6322}}}\ignorespaces
\bibitem{Aganagic:2003db}
M.~Aganagic, A.~Klemm, M.~Marino \& C.~Vafa,
\textit{``{The Topological vertex}''},
\doiref{10.1007/s00220-004-1162-z}{Commun.~Math.~Phys. \textbf{254}, 425
  (2005)\ignorespaces}\ignorespaces,
\normalsize{\texttt{\arxivref{hep-th/0305132}{hep-th/0305132}}}\ignorespaces
\bibitem{Iqbal:2007ii}
A.~Iqbal, C.~Kozcaz \& C.~Vafa,
\textit{``{The Refined topological vertex}''},
\doiref{10.1088/1126-6708/2009/10/069}{JHEP \textbf{0910}, 069
  (2009)\ignorespaces}\ignorespaces,
\normalsize{\texttt{\arxivref{hep-th/0701156}{hep-th/0701156}}}\ignorespaces
\bibitem{Beem:2014kka}
C.~Beem, L.~Rastelli \& B.~C. van~Rees,
\textit{``{$ \mathcal{W} $ symmetry in six dimensions}''},
\doiref{10.1007/JHEP05(2015)017}{JHEP \textbf{1505}, 017
  (2015)\ignorespaces}\ignorespaces,
\normalsize{\texttt{\arxivref{1404.1079}{arXiv:1404.1079}}}\ignorespaces
\bibitem{Beem:2015aoa}
C.~Beem, M.~Lemos, L.~Rastelli \& B.~C. van~Rees,
\textit{``{The (2, 0) superconformal bootstrap}''},
\doiref{10.1103/PhysRevD.93.025016}{Phys.~Rev. \textbf{D93}, 025016
  (2016)\ignorespaces}\ignorespaces,
\normalsize{\texttt{\arxivref{1507.05637}{arXiv:1507.05637}}}\ignorespaces
\bibitem{Aharony:1997an}
O.~Aharony, M.~Berkooz \& N.~Seiberg,
\textit{``{Light cone description of (2,0) superconformal theories in
  six-dimensions}''},
Adv.~Theor.~Math.~Phys. \textbf{2}, 119 (1998)\ignorespaces\ignorespaces,
\normalsize{\texttt{\arxivref{hep-th/9712117}{hep-th/9712117}}}\ignorespaces
\bibitem{ArkaniHamed:2001ie}
N.~Arkani-Hamed, A.~G. Cohen, D.~B. Kaplan, A.~Karch \& L.~Motl,
\textit{``{Deconstructing (2,0) and little string theories}''},
\doiref{10.1088/1126-6708/2003/01/083}{JHEP \textbf{0301}, 083
  (2003)\ignorespaces}\ignorespaces,
\normalsize{\texttt{\arxivref{hep-th/0110146}{hep-th/0110146}}}\ignorespaces
\bibitem{Lambert:2012qy}
N.~Lambert, C.~Papageorgakis \& M.~Schmidt-Sommerfeld,
\textit{``{Deconstructing (2,0) Proposals}''},
\doiref{10.1103/PhysRevD.88.026007}{Phys.~Rev. \textbf{D88}, 026007
  (2013)\ignorespaces}\ignorespaces,
\normalsize{\texttt{\arxivref{1212.3337}{arXiv:1212.3337}}}\ignorespaces
\bibitem{Benvenuti:2006qr}
S.~Benvenuti, B.~Feng, A.~Hanany \& Y.-H. He,
\textit{``{Counting BPS Operators in Gauge Theories: Quivers, Syzygies and
  Plethystics}''},
\doiref{10.1088/1126-6708/2007/11/050}{JHEP \textbf{0711}, 050
  (2007)\ignorespaces}\ignorespaces,
\normalsize{\texttt{\arxivref{hep-th/0608050}{hep-th/0608050}}}\ignorespaces
\bibitem{Howe:1997ue}
P.~S. Howe, N.~D. Lambert \& P.~C. West,
\textit{``{The Selfdual string soliton}''},
\doiref{10.1016/S0550-3213(97)00750-5}{Nucl.~Phys. \textbf{B515}, 203
  (1998)\ignorespaces}\ignorespaces,
\normalsize{\texttt{\arxivref{hep-th/9709014}{hep-th/9709014}}}\ignorespaces
\bibitem{Baggio:2014sna}
M.~Baggio, V.~Niarchos \& K.~Papadodimas,
\textit{``{Exact correlation functions in $SU(2)$ $\mathcal N=2$ superconformal
  QCD}''},
\doiref{10.1103/PhysRevLett.113.251601}{Phys.~Rev.~Lett. \textbf{113}, 251601
  (2014)\ignorespaces}\ignorespaces,
\normalsize{\texttt{\arxivref{1409.4217}{arXiv:1409.4217}}}\ignorespaces
\bibitem{Baggio:2015vxa}
M.~Baggio, V.~Niarchos \& K.~Papadodimas,
\textit{``{On exact correlation functions in SU(N) $ \mathcal{N}=2 $
  superconformal QCD}''},
\doiref{10.1007/JHEP11(2015)198}{JHEP \textbf{1511}, 198
  (2015)\ignorespaces}\ignorespaces,
\normalsize{\texttt{\arxivref{1508.03077}{arXiv:1508.03077}}}\ignorespaces
\bibitem{Gerchkovitz:2016gxx}
E.~Gerchkovitz, J.~Gomis, N.~Ishtiaque, A.~Karasik, Z.~Komargodski \& S.~S.
  Pufu,
\textit{``{Correlation Functions of Coulomb Branch Operators}''},
\normalsize{\texttt{\arxivref{1602.05971}{arXiv:1602.05971}}}\ignorespaces
\bibitem{Baggio:2016skg}
M.~Baggio, V.~Niarchos, K.~Papadodimas \& G.~Vos,
\textit{``{Large-N correlation functions in ${\cal N} = 2$ superconformal
  QCD}''},
\normalsize{\texttt{\arxivref{1610.07612}{arXiv:1610.07612}}}\ignorespaces
\bibitem{Rodriguez-Gomez:2016ijh}
D.~Rodriguez-Gomez \& J.~G. Russo,
\textit{``{Large N Correlation Functions in Superconformal Field Theories}''},
\doiref{10.1007/JHEP06(2016)109}{JHEP \textbf{1606}, 109
  (2016)\ignorespaces}\ignorespaces,
\normalsize{\texttt{\arxivref{1604.07416}{arXiv:1604.07416}}}\ignorespaces
\bibitem{Pini:2017ouj}
A.~Pini, D.~Rodriguez-Gomez \& J.~G. Russo,
\textit{``{Large $N$ correlation functions in $\mathcal{N}=2$ superconformal
  quivers}''},
\normalsize{\texttt{\arxivref{1701.02315}{arXiv:1701.02315}}}\ignorespaces
\bibitem{Mitev:2014yba}
V.~Mitev \& E.~Pomoni,
\textit{``{Exact effective couplings of four dimensional gauge theories with
  $\mathcal N=$ 2 supersymmetry}''},
\doiref{10.1103/PhysRevD.92.125034}{Phys.~Rev. \textbf{D92}, 125034
  (2015)\ignorespaces}\ignorespaces,
\normalsize{\texttt{\arxivref{1406.3629}{arXiv:1406.3629}}}\ignorespaces
\bibitem{Fraser:2015xha}
B.~Fraser,
\textit{``{Higher rank Wilson loops in the ${\mathcal{N}}=2{SU}(N)\times
  {SU}(N)$ conformal quiver}''},
\doiref{10.1088/1751-8113/49/2/02LT03}{J.~Phys. \textbf{A49}, 02LT03
  (2016)\ignorespaces}\ignorespaces,
\normalsize{\texttt{\arxivref{1503.05634}{arXiv:1503.05634}}}\ignorespaces
\bibitem{Fiol:2015spa}
B.~Fiol, E.~Gerchkovitz \& Z.~Komargodski,
\textit{``{Exact Bremsstrahlung Function in $N=2$ Superconformal Field
  Theories}''},
\doiref{10.1103/PhysRevLett.116.081601}{Phys.~Rev.~Lett. \textbf{116}, 081601
  (2016)\ignorespaces}\ignorespaces,
\normalsize{\texttt{\arxivref{1510.01332}{arXiv:1510.01332}}}\ignorespaces
\bibitem{Mitev:2015oty}
V.~Mitev \& E.~Pomoni,
\textit{``{Exact Bremsstrahlung and Effective Couplings}''},
\doiref{10.1007/JHEP06(2016)078}{JHEP \textbf{1606}, 078
  (2016)\ignorespaces}\ignorespaces,
\normalsize{\texttt{\arxivref{1511.02217}{arXiv:1511.02217}}}\ignorespaces
\bibitem{Bourdier:2015sga}
J.~Bourdier, N.~Drukker \& J.~Felix,
\textit{``{The $\mathcal{N}=2$ Schur index from free fermions}''},
\doiref{10.1007/JHEP01(2016)167}{JHEP \textbf{1601}, 167
  (2016)\ignorespaces}\ignorespaces,
\normalsize{\texttt{\arxivref{1510.07041}{arXiv:1510.07041}}}\ignorespaces
\bibitem{Pomoni:2013poa}
E.~Pomoni,
\textit{``{Integrability in N=2 superconformal gauge theories}''},
\doiref{10.1016/j.nuclphysb.2015.01.006}{Nucl.~Phys. \textbf{B893}, 21
  (2015)\ignorespaces}\ignorespaces,
\normalsize{\texttt{\arxivref{1310.5709}{arXiv:1310.5709}}}\ignorespaces
\bibitem{Douglas:1996sw}
M.~R. Douglas \& G.~W. Moore,
\textit{``{D-branes, Quivers, and ALE Instantons}''},
\normalsize{\texttt{\arxivref{hep-th/9603167}{hep-th/9603167}}}\ignorespaces
\bibitem{Ooguri:1995wj}
H.~Ooguri \& C.~Vafa,
\textit{``{Two-dimensional black hole and singularities of CY manifolds}''},
\doiref{10.1016/0550-3213(96)00008-9}{Nucl.~Phys. \textbf{B463}, 55
  (1996)\ignorespaces}\ignorespaces,
\normalsize{\texttt{\arxivref{hep-th/9511164}{hep-th/9511164}}}\ignorespaces
\bibitem{Gregory:1997te}
R.~Gregory, J.~A. Harvey \& G.~W. Moore,
\textit{``{Unwinding strings and t duality of Kaluza-Klein and h monopoles}''},
Adv.~Theor.~Math.~Phys. \textbf{1}, 283 (1997)\ignorespaces\ignorespaces,
\normalsize{\texttt{\arxivref{hep-th/9708086}{hep-th/9708086}}}\ignorespaces
\bibitem{Tong:2002rq}
D.~Tong,
\textit{``{NS5-branes, T duality and world sheet instantons}''},
\doiref{10.1088/1126-6708/2002/07/013}{JHEP \textbf{0207}, 013
  (2002)\ignorespaces}\ignorespaces,
\normalsize{\texttt{\arxivref{hep-th/0204186}{hep-th/0204186}}}\ignorespaces
\bibitem{Witten:2009xu}
E.~Witten,
\textit{``{Branes, Instantons, And Taub-NUT Spaces}''},
\doiref{10.1088/1126-6708/2009/06/067}{JHEP \textbf{0906}, 067
  (2009)\ignorespaces}\ignorespaces,
\normalsize{\texttt{\arxivref{0902.0948}{arXiv:0902.0948}}}\ignorespaces
\bibitem{Witten:1997sc}
E.~Witten,
\textit{``{Solutions of four-dimensional field theories via M theory}''},
\doiref{10.1016/S0550-3213(97)00416-1}{Nucl.~Phys. \textbf{B500}, 3
  (1997)\ignorespaces}\ignorespaces,
\normalsize{\texttt{\arxivref{hep-th/9703166}{hep-th/9703166}}}\ignorespaces
\bibitem{Giveon:1997sn}
A.~Giveon \& O.~Pelc,
\textit{``{M theory, type IIA string and 4-D N=1 SUSY SU(N(L)) $\times$
  SU(N(R)) gauge theory}''},
\doiref{10.1016/S0550-3213(97)00687-1}{Nucl.~Phys. \textbf{B512}, 103
  (1998)\ignorespaces}\ignorespaces,
\normalsize{\texttt{\arxivref{hep-th/9708168}{hep-th/9708168}}}\ignorespaces
\bibitem{Gaiotto:2012xa}
D.~Gaiotto, L.~Rastelli \& S.~S. Razamat,
\textit{``{Bootstrapping the superconformal index with surface defects}''},
\doiref{10.1007/JHEP01(2013)022}{JHEP \textbf{1301}, 022
  (2013)\ignorespaces}\ignorespaces,
\normalsize{\texttt{\arxivref{1207.3577}{arXiv:1207.3577}}}\ignorespaces
\bibitem{Bhattacharyya:2007sa}
S.~Bhattacharyya \& S.~Minwalla,
\textit{``{Supersymmetric states in M5/M2 CFTs}''},
\doiref{10.1088/1126-6708/2007/12/004}{JHEP \textbf{0712}, 004
  (2007)\ignorespaces}\ignorespaces,
\normalsize{\texttt{\arxivref{hep-th/0702069}{hep-th/0702069}}}\ignorespaces
\bibitem{Kim:2012tr}
H.-C. Kim \& K.~Lee,
\textit{``{Supersymmetric M5 Brane Theories on $\mathbb{R}$ $\times$
  $\mathbb{CP}^2$}''},
\doiref{10.1007/JHEP07(2013)072}{JHEP \textbf{1307}, 072
  (2013)\ignorespaces}\ignorespaces,
\normalsize{\texttt{\arxivref{1210.0853}{arXiv:1210.0853}}}\ignorespaces
\bibitem{Minwalla:1997ka}
S.~Minwalla,
\textit{``{Restrictions imposed by superconformal invariance on quantum field
  theories}''},
Adv.~Theor.~Math.~Phys. \textbf{2}, 781 (1998)\ignorespaces\ignorespaces,
\normalsize{\texttt{\arxivref{hep-th/9712074}{hep-th/9712074}}}\ignorespaces
\bibitem{Buican:2016hpb}
M.~Buican, J.~Hayling \& C.~Papageorgakis,
\textit{``{Aspects of Superconformal Multiplets in ${\mathrm D}>4$}''},
\doiref{10.1007/JHEP11(2016)091}{JHEP \textbf{1611}, 091
  (2016)\ignorespaces}\ignorespaces,
\normalsize{\texttt{\arxivref{1606.00810}{arXiv:1606.00810}}}\ignorespaces
\bibitem{Dolan:2002zh}
F.~A. Dolan \& H.~Osborn,
\textit{``{On short and semi-short representations for four-dimensional
  superconformal symmetry}''},
\doiref{10.1016/S0003-4916(03)00074-5}{Annals~Phys. \textbf{307}, 41
  (2003)\ignorespaces}\ignorespaces,
\normalsize{\texttt{\arxivref{hep-th/0209056}{hep-th/0209056}}}\ignorespaces
\bibitem{Benvenuti:2010pq}
S.~Benvenuti, A.~Hanany \& N.~Mekareeya,
\textit{``{The Hilbert Series of the One Instanton Moduli Space}''},
\doiref{10.1007/JHEP06(2010)100}{JHEP \textbf{1006}, 100
  (2010)\ignorespaces}\ignorespaces,
\normalsize{\texttt{\arxivref{1005.3026}{arXiv:1005.3026}}}\ignorespaces
\bibitem{Gadde:2011uv}
A.~Gadde, L.~Rastelli, S.~S. Razamat \& W.~Yan,
\textit{``{Gauge Theories and Macdonald Polynomials}''},
\doiref{10.1007/s00220-012-1607-8}{Commun.~Math.~Phys. \textbf{319}, 147
  (2013)\ignorespaces}\ignorespaces,
\normalsize{\texttt{\arxivref{1110.3740}{arXiv:1110.3740}}}\ignorespaces
\bibitem{Gaiotto:2012uq}
D.~Gaiotto \& S.~S. Razamat,
\textit{``{Exceptional Indices}''},
\doiref{10.1007/JHEP05(2012)145}{JHEP \textbf{1205}, 145
  (2012)\ignorespaces}\ignorespaces,
\normalsize{\texttt{\arxivref{1203.5517}{arXiv:1203.5517}}}\ignorespaces
\bibitem{Argyres:1996eh}
P.~C. Argyres, M.~R. Plesser \& N.~Seiberg,
\textit{``{The Moduli space of vacua of N=2 SUSY QCD and duality in N=1 SUSY
  QCD}''},
\doiref{10.1016/0550-3213(96)00210-6}{Nucl.~Phys. \textbf{B471}, 159
  (1996)\ignorespaces}\ignorespaces,
\normalsize{\texttt{\arxivref{hep-th/9603042}{hep-th/9603042}}}\ignorespaces
\bibitem{Dey:2013fea}
A.~Dey, A.~Hanany, N.~Mekareeya, D.~Rodríguez-Gómez \& R.-K. Seong,
\textit{``{Hilbert Series for Moduli Spaces of Instantons on $C^2/Z_n$}''},
\doiref{10.1007/JHEP01(2014)182}{JHEP \textbf{1401}, 182
  (2014)\ignorespaces}\ignorespaces,
\normalsize{\texttt{\arxivref{1309.0812}{arXiv:1309.0812}}}\ignorespaces
\bibitem{Kim:2014kta}
J.~Kim, S.~Kim, K.~Lee \& J.~Park,
\textit{``{Super-Yang-Mills theories on $S^{4} \times \mathbb{R}$}''},
\doiref{10.1007/JHEP08(2014)167}{JHEP \textbf{1408}, 167
  (2014)\ignorespaces}\ignorespaces,
\normalsize{\texttt{\arxivref{1405.2488}{arXiv:1405.2488}}}\ignorespaces
\bibitem{Nekrasov:2003rj}
N.~Nekrasov \& A.~Okounkov,
\textit{``{Seiberg-Witten theory and random partitions}''},
\normalsize{\texttt{\arxivref{hep-th/0306238}{hep-th/0306238}}}\ignorespaces
\bibitem{Nekrasov:2002qd}
N.~A. Nekrasov,
\textit{``{Seiberg-Witten prepotential from instanton counting}''},
\doiref{10.4310/ATMP.2003.v7.n5.a4}{Adv.~Theor.~Math.~Phys. \textbf{7}, 831
  (2004)\ignorespaces}\ignorespaces,
\normalsize{\texttt{\arxivref{hep-th/0206161}{hep-th/0206161}}}\ignorespaces
\bibitem{Alday:2009aq}
L.~F. Alday, D.~Gaiotto \& Y.~Tachikawa,
\textit{``{Liouville Correlation Functions from Four-dimensional Gauge
  Theories}''},
\doiref{10.1007/s11005-010-0369-5}{Lett.~Math.~Phys. \textbf{91}, 167
  (2010)\ignorespaces}\ignorespaces,
\normalsize{\texttt{\arxivref{0906.3219}{arXiv:0906.3219}}}\ignorespaces
\bibitem{Hama:2012bg}
N.~Hama \& K.~Hosomichi,
\textit{``{Seiberg-Witten Theories on Ellipsoids}''},
\doiref{10.1007/JHEP09(2012)033, 10.1007/JHEP10(2012)051}{JHEP \textbf{1209},
  033 (2012)\ignorespaces}\ignorespaces,
\normalsize{\texttt{\arxivref{1206.6359}{arXiv:1206.6359}}}\ignorespaces,
[Addendum: JHEP10,051(2012)]\ignorespaces
\bibitem{Iqbal:2002we}
A.~Iqbal,
\textit{``{All genus topological string amplitudes and five-brane webs as
  Feynman diagrams}''},
\normalsize{\texttt{\arxivref{hep-th/0207114}{hep-th/0207114}}}\ignorespaces
\bibitem{Awata:2005fa}
H.~Awata \& H.~Kanno,
\textit{``{Instanton counting, Macdonald functions and the moduli space of
  D-branes}''},
\doiref{10.1088/1126-6708/2005/05/039}{JHEP \textbf{0505}, 039
  (2005)\ignorespaces}\ignorespaces,
\normalsize{\texttt{\arxivref{hep-th/0502061}{hep-th/0502061}}}\ignorespaces
\bibitem{Iqbal:2012xm}
A.~Iqbal \& C.~Vafa,
\textit{``{BPS Degeneracies and Superconformal Index in Diverse Dimensions}''},
\doiref{10.1103/PhysRevD.90.105031}{Phys.~Rev. \textbf{D90}, 105031
  (2014)\ignorespaces}\ignorespaces,
\normalsize{\texttt{\arxivref{1210.3605}{arXiv:1210.3605}}}\ignorespaces
\bibitem{Nieri:2013vba}
F.~Nieri, S.~Pasquetti, F.~Passerini \& A.~Torrielli,
\textit{``{5D partition functions, q-Virasoro systems and integrable
  spin-chains}''},
\doiref{10.1007/JHEP12(2014)040}{JHEP \textbf{1412}, 040
  (2014)\ignorespaces}\ignorespaces,
\normalsize{\texttt{\arxivref{1312.1294}{arXiv:1312.1294}}}\ignorespaces
\bibitem{Mitev:2014jza}
V.~Mitev, E.~Pomoni, M.~Taki \& F.~Yagi,
\textit{``{Fiber-Base Duality and Global Symmetry Enhancement}''},
\doiref{10.1007/JHEP04(2015)052}{JHEP \textbf{1504}, 052
  (2015)\ignorespaces}\ignorespaces,
\normalsize{\texttt{\arxivref{1411.2450}{arXiv:1411.2450}}}\ignorespaces
\bibitem{Dimofte:2010tz}
T.~Dimofte, S.~Gukov \& L.~Hollands,
\textit{``{Vortex Counting and Lagrangian 3-manifolds}''},
\doiref{10.1007/s11005-011-0531-8}{Lett.~Math.~Phys. \textbf{98}, 225
  (2011)\ignorespaces}\ignorespaces,
\normalsize{\texttt{\arxivref{1006.0977}{arXiv:1006.0977}}}\ignorespaces
\bibitem{Hayashi:2013qwa}
H.~Hayashi, H.-C. Kim \& T.~Nishinaka,
\textit{``{Topological strings and 5d $T_N$ partition functions}''},
\doiref{10.1007/JHEP06(2014)014}{JHEP \textbf{1406}, 014
  (2014)\ignorespaces}\ignorespaces,
\normalsize{\texttt{\arxivref{1310.3854}{arXiv:1310.3854}}}\ignorespaces
\bibitem{Mac2}
D.~R. Grayson \& M.~E. Stillman,
\textit{``Macaulay2, a software system for research in algebraic geometry''}
\bibitem{Bao:2013pwa}
L.~Bao, V.~Mitev, E.~Pomoni, M.~Taki \& F.~Yagi,
\textit{``{Non-Lagrangian Theories from Brane Junctions}''},
\doiref{10.1007/JHEP01(2014)175}{JHEP \textbf{1401}, 175
  (2014)\ignorespaces}\ignorespaces,
\normalsize{\texttt{\arxivref{1310.3841}{arXiv:1310.3841}}}\ignorespaces
\bibitem{Iqbal:2012mt}
A.~Iqbal \& C.~Kozcaz,
\textit{``{Refined Topological Strings and Toric Calabi-Yau Threefolds}''},
\normalsize{\texttt{\arxivref{1210.3016}{arXiv:1210.3016}}}\ignorespaces
\bibitem{Taki:2007dh}
M.~Taki,
\textit{``{Refined Topological Vertex and Instanton Counting}''},
\doiref{10.1088/1126-6708/2008/03/048}{JHEP \textbf{0803}, 048
  (2008)\ignorespaces}\ignorespaces,
\normalsize{\texttt{\arxivref{0710.1776}{arXiv:0710.1776}}}\ignorespaces
\bibitem{Bao:2011rc}
L.~Bao, E.~Pomoni, M.~Taki \& F.~Yagi,
\textit{``{M5-Branes, Toric Diagrams and Gauge Theory Duality}''},
\doiref{10.1007/JHEP04(2012)105}{JHEP \textbf{1204}, 105
  (2012)\ignorespaces}\ignorespaces,
\normalsize{\texttt{\arxivref{1112.5228}{arXiv:1112.5228}}}\ignorespaces
\bibitem{Mitev:2014isa}
V.~Mitev \& E.~Pomoni,
\textit{``{Toda 3-Point Functions From Topological Strings}''},
\doiref{10.1007/JHEP06(2015)049}{JHEP \textbf{1506}, 049
  (2015)\ignorespaces}\ignorespaces,
\normalsize{\texttt{\arxivref{1409.6313}{arXiv:1409.6313}}}\ignorespaces
\end{thebibliography}

\end{document}